\documentclass[a4paper,11pt,hyper]{JHEP3}

\title{Special Geometry of Euclidean Supersymmetry III:
the local r-map, instantons and black holes}

\author{Vicente Cort\'es \\Department of Mathematics and \\
Center for Mathematical Physics\\
University of Hamburg\\E-mail: \email{cortes@math.uni-hamburg.de}}
\author{Thomas Mohaupt\\Theoretical Physics Division\\
Department of Mathematical Sciences \\
University of Liverpool\\E-mail: 
\email{Thomas.Mohaupt@liv.ac.uk}}

\abstract{We define and study projective special para-K\"ahler manifolds
and show that they appear as target manifolds when reducing
five-dimensional vector multiplets coupled to supergravity with 
respect to time. The dimensional reductions with respect to time
and space are carried out in a uniform way using an $\epsilon$-complex
notation. We explain the relation of our formalism to other 
formalisms of special geometry used in the literature. In the 
second part of the paper we investigate instanton solutions and
their dimensional lifting to black holes. We show that the instanton
action, which can be defined after dualising axions into tensor fields,
agrees with the ADM mass of the corresponding black hole. The relation
between actions via Wick rotation, Hodge dualisation and analytic
continuation of axions is discussed.} 


\usepackage{amsmath,amssymb,bbm}
\usepackage[all,cmtip]{xy}

\def\der{\partial}

\newcommand{\ra}{\rangle}

\newcommand{\cL}{{\cal L}}

\newcommand{\cF}{{\cal F}}
\newcommand{\cA}{{\cal A}}

\newcommand{\cV}{{\cal V}}

\newcommand{\prt}{\partial}

\newcommand{\mscr}[1]{\mbox{\scriptsize #1}}

\newcommand{\nnu}{\nonumber}
\newcommand{\ft}[2]{{\textstyle\frac{#1}{#2}}}
\newcommand{\commentout}[1]{{}}

\let\Bbb\relax
\newfont{\Bb }{msbm10 scaled 1000}
\newfont{\Bbb}{msbm10 scaled 1200}
\newfam\eufam%
\font\euzw=eufm10 scaled 1200%
\font\euac=eufm9%
\textfont\eufam=\euzw\scriptfont\eufam=\euac%

\newcommand{\bR}{\mathbb{R}}
\newcommand{\bC}{\mathbb{C}}
\def\bt{\begin{thm}}
\def\et{\end{thm}}
\def\bp{\begin{prop}}
\def\ep{\end{prop}}
\def\bc{\begin{cor}}
\def\ec{\end{cor}}
\def\bl{\begin{lemma}}
\def\el{\end{lemma}}
\def\bd{\begin{dof}}
\def\ed{\end{dof}}

\def\square{\kern1pt\vbox
                 {\hrule height 0.6pt\hbox{\vrule width 0.6pt\hskip 3pt
      \vbox{\vskip 6pt}\hskip 3pt\vrule width 0.6pt}\hrule height 
0.6pt}\kern1pt}

\def\pf{\noindent{\it Proof:\ }}
\def\qed{\hfill\square}

\def\a{\alpha}

\def\e{\epsilon}

\def\g{\gamma}

\def\l{\lambda}
\def\o{\omega}

\def\t{\tau}

\def\O{\Omega}

\newfont{\mcal}{eusm10 scaled \magstep1}

\newfont{\goth}{eufm10 scaled \magstep1}

\font\cmssl=cmss10 at 11 pt

\def\n{\nabla}

\newtheorem{thm}{Theorem}
\newtheorem{prop}{Proposition}
\newtheorem{cor}{Corollary}
\newtheorem{lemma}{Lemma}
\newtheorem{dof}{Definition}

\def\ol{\overline}
\def\ra{\rightarrow}

\def\be{\begin{equation}}
\def\ee{\end{equation}}

\def\re#1{(\ref{#1})}
\def\arr{\begin{array}{rlll}}
\def\ea{\end{array}}
\def\bea{\begin{eqnarray}}
\def\eea{\end{eqnarray}}

\preprint{LTH 830}
\begin{document}

\section{Introduction}

This is the third in a series of papers on the special geometry
of Euclidean supersymmetry. The first two papers \cite{CMMS,CMMS2}
explored the
geometries of rigid vector and hypermultiplets, respectively. 
This paper is devoted to vector multiplets coupled to Euclidean
supergravity. We address three main topics: scalar geometry,
dimensional reduction of five-dimensional supergravity,
and instanton solutions for vector multiplets in four 
dimensions.

In the first part of the paper we introduce projective special para-K\"ahler 
manifolds as quotients of conical (affine) special para-K\"ahler manifolds.
These will turn out later to be the target geometries of Euclidean
vector multiplets coupled to supergravity. Affine special para-K\"ahler 
manifolds were introduced in \cite{CMMS}, where it was shown that
they are precisely the target spaces for rigid Euclidean vector 
multiplets. A {\em conical} special para-K\"ahler manifold is an affine
special para-K\"ahler manifold together with a vector field $\xi$, such
that 
\[
\nabla \xi = D \xi = \mbox{Id} \;,
\]
where $D$ is the Levi-Civita connection, and $\nabla$ is the flat
special connection. The main result of the
first part is Theorem \ref{conicThm}, which provides a canonical realisation
of (simply connected) 
conical special para-K\"ahler manifolds as certain Lagrangian cones.
As a corollary we obtain that the geometry of any conical 
special para-K\"ahler manifold and, hence, of any projective special 
K\"ahler manifold is locally encoded in a para-holomorphic function 
which is homogenous of degree 2. 
Throughout the paper we use a notation involving $\epsilon=\pm 1$,
which allows to treat the scalar
geometries of Euclidean ($\epsilon=+1$)
and Minkowskian ($\epsilon=-1$) supergravity in parallel.

In the second part we work out the dimensional reduction of the
bosonic part of the Lagrangian of vector multiplets coupled to
five-dimensional supergravity. We find that the resulting scalar
manifold of the four-dimensional theory is projective special
K\"ahler for reduction over a space-like direction, and 
projective special {\em para}-K\"ahler for reduction over time.
The projective special $\epsilon$-K\"ahler manifolds obtained in this
way are not generic, because they are fully captured by the homogenous 
{\em cubic} polynomial which defines the five-dimensional theory. In the
case $\epsilon=-1$, it is known that {\em any} choice of a holomorphic
prepotential which is homogenous of degree two and gives rise to a 
non-degenerate metric defines a consistent Minkowskian supergravity 
theory \cite{dWvP84,deWLauVP85}. Starting from a general homogeneous
para-holomorphic prepotential, we derive
the corresponding bosonic Euclidean Lagrangian, which is then found
to be related to the Minkowskian Lagrangian through replacing special 
holomorphic coordinates by special para-holomorphic coordinates and
the holomorphic prepotential by a para-holomorphic prepotential. 
We then show that a non-linear sigma model with projective special
$\epsilon$-K\"ahler target is equivalent to a gauged sigma model
with conical special $\epsilon$-K\"ahler target. For the case $\epsilon=-1$
this construction is part of the superconformal quotient which 
we expect to have a counterpart for Euclidean theories. Finally
we reformulate our constructions in the language of line bundles.
This allows to compare our formulae, which hold for both $\epsilon=1$
and $\epsilon=-1$ to formulae  obtained in the supergravity literature 
for $\epsilon=-1$.

In the third part we investigate solutions of the Euclidean 
field equations for the scalars and the metric
in four dimensions. We start by a general
analysis which is valid for any projective special $\epsilon$-K\"ahler
target. The field equations consist of the harmonic map equation 
for the scalars, and the Einstein equation with the energy-momentum
tensor of the scalars as source. We discuss the relation between
the harmonic map equation and totally geodesic submanifolds of the
target and derive some consequences of the Einstein equation. 
For symmetric target manifolds the description of totally geodesic
submanifolds reduces to an algebraic problem. We illustrate this
method for the projective special para-K\"ahler manifold 
\[
\frac{SL_2(\bR)}{SO_0(1,1)} \times
\frac{SO_0(p+1,q+1)}{SO_0(1,1) \times SO_0(p,q)} \;.
\]
For the rest of the paper we specialise to the case 
$p=q=1$, which is the Euclidean STU model \cite{MohProc08,BCPTvR08}. 
As the simplest
example for our method we construct a solution involving 
only the four-dimensional heterotic dilaton-axion field. This solution
is used to explore features of vector multiplet instanton solutions. 
We find that vector multiplet instantons are quite similar to instanton 
solutions for hypermultiplets \cite{BehGaiLueMahMoh:97,GutSpa:00,VanEtAl1,
ChiGut:09}. 
The most pronounced feature is that the action obtained by dimensional
reduction vanishes when evaluated on instanton solutions. 
A non-zero finite action, is found after dualising 
the axion into an antisymmetric tensor field. Instanton solutions
are charged under the axion and, hence, under the dual antisymmetric
tensor field. The instanton action is proportional to the absolute
value of the instanton charge, and inversely proportional to 
the square of the coupling constant. Moreover, the action of our 
instanton solution is the minimal action for given charge. 
 
When dualising the antisymmetric tensor field back into an axion,
one obtains a boundary term, which we keep as part of the action.
When this boundary is evaluated on instanton solutions, it gives
precisely the instanton action found in the scalar-tensor formulation
of the theory. 
We show that the instanton solution lifts to 
a five-dimensional extremal black hole, and we 
find that the ADM mass of this black hole 
equals the action of the corresponding instanton.
The ADM mass is a boundary term, which is different from the
boundary term obtained by dualising the antisymmetric tensor field,
but which takes the same value when evaluated on solutions.

The Euclidean action obtained by dimensional reduction is not
positive definite, while the dual Euclidean action, where the
axion has been dualised into an antisymmetric tensor field 
is positive definite. We determine all Euclidean and Minkowskian
actions which can be obtained by composing the operations
of dimensional reduction, Wick rotation and Hodge dualisation. 
A detailed discussion of the properties and physical interpretation
of these actions is given.

Finally we show that our explicit instanton solution can be lifted
to a five-brane solution in ten dimensions. Therefore this solution
is relevant for five-brane instanton effects in heterotic string
theory compactified on $K3 \times T^2$.

\section{Affine special $\e$-K\"ahler manifolds \label{ASeKM}}
In this section we briefly review affine special pseudo-K\"ahler manifolds
and affine special para-K\"ahler manifolds, see \cite{ACD,CMMS} and references
therein for more details. We will use the following unified terminology:   
\bd An {\sf $\e$-K\"ahler manifold} $(M,J,g)$ is a pseudo-Riemannian
manifold $(M,g)$ endowed with a parallel skew-symmetric endomorphism
field $J\in \Gamma ({\rm End}\, TM)$ such that $J^2 = \e {\rm Id}$, 
where $\e \in \{ -1,1\}$. 
\ed 
$-1$-K\"ahler manifolds are usually called {\it pseudo-K\"ahler manifolds},
whereas $+1$-K\"ahler manifolds are known as {\it para-K\"ahler manifolds}. 
The signature of the pseudo-Riemannian metric 
$g$ is of the form $(2p,2q)$, in the former case and is  
$(n,n)$ in the latter case, where $2n=\dim M$. In both cases, we have 
a symplectic form $\o$, which is defined by 
\be g = \o (J\cdot ,\cdot )\, ,\; \mbox{i.e.}\quad 
\o = \e g(J\cdot ,\cdot )\, .\ee  
It is called the {\it K\"ahler form}.  
The endomorphism field $J$ has vanishing Nijenhuis
tensor and defines 
on $M$ the structure of an {\it $\e$-complex manifold}, i.e.\ 
complex or para-complex manifold for 
$\e = \pm 1$, respectively. In both cases, we can define the 
notion of a holomorphic function $f : M \ra \bC_\e$ 
with values in the ring of {\it $\e$-complex numbers}   
\be \bC_\e 
:= \bR[i_\e]\, , \quad i_\e^2 = 
\e\, , \ee  
(complex or para-complex numbers for $\e = \pm 1$, respectively). 
A function $f : M \ra \bC_\e$ is called {\it $\e$-holomorphic}, or simply
{\it holomorphic}, if 
$df J =  i_\e df$. More generally, a differentiable map 
$f : (M,J) \ra (M',J')$ between $\e$-complex manifolds is called
{\it holomorphic} if $df J = J'df$.  

\bd An {\sf affine special $\e$-K\"ahler manifold} $(M,J,g,\n )$ is an  
$\e$-K\"ahler manifold $(M,J,g)$ endowed with a flat torsion-free
connection $\n$ such that
\begin{enumerate}
\item[(i)] $\n$ is symplectic with respect to the $\e$-K\"ahler form, 
i.e.\ $\n \o = 0$ and 
\item[(ii)] $\n J$ is a symmetric (1,2)-tensor field, i.e.\  
$(\n_XJ)Y = (\n_YJ)X$ for all $X,Y$.   
\end{enumerate} 
\ed

Let us now recall how such manifolds can be constructed from
suitable immersions into $V = \bC_{\e}^{2n}$. Here $V$ is endowed with:  
\begin{enumerate} 
\item[(i)] the  
standard holomorphic symplectic form 
\be \O  = \sum dz^i\wedge dw_i\, ,\ee
where  
\be \label{zwEqu} (z^i,w_i) = (x^i+i_\e u^i,y_i+ i_\e v_i)\ee
are the standard linear holomorphic coordinates, and
\item[(ii)]  the standard 
real structure, i.e.\ anti-linear involution $\t : V\ra V$, $v\mapsto 
\tau v = \bar{v}$, for which 
$V^\t=\bR^{2n} \subset \bC_{\e}^{2n}$ is 
the subset of real points, i.e.\ fixed points of $\t$. 
\end{enumerate} 
Combining these two data 
one obtains the sesquilinear form 
\be \g:=i_\e \O(\cdot , \tau \cdot )\label{sesquiEqu}\ee 
which is Hermitian-symmetric, i.e.\ 
\be \g (Y,X)= \overline{\g (X,Y)}\, ,\ee
where the overline stands for the $\e$-complex conjugation:
\be \overline{a+i_\e b} = a-i_\e b\, ,\quad a,b \in \bR\, .\ee 
Its real part $g_V:={\rm Re}\, \g$ is an $\e$-K\"ahler metric 
of split signature $(2n,2n)$.
\noindent  

\bd Let $(M,J)$ be a connected $\e$-complex manifold of real dimension $2n$. 
A holomorphic immersion $\phi : M \ra V$ is called {\sf $\e$-K\"ahlerian} 
(respectively, {\sf Lagrangian}) if    
$\phi^*\g_V$ is non-degenerate (respectively, if   $\phi^*\O=0$). 
\ed 

The following results are proven in \cite{ACD,CMMS}:  
\bp 
Let $\phi : M \ra V$ be an $\e$-K\"ahlerian Lagrangian immersion. 
It induces the following data on the $\e$-complex manifold $(M,J)$:
\begin{enumerate}
\item[(i)]  an $\e$-K\"ahler metric $g := \phi^*g_V$ with 
the K\"ahler form
\be \label{omegaEqu} \o = 2\sum d\tilde{x}^i\wedge d\tilde{y}_i \, ,\ee
where 
\be \label{xyEqu} 
\tilde{x}^i := x^i \circ \phi \, ,\quad \tilde{y}_i := y_i\circ  \phi \, ,
\ee 
see \re{zwEqu}, and  
\item[(ii)] a flat torsion-free connection $\n$  such that
the globally defined functions $(\tilde{x}^i,\tilde{y}_i)$  
form a system of $\n$-affine local coordinates near any point of $M$.    
\end{enumerate} 
\ep 

\bt \label{ACDCMMSThm} 
Let $\phi : M \ra V$ be an $\e$-K\"ahlerian Lagrangian immersion
of a connected $\e$-complex manifold $(M,J)$ with induced data $(g,\n )$. 
Then $(M,J,g,\n )$ is an affine special $\e$-K\"ahler manifold. 
Conversely, let $(M,J,g,\n )$ be a simply connected affine special 
$\e$-K\"ahler manifold. Then there exists an $\e$-K\"ahlerian Lagrangian 
immersion $\phi : M \ra V$ which induces the special geometric structures
on $M$. Moreover, the immersion $\phi$ is unique up to an affine 
transformation of $\bC^{2n}_\e$ with linear part in the real 
symplectic group ${\rm Sp}(2n,\bR )$. 
\et 
Given a simply connected affine special 
$\e$-K\"ahler manifold $(M,J,g,\n )$ and a point $p\in M$, 
one can choose the $\e$-K\"ahlerian Lagrangian immersion
$\phi : M\ra V$ in such a way that 
the image $\phi (U)$ of some neighborhood 
$U\subset M$ of $p$ is defined by a system of equations of the form
\be w_i = F_i := \frac{\partial F}{\partial {z^i}}\, ,\ee
where $F=F(z^1, \ldots, z^n)$ is a (locally defined) 
$\e$-holomorphic function of $n$ $\e$-complex variables. 
$F$ is called 
the holomorphic {\it prepotential}.    
The holomorphic functions 
\be \label{zEqu}\tilde{z}^i := z^i \circ \phi|_U : 
U \ra \bC_\e\, ,\quad i = 1,\ldots  ,n\, ,\ee 
form a system of local holomorphic coordinates. 
Such coordinates are called {\it special holomorphic coordinates}, 
whereas the $\n$-affine local coordinates  
$(\tilde{x}^i, \tilde{y}_i)$ are called {\it special affine coordinates}. 
\bp \label{sKsKProp} 
Let $(M,J,g,\n )$ be an affine special $\e$-K\"ahler manifold.  
Then $(M,J,g,\n^J )$ is an affine special $\e$-K\"ahler manifold,
where the connection $\n^J$ is defined by 
\be \n^J := J\circ \n \circ J^{-1} \, .\ee
Moreover, given an $\e$-K\"ahlerian Lagrangian immersion $\phi : M\ra V$, 
which induces the special geometric data on $M$, the functions 
\be \tilde{u}^i := u^i \circ \phi \, ,\quad \tilde{v}_i := v_i\circ  \phi\ee
are special  affine coordinates for the affine special $\e$-K\"ahler
manifold $(M,J,g,\n^{J})$. 
\ep   
\section{Conical special $\e$-K\"ahler manifolds \label{CSeKM}}
\bd A {\sf conical} affine special $\e$-K\"ahler manifold $(M,J,g,\n, \xi )$ 
is an affine special $\e$-K\"ahler manifold $(M,J,g,\n )$ 
endowed with a  
vector field $\xi$ such that
\be \nabla \xi = D\xi = {\rm Id}\, , \ee
where $D$ is the Levi-Civita connection.  
\ed

\bp \label{con1-3Prop} 
Let $(M,J,g,\n, \xi )$ be a conical affine special $\e$-K\"ahler manifold.
Then the following holds: 
\begin{enumerate}
\item[(i)] $L_\xi X = -X$ and  $L_\xi (JX) = -JX$ for 
all $\nabla$-parallel local vector fields $X$,
\item[(ii)] $L_\xi \a = \a$ and $L_\xi (J^*\a) = J^*\a$ for all 
$\nabla$-parallel local $1$-forms $\a$,
\item[(iii)] $L_\xi \o = 2\o$, $L_\xi g = 2g$ and $L_\xi J =0$.
\item[(iv)] $L_{J\xi} \o = 0$, $L_{J\xi} g = 0$ and $L_{J\xi} J =0$.
\end{enumerate} 
\ep
\pf 
To prove the first part of (i), we calculate
\[ L_\xi X = \n_\xi X -\n_X\xi = -\n_X\xi= -X\, .\]
For the second part, we observe that the flat torsionfree connection
$\n^J = J\circ \n \circ J^{-1}$ is related to the connection 
$D$ by the equation
\be \label{nnJDEqu}\n^J = D -S\, ,\ee
where $S=D-\n^J=\n-D$.  
This shows that 
\[ L_\xi (JX) =  \n^J_\xi (JX) -\n^J_{JX}\xi = -\n^J_{JX}\xi = 
-D_{JX}\xi +S_{JX}\xi = -JX \, .\] 
Here we have used that $S\xi = \n \xi -D\xi =0$.  
Item (ii) follows immediately from (i), by calculating 
the Lie derivative of the constant functions $\a (X)$ and
$(J^*\a ) (JX)$, e.g.
\[ 0 = L_\xi(\a (X)) =  (L_\xi\a )(X) +  \a (L_\xi X) = (L_\xi\a )(X)-
\a (X)\, .\]
This shows that $L_\xi\a =\a$ for all $\n$-parallel 1-forms $\a$.
In particular,
\be L_\xi d\tilde{x}^i = d\tilde{x}^i \quad \mbox{and}\quad 
L_\xi d\tilde{y}_i = d\tilde{y}_i\, .\ee
Using \re{omegaEqu}, we obtain 
\[ L_\xi \o = 2 \sum L_\xi  (d\tilde{x}^i\wedge d\tilde{y}_i) = 
2\sum L_\xi  (d\tilde{x}^i)\wedge d\tilde{y}_i + 2\sum d\tilde{x}^i\wedge 
L_\xi d\tilde{y}_i = 2\o \, .\] 
Next we calculate $(L_\xi g)(X,Y)$, with the help of (i) and (ii), 
for two $\n$-parallel vector fields $X$
and $Y$:  
\begin{eqnarray*} (L_\xi g)(X,Y) &=& 
L_\xi (g(X,Y)) - g(L_\xi X,Y) - g(X,L_\xi Y) = 
L_\xi (\o (JX,Y)) +2g(X,Y)\\
 &=&  
(2-1-1)\o (JX,Y) + 2g(X,Y) = 2g(X,Y)\, .\end{eqnarray*} 
This proves (iii), since the Lie derivative of $J=\o^{-1}g$ is 
determined by that of $g$ and $\o$: $L_\xi J = L_\xi (\o^{-1})g + 
\o^{-1}L_\xi g = -2J + 2J = 0$. 
To prove (iv), we observe that the vector field $J\xi$ satisfies  
\[ D (J\xi )= JD\xi = J\]
and is therefore a Killing field, i.e.\ $L_{J\xi} g = 0$.
Similarly, 
\[ \nabla (J\xi )= (D+S)(J\xi )= JD\xi -JS\xi = J\]  
implies that $L_{J\xi }\omega =0$ and, hence, $L_{J\xi }J=0$.     
\qed

\bp \label{conconProp}
Let $(M,J,g,\n, \xi )$ be a conical affine special $\e$-K\"ahler manifold. 
Then $(M,J,g,\n^J, \xi )$ is a conical 
affine special $\e$-K\"ahler manifold.
\ep  

\pf It is sufficient to check that $\n^J\xi = {\rm Id}$. This follows from
\re{nnJDEqu}. \qed 

\bp \label{concoordProp} Let $(M,J,g,\n, \xi )$ be a 
conical affine special 
$\e$-K\"ahler manifold. Then near any point $p\in M$ there 
exists a system of special affine coordinates 
$(q^a) = (\tilde{x}^i,\tilde{y}_i)$, 
$a=1,\ldots, 2n$,    
such that $\xi$ takes the form
\be  \xi = \sum q^a\frac{\partial}{\partial q^a} = \sum \tilde{x}^i 
\frac{\partial}{\partial \tilde{x}^i} + \sum \tilde{y}_i 
\frac{\partial}{\partial \tilde{y}_i}\, .\ee
The special affine coordinates $(q^a)$ are unique up to a linear symplectic
transformation. \label{conProp}
\ep 

\pf 
Let $\xi = \sum \xi^a\partial/\partial q^a$ be the expression for $\xi$ 
with respect to some system of special affine coordinates $(q^a)$. 
{}From Proposition \ref{con1-3Prop} (i), we have that
\[ \sum \frac{\partial \xi^a}{\partial q^b}\frac{\partial}{\partial q^a} = 
\left[\frac{\partial}{\partial q^b},\xi \right] = 
\frac{\partial}{\partial q^b}\, .\]
Therefore, $\xi^a = q^a + c^a$ for some constants $c^a\in \bR$ and
putting ${q'}^{a} := q^a + c^a$ yields special affine coordinates
such that $\xi = \sum {q'}^{a}\partial/\partial {q'}^{a}$. The uniqueness
statement is clear, since, in virtue of Theorem \ref{ACDCMMSThm},  
the special affine coordinates $(q^a)$ we started 
with are unique 
up to an affine transformation  of $\bR^{2n}$ with linear part in the real 
symplectic group ${\rm Sp}(2n,\bR )$. \qed   

\bd Special affine coordinates $(q^a) = (\tilde{x}^i,\tilde{y}_i)$
as in Proposition \ref{conProp} 
are called {\sf conical special affine coordinates}. 
\ed

Let us denote by $\xi^V$ the position 
vector field in the vector space $V=\bC^{2n}_\e$: 
\be \xi^V_p = p\in V \cong T_pV\, .\ee     
\bd Let $(M,J)$ be a connected $\e$-complex manifold of real dimension $2n$. 
A holomorphic immersion is called {\sf conical} if the vector field
$\xi^V$ is tangent along $\phi$, i.e.\ if 
\be \xi^V_{\phi(p)} \in d\phi_p T_pM\ee
for all $p\in M$.   
\ed 
A conical $\e$-K\"ahlerian Lagrangian immersion 
$\phi : M \ra V$ 
induces a smooth vector field $\xi$ on $M$ such that  
\be \label{xiEqu} d\phi_p\xi_p=\xi^V_{\phi(p)}\, .\ee

\bl \label{1stLemma} 
Let $(M,J,g,\n )$ be an affine special $\e$-K\"ahler manifold and 
$\phi : M \ra V$ an $\e$-K\"ahlerian Lagrangian immersion inducing the
data $(g,\n )$ on $M$. If  $\phi$ is conical and $\xi$ is the induced vector
field on $M$, then 
$\xi = \sum \tilde{x}^i\partial /\partial \tilde{x}^i + \sum \tilde{y}_i 
\partial /\partial \tilde{y}_i$, in the special affine coordinates  
$(\tilde{x}^i,\tilde{y}_i)$ and 
\be \label{xiuvEqu} \xi = \sum \tilde{u}^i 
\frac{\partial}{\partial \tilde{u}^i} + \sum \tilde{v}_i 
\frac{\partial}{\partial \tilde{v}_i}\, ,\ee
in the special $\n^J$-affine 
coordinates $(\tilde{u}^i,\tilde{v}_i)$, see Proposition 
\ref{sKsKProp}. 
\el 

\pf These expressions for the induced vector field $\xi$ follow from 
\re{xiEqu}. 
 \qed

\bt \label{conicThm} Let $\phi : M \ra V$ be a conical  $\e$-K\"ahlerian Lagrangian immersion
of a connected $\e$-complex manifold $(M,J)$ with induced 
data $(g,\n , \xi )$. 
Then $(M,J,g,\n ,\xi )$ is a conical affine special $\e$-K\"ahler manifold. 
Moreover, the special affine coordinates $(\tilde{x}^i,\tilde{y}_i)$, defined
in  \re{xyEqu}, are conical and the special $\n^J$-affine coordinates 
$(\tilde{u}^i,\tilde{v}_i)$ are also conical, cf.\ 
Proposition \ref{conconProp}.
Conversely, let $(M,J,g,\n , \xi )$ be a simply connected conical 
affine special 
$\e$-K\"ahler manifold. Then there exists a conical  
$\e$-K\"ahlerian Lagrangian 
immersion $\phi : M \ra V$ which induces the special geometric structures
on $M$. Moreover, the immersion $\phi$ is unique up to a 
linear transformation from the group ${\rm Sp}(2n,\bR )$. 
\et 

\pf Let $\phi : M \ra V$ be a conical  $\e$-K\"ahlerian Lagrangian immersion
of a connected manifold with induced data $(g,\n , \xi )$. According to 
Theorem \ref{ACDCMMSThm}, $(M,J,g,\n)$ is an affine special $\e$-K\"ahler
manifold. By Lemma \ref{1stLemma}, we have that 
$\xi = \sum \tilde{x}^i\partial /\partial \tilde{x}^i + \sum \tilde{y}_i 
\partial /\partial \tilde{y}_i$ with respect to the 
$\n$-affine coordinates $(\tilde{x}^i,\tilde{y}_i)$. 
This shows that $\n\xi={\rm Id}$. 
Similarly, \re{xiuvEqu} shows that $\n^J\xi={\rm Id}$ and, hence, 
by \re{nnJDEqu}, 
\[ D\xi = \frac{1}{2}(\n \xi+ \n^J\xi ) = {\rm Id}\, .\]  
This proves that  $(M,J,g,\n, \xi )$ is a conical affine special $\e$-K\"ahler
manifold, that $(\tilde{x}^i,\tilde{y}_i)$ are conical special affine 
coordinates and that $(\tilde{u}^i,\tilde{v}_i)$ are conical $\n^J$-special 
affine coordinates. 

To prove the converse, let $(M,J,g,\n , \xi )$ be a simply connected conical 
affine special $\e$-K\"ahler manifold. By Theorem \ref{ACDCMMSThm}, there
exists an $\e$-K\"ahlerian Lagrangian 
immersion $\phi : M \ra V$ which induces the special geometric structures
on $M$. Moreover, the immersion $\phi$ is unique up to an affine 
transformation of $\bC^{2n}_\e$ with linear part in the real 
symplectic group ${\rm Sp}(2n,\bR )$. The argument in the proof of 
Proposition \ref{concoordProp}, shows that there exists a translation
$t_v : V \ra V$ by a vector $v\in V$ such that the special
affine coordinates ($x^i\circ \phi_v$, $y_i\circ \phi_v$) associated 
with the $\e$-K\"ahlerian Lagrangian immersion  $\phi_v = t_v\circ \phi =
\phi + v$ are conical. Moreover, the real part 
\be {\rm Re}\, v = \frac{1}{2}(v + \bar{v})\ee
of $v$ is uniquely determined,
whereas the imaginary part
\be {\rm Im}\, v = \frac{1}{2i_\e}(v - \bar{v})\ee   
is arbitrary. By the same argument, there is a unique choice of the 
imaginary part ${\rm Im}\, v$ for which the $\n^J$-affine functions 
($u^i\circ \phi_v$, $v_i\circ \phi_v$) 
are conical special affine coordinates for the conical 
affine special $\e$-K\"ahler manifold
$(M,J,g,\n^J,\xi )$. These conditions mean precisely that the vector field
$d(\phi_v) \xi$ along $\phi_v$ has the components
\[(x^i\circ \phi_v,
y_i\circ \phi_v,
u^i\circ \phi_v,v_i\circ \phi_v)\] 
with respect to the standard basis of the
real vector space $V= \bC_\e^{2n}=\bR^{4n}$, i.e.\  
$d(\phi_v) \xi = \xi^V\circ \phi_v$. 
In other words, there is a unique vector $v\in V$
such that $\phi_v : M \ra V$ is a conical $\e$-K\"ahlerian Lagrangian 
immersion. This shows that a conical $\e$-K\"ahlerian Lagrangian 
immersion exists and 
is unique up to a linear transformation in ${\rm Sp}(2n,\bR )$.  
\qed 

Special holomorphic coordinates $\tilde{z}^i := z^i \circ \phi$ 
associated to a conical $\e$-K\"ahlerian Lagrangian 
immersion $\phi : U \ra V$ of some connected open subset $U \subset M$ 
will be called {\it conical} 
special holomorphic coordinates, cf.\ \re{zEqu}. Let us denote by 
$\tilde{U} \subset \bC_\e^n$ the open 
subset which corresponds to $U\subset M$ under a system of special holomorphic 
coordinates $(z^i)$ and   
and let $F : \tilde{U} \ra \bC_\e$ be a  
corresponding holomorphic prepotential such that  
\be \label{phEqu} \phi (U) = \{ (z,w) \in \bC^{2n}_\e \;|\; z \in \tilde{U}
\quad\mbox{and}\quad w_i = F_i(z)\quad\mbox{for}\quad i = 1,\ldots, n\}\, ,\ee
where $z=(z^1,\ldots,z^n)$ and $w=(w_1,\ldots ,w_n)$. Notice that
$F$ is determined only up to an additive constant. 
\bp 
The holomorphic prepotential $F : \tilde{U}\ra \bC_\e$ 
associated to a system of 
special holomorphic coordinates $(\tilde{z}^i)$ 
can be chosen homogeneous of degree $2$ if and only if the special holomorphic 
coordinates are conical. 
\ep 

\pf It is easy to see that an $\e$-K\"ahlerian Lagrangian 
immersion $\phi : U \ra V$ is conical if and only if for all 
$(z,w)\in \phi (U)$ there exists a neighborhood $W \subset \bC_\e$ of
$1\in\bC_\e$ such that $(\l z, \l w) \in \phi (U)$ for all $\l\in W$.   
This is true if and only if $F_i(\l z)= \l F_i(z)$ for all $\l\in W$, 
see \re{phEqu}, which means that $F_i$ is homogeneous of degree 1. In that case,  
\be \tilde{F} := \frac{1}{2}\sum z^i F_i\ee
is homogeneous of degree 2 and differs from $F$  by a constant. In fact, 
\be \frac{\partial}{\partial z^j} (F-\tilde{F}) = F_j -\frac{1}{2}(F_j + 
\sum z^iF_{ij}) = F_j - \frac{1}{2}(F_j + F_j) = 0\, .\ee
So $\tilde{F}$ is a prepotential which is homogeneous of degree 2. 
Conversely, if $F$ is  homogeneous of degree 2 then the $F_i$ are 
homogeneous of degree 1 and $\phi : U \ra V$ is conical. 
\qed 

\section{Projective special $\e$-K\"ahler manifolds \label{PSeKM}}
Let $(M,J,g,\n , \xi )$ be a conical 
affine special $\e$-K\"ahler manifold of real dimension 2n+2. 
At any point $p\in M$ we consider the subspace 
\be {\cal D}_p = {\rm span}\{ \xi_p , J\xi_p \} \subset T_pM\, .\ee 
The vector fields
$\xi$ and $J\xi$ commute:
\be [\xi ,J\xi ] = L_\xi (J)\xi = 0\, ,\ee   
see Proposition \ref{con1-3Prop} (iii). Therefore ${\cal D}\subset TM$ is an 
integrable distribution of $\e$-complex subspaces, provided that 
$\dim {\cal D}_p=2$ for all $p\in M$. 
In that case, we consider the space of leaves (i.e.\ the space of 
integral surfaces) 
$\bar{M}$ of ${\cal D}$ endowed with the topology induced by the canonical 
quotient map
$\pi : M\ra \bar{M}$. We will assume that  
$\pi : M\ra \bar{M}$ is a holomorphic
submersion onto a {\it Hausdorff} $\e$-complex manifold of real dimension
$2n$. The $\e$-complex structure of $\bar{M}$ is again denoted by $J$. 
The following definition will ensure that ${\cal D}$ is a two-dimensional
distribution and that $\bar{M}$ inherits an $\e$-K\"ahler metric
$\bar{g}$ from the affine special K\"ahler metric $g$. 
\bd \label{regDef} A conical special $\e$-K\"ahler manifold $(M,J,g,\n, \xi )$ is called 
{\sf regular} if the function $g(\xi ,\xi )$ does not vanish
on $M$ and $\pi : M \ra \bar{M}$ is a holomorphic  
submersion (onto a Hausdorff manifold). 
\ed 
The regularity condition implies the orthogonal
decomposition  
$T_pM={\cal D}_p\oplus {\cal D}_p^\perp$ 
 for all $p\in M$.
In particular, $d\pi_p$ maps ${\cal D}_p^\perp$ isomorphically onto 
$T_{\pi (p)}\bar{M}$. 

\bp The $(0,2)$-tensor field
\be \label{h-tensor} h= \frac{g}{g(\xi ,\xi )}-\frac{g(\cdot , \xi )\otimes 
g(\cdot , \xi ) 
- \epsilon g(\cdot , J\xi )\otimes g(\cdot , J\xi )}{g(\xi ,\xi )^2}\ee
on $M$ induces an $\e$-K\"ahler metric $\bar{g}$ on $\bar{M}$, such that
$\pi^*\bar{g}=h$.   
\ep  

\pf The Proposition \ref{con1-3Prop} easily implies that $L_\xi h=L_{J\xi }h=0$. 
This shows that $h=\pi^*\bar{g}$ for a pseudo-Riemannian scalar product 
$\bar{g}$ on $\bar{M}$. Since $J$ is skew-symmetric with respect to
$h$, the induced $\e$-complex structure $J$ on $\bar{M}$ is 
skew-symmetric with respect to the induced metric $\bar{g}$ on $\bar{M}$. 
To prove that $(\bar{M},\bar{g})$ is $\e$-K\"ahler  it suffices to check
that the two-form $\bar{\omega }=\e \bar{g}(J\cdot, \cdot )$ is closed. 
Let $c\in \bR^*$ be a value of the function $g(\xi ,\xi )$.
The equation $g(\xi ,\xi )= c$ defines a smooth hypersurface $S\subset M$,
as we see from 
$$dg(\xi ,\xi )= 2 g(D\xi ,\xi )= 2g(\cdot ,\xi ).$$ 
Since $TS=\xi^\perp \supset {\cal D}^\perp$, it is sufficient to check that 
$\pi^*\bar{\omega}=\e h(J\cdot ,\cdot )$ restricts to a closed form on $S$.
The restriction of $\e h(J\cdot ,\cdot )$ to a two-form on $S$ coincides with 
the restriction of $\frac{1}{c}\omega$, which is closed since $\omega$ is the 
K\"ahlerform of $M$.     
\qed 

The $\e$-K\"ahler manifold $(\bar{M},J, \bar{g})$ is called a 
{\it projective special $\e$-K\"ahler manifold}.   


\section{The universal bundle of a projective special\\ 
$\e$-K\"ahler manifold \label{UBdl}}
\subsection{The Chern connection of the universal bundle 
${\cal U}\ra P(V')$} 
Let us consider the $\e$-complex symplectic vector space 
$V=T^*\bC^{n+1}_\e$ endowed with the $\e$-Hermitian metric (\ref{sesquiEqu}). 
We denote by $V':=\{ v\in V| \gamma (v,v) \neq 0\}\subset V$ the open subset
of non-isotropic vectors and by $P(V')$ the set of $\e$-complex lines
$\bC_\e v$, $v\in V'$.  

Let us first discuss the universal bundle $\pi_{\cal U} : 
{\cal U}\ra P(V')$. The fiber 
${\cal U}_p$ over $p=\bC_\epsilon v \in P(V')$ 
is given by the line $\bC_\epsilon v \subset V$. 
This defines a line subbundle ${\cal U} \subset \underline{V}$ 
of the trivial bundle $\underline{V}:=P(V')\times V \ra P(V')$. 
The $\e$-Hermitian metric $\gamma$ 
on $V$ induces an $\e$-Hermitian metric on ${\cal U}$.

\bl \label{ChernL} 
There exists a unique connection ${\cal D}$ on ${\cal U}$ which satisfies
the following constraints:
\begin{itemize}
\item[(i)] ${\cal D}$ is metric, that is 
$$X\g (v,w)=\g ({\cal D}_Xv,w)+ \g (v,{\cal D}_{\ol{X}}w),$$
for all sections $v,w\in \Gamma ({\cal U})$ of $\cal U$ and all 
$\e$-complex valued 
vector fields $X\in \Gamma (TP(V')\otimes \bC_\e )$ on $P(V')$. 
\item[(ii)] For all $\e$-holomorphic sections $v\in {\cal O}({\cal U})$
and all $Z\in T^{1,0}M$ we have 
$${\cal D}_{\ol{Z}}v=0.$$
\end{itemize}
\el 

The above connection will be called the {\it Chern connection}. 

\pf We give a geometric description of the connection $\cal D$.
Let us denote by $d_Xv$ the ordinary derivative of
a section $v$ of the trivial bundle $\underline{V}$ and by
$\pi_{\cal U}^V$ the orthogonal projection $\underline{V}\ra 
{\cal U}\subset \underline{V}$ with respect to the 
$\e$-Hermitian scalar product $\g$ on $V$. Then $\cal D$
is given by
\be\label{DEqu} {\cal D}_X v := \pi_{\cal U}^Vd_Xv,\ee
where $X$ is a vector field on $P(V')$ and $v$ is a section of
${\cal U}\subset \underline{V}$.  Let us check that
$\cal D$ satisfies (i-ii).\\
(i) For all $v,w\in\Gamma (\underline{V})$ and all $X\in 
\Gamma (TP(V')\otimes \bC_\e)$
we have 
$$d_X\g (v,w) =\g (d_Xv,w)+ \g (v,d_{\ol{X}}w).$$
For $v,w\in \Gamma ({\cal U})$ we may replace $d$ by $\cal D$ in that
formula. This proves (i).\\
(ii) For all $v\in {\cal O}(\underline{V})$ and $Z\in T^{1,0}M$ we have 
$d_{\ol{Z}}v=0$. In particular, ${\cal D}_{\ol{Z}}v=\pi_{\cal U}^Vd_{\ol{Z}}v=0$
for all $v\in  {\cal O}({\cal U})$.  

To prove the uniqueness we consider the difference 
$\Theta := {\cal D}-{\cal D}'$ of two connections ${\cal D}, {\cal D}'$
satisfying (i-ii). The tensor field $\Theta$ verifies
$$\g (\Theta (Z)v,w)= -\g (v,\Theta (\ol{Z})w)$$
for all $Z\in T^{1,0}M$ and $v, w\in \Gamma ({\cal U})$ 
and  
$$\Theta (\ol{Z})u=0$$ 
for all  $Z\in T^{1,0}M$ and $u\in {\cal O}({\cal U})$. 
The second condition implies $\Theta (\ol{Z})=0$, since 
$\Theta (\ol{Z})$ is tensorial. Then the first condition 
implies $\Theta=0$. 
\qed 

\subsection{The pull back of $({\cal U},{\cal D})$ to $M$} 
Now let $(M,J,g,\n ,\xi )$ be a regular conical affine
special $\e$-K\"ahler manifold. Then we have the 
following commutative diagram:
\be \label{diagrEqu}
\xymatrix{
M \ar[r]^{\phi} \ar[d]^{\pi} & V' \ar[d]^{\pi_V} \\
\bar{M} \ar[r]^{\bar{\phi}}& P(V') 
}
\end{equation}
where $\phi$ is a conical $\e$-K\"ahlerian Lagrangian
immersion inducing the special geometric data on $M$ and
$\ol{\phi}$ is the corresponding $\e$-holomorphic Legendrian immersion.  

We denote by ${\cal U}^M := (\ol{\phi}\circ \pi)^*{\cal U}
=(\pi_V\circ \phi )^*{\cal U}$ the 
pull back of the universal bundle under the map $M \ra P(V')$. 
Let us recall that given a smooth map $f : M \ra N$ between
smooth manifolds $M$ and $N$ we can pull back any vector
bundle $\pi_E : E \ra N$ on $N$ to a vector bundle $f^*E$ on $M$. 
The total space of $f^*E$ is defined by
$$f^*E := \{ (e,m)\in E\times M|  \pi_E (e)=f(m)\}$$
and the bundle projection $f^*E \ra M$ is the restriction
of the canonical projection\linebreak[4] 
$E\times M \ra M$ to $f^*E \subset E\times M$. Any section
$s\in \Gamma (E)$ gives rise to a section 
$f^*s\in \Gamma (f^*E)$ defined by
$$(f^*s)(m)=s(f(m)).$$ In particular, the pull back of any trivial
bundle is again trivial. Given a connection $D$ in $E$, the pull back
connection $f^*D$ in $f^*E$ is defined by
$$(f^*D)_Xf^*s := D_{dfX}s.$$
Notice that $(f^*E)_m=E_{f(m)}\times \{ m\} \cong E_{f(m)}$ for all
$m\in M$. 

We can consider $\phi : M \ra V$ as an $\e$-holomorphic 
section of ${\cal U}^M$. This follows from 
$$\phi (m) \in V',\quad \pi_V \phi (m) = \ol{\phi} (\pi (m)),$$  
since $\pi_{\cal U}$ coincides with $\pi_V$ on the 
complement $V'={\cal U}\setminus  \mathbf{0}$ of the 
zero section in $\pi_{\cal U} :
{\cal U} \ra P(V')$. (${\cal U}$ is precisely the blow up of the open 
cone $V'$ at the origin.)  

Next we consider the pull back via 
$\pi_V \circ \phi =
\ol{\phi }\circ \pi : M \ra P(V')$ 
of the connection ${\cal D}$ on ${\cal U}\ra P(V')$ to a connection
on ${\cal U}^M\ra M$. We shall denote all pull backs of $\cal D$ 
by the same letter $\cal D$. 
Since $\phi$ is an $\e$-holomorphic section
the pull back connection satisfies 
$${\cal D}_i\phi = i_\e A_i^h\phi ,\quad {\cal D}_{\ol{i}}\phi =0,$$
where ${\cal D}_i:= 
{\cal D}_{\partial_i}$ and ${\cal D}_{\ol{i}}:= 
{\cal D}_{\partial_{\ol{i}}}$ are derivatives with respect to  
holomorphic and anti-holomorphic coordinates.
\bp The connection one-form $i_\e\sum A_i^hdz^i$ 
of the  pull back connection on ${\cal U}^M$ with respect to the 
$\e$-holomorphic section $\phi$ is given by
$$i_\e A_i^h = \frac{\g (\partial_i\phi ,\phi )}{\g (\phi ,\phi )}=
\frac{\sum (\partial_iz^j\ol{F}_j-\partial_iF_j\ol{z}^j)}{\sum (z^j\ol{F}_j
-F_j\ol{z}^j)},\quad A_{\ol{i}}=0,$$
where $(z^i)$, $i=1,2,\ldots, n+1$, are conical 
special $\e$-holomorphic coordinates and $F$ is the 
corresponding prepotential. 
\ep 

\pf This follows from (\ref{DEqu}). \qed

For future use we express the above pullback connection  
also with respect to the unit section
$\phi_1 := \frac{\phi}{\| \phi \|}$, where $\| \phi \| := 
\sqrt{| \g (\phi ,\phi )|}$.   

\bp The connection one-form $i_\e\sum A_idz^i+i_\e\sum A_{\ol{i}}d\ol{z}^i$ 
of the  pull back connection on ${\cal U}^M$ 
with respect to the unitary section $\phi_1$ is given by
$$A_i = \frac{1}{2}A_i^h,\quad A_{\ol{i}} = \frac{1}{2}
\ol{A_i^h}.$$
\ep 

\pf We compute 
$${\cal D}_i\phi_1 = \partial_i \left( \frac{1}{\| \phi \|}\right) \phi 
+ \frac{1}{\| \phi \|} {\cal D}_i \phi = -\frac{i_\e}{2}A_i^h\phi_1 +i_ \e 
A_i^h\phi_1
=\frac{i_\e}{2}A_i^h\phi_1,$$
$${\cal D}_{\ol{i}}\phi_1 = \partial_{\ol{i}} \left( \frac{1}{\| \phi \|}\right) \phi 
+ \frac{1}{\| \phi \|} {\cal D}_{\ol{i}} \phi = 
-\ol{\frac{i_\e}{2}A_i^h}\phi_1 +0 = \frac{i_\e}{2}\ol{A_i^h}\phi_1.$$
\qed 
\subsection{The pull back of $({\cal U}^M,{\cal D})$ under a
smooth map $f: N \ra M$}
Let $N$ be a smooth manifold with local coordinates $(x^\mu )$ and
$f: N \ra M$ a smooth map into the regular conical affine
special $\e$-K\"ahler manifold $M$. 

\bp 
\begin{itemize}
\item[(i)] 
The connection one-form $i_\e\sum A_\mu^hdx^\mu$ 
of the  pull back connection on $f^*{\cal U}^M$ with respect to the 
pull back $\phi^N=f^*\phi$ of the $\e$-holomorphic section $\phi$ is given by
$$A_\mu^h = \sum \partial_\mu z^iA^h_i.$$ 
Here $\partial_\mu z^j$ stands for $\partial_\mu 
(z^j \circ \phi \circ f)$ and $A^h_i$ is evaluated along $f$. 
\item[(ii)] The connection one-form $i_\e\sum A_\mu dx^\mu$ 
of the  pull back connection on $f^*{\cal U}^M$ 
with respect to the pull back $\phi^N_1=f^*\phi_1$ of the unitary section 
$\phi_1$ is given by
$$A_\mu = \sum \partial_\mu z^iA_i+  \sum \partial_\mu z^{\ol{i}}A_{\ol{i}}.$$
\end{itemize}
\ep 

Next we consider the special case $N=\bar{M}$ and $f=s : \bar{M}\ra M$ 
a section of $\pi : M \ra \bar{M}$. Let us first observe that
${\cal U}^{\bar{M}}:=\ol{\phi}^*{\cal U}= s^*{\cal U}^M$, since 
$\ol{\phi}= \pi_V \circ \phi \circ s$, see (\ref{diagrEqu}). 
\bc The pull back connection ${\cal D}$ on ${\cal U}^{\bar{M}}$ 
satisfies:  
\begin{itemize}
\item[(i)] For every holomorphic section $s$ 
\begin{equation}
\label{HolConn}
{\cal D}_as=i_\e \underset{A^h_a:=}{\underbrace{\partial_a z^iA^h_i}}s
=\frac{\g (\partial_a\phi ,\phi )}{\g (\phi ,\phi )}s=\frac{\sum (\partial_az^j\ol{F}_j-\partial_a F_j\ol{z}^j)}{\sum (z^j\ol{F}_j
-F_j\ol{z}^j)}s,\quad {\cal D}_{\ol{a}}s=0,
\end{equation}
where the derivative $\partial_a = \frac{\partial}{\partial \zeta^a}$
is with respect to local $\e$-holomorphic coordinates on
$\bar{M}$, $\phi$ and $A^h_i$ are evaluated on $s$ and 
the functions $z^j$ and $F_j$ are evaluated
on $\phi \circ s$.   
\item[(ii)] For every unitary section $s_1$ we have
\begin{equation}\label{UniConn}{\cal D}_as_1=: i_\e A_as_1 = 
\frac{\g (\partial_a\phi ,\phi )-\g (\phi ,\partial_{\ol{a}}\phi )}{2\g (\phi ,\phi )}s_1,
\quad  {\cal D}_{\ol{a}}s_1=: i_\e A_{\ol{a}}s_1 =
\frac{\g (\partial_{\ol{a}}\phi ,\phi )
-\g (\phi ,\partial_{a}\phi )}{2\g (\phi ,\phi )}s_1,
\end{equation}
where $\gamma (\phi ,\phi )=\gamma (\phi (s_1) ,\phi (s_1))=\pm 1$ and 
\begin{eqnarray*}&&
\g (\partial_a\phi ,\phi )-\g (\phi ,\partial_{\ol{a}}\phi ) =
-\ol{(\g (\partial_{\ol{a}}\phi ,\phi )
-\g (\phi ,\partial_{a}\phi ))}\\ 
&=& i_\e 
\sum (\partial_az^j\ol{F}_j-\partial_a F_j\ol{z}^j -z^j\partial_a 
\ol {F}_j+F_j\partial_a\ol{z}^j)
=  -i_\e \sum (z^j\stackrel{\leftrightarrow}{\partial_a}\ol{F}_j-F_j\stackrel{\leftrightarrow}{\partial_a}\ol{z}^j)\;,
\end{eqnarray*}
Here we use the notation $a\stackrel{\leftrightarrow}{\partial_\mu} b :=
a \partial_\mu b - (\partial_\mu a) b$. 
(Notice that  $A_{\bar{a}}=\ol{A_a}$ and that 
these formulas can be rewritten in various
ways using that for a unitary section 
$\g (\partial_a\phi ,\phi )=
-\g (\phi ,\partial_{\ol{a}}\phi )$.) 
\end{itemize}
\ec 
Now let $f : N \ra \bar{M}$ be any smooth map from a manifold
$N$ with local coordinates $(x^\mu)$ into the projective 
special $\e$-K\"ahler manifold $\bar{M}$. Pulling back the 
connection $\cal D$ on ${\cal U}^{\bar{M}}$  we get with the above 
notation 
\begin{eqnarray} {\cal D}_\mu f^*s &=:&  i_\e A_\mu^h f^*s 
=  i_\e \partial_\mu \zeta^a A_a^hf^*s\\
 {\cal D}_\mu f^*s_1 &=:&  i_\e A_\mu f^*s_1 
=  i_\e (\partial_\mu \zeta^a A_a + \partial_\mu 
\ol{\zeta}^a A_{\ol{a}})f^*s_1.
\label{U1connST}
\end{eqnarray}

\section{Dimensional reduction of five-dimensional supergravity}

\subsection{The five-dimensional theory}

In \cite{GST} the general Lagrangian for  vector multiplets
coupled 
to five-dimensional supergravity was derived. By dimensional reduction 
on a space-like circle they obtained  four-dimensional 
${\cal N}=2$ vector multiplets coupled to supergravity.
Whereas the five-dimensional couplings are determined 
by very special real geometry, the four-dimensional 
couplings are determined by projective special K\"ahler geometry. 
We will generalise the analysis of \cite{GST} to the case
where the compactification circle is time-like, which 
leads to a theory with Euclidean space-time  signature. 
To compare the effects of space-like and time-like 
dimensional reduction we perform both types of reduction
in parallel. Then,
it is convenient to introduce a parameter $\epsilon$,
which takes the value
$\epsilon =-1$ for reduction over space and $\epsilon =1$
for reduction over time. As we will see in due course,
the geometry of the scalar target space of the four-dimensional
theory is (projective special) $\epsilon$-K\"ahler, and the 
$\epsilon$ introduced
above will turn out to be identical to the one defined 
in section \ref{ASeKM}.


The fields of the five-dimensional theory organise themselves
into the following supermultiplets:
\begin{itemize}
\item
The gravity supermultiplet $(e_{\hat{\mu}}^{\;\;\hat{m}}, 
\psi_{\hat{\mu}}^A, {\cal A}_{\hat{\mu}})$ contains the
f\"unfbein (graviton), two gravitini and the graviphoton. 
\item
A vector multiplet $({\cal A}_{\hat{\mu}}, \Lambda^{A}, \phi)$
consists of a gauge field, a pair of symplectic Majorana spinors and a real
scalar field. We consider a theory with an arbitrary number of
vector multiplets, labeled by the index $x=1,\ldots, n_V^{(5)}$. 
\end{itemize}
The other indices have the following ranges:
$\hat{\mu}, \hat{\nu}, \ldots = 0, \ldots, 4$ 
are five-dimensional world indices,
$\hat{m}, \hat{n}, \ldots = 0, \ldots, 4$ are five-dimensional tangent
space indices and $A=1,2$ is the index of R-symmetry group 
$SU(2)_R$. 
Since the gravity multiplet contributes
an additional gauge field, there are $n_V^{(5)} +1$ gauge fields,
which we denote by 
${\cal A}^i_{\hat{\mu}}$, with $i = 0,\ldots, n_V^{(5)}$.
The corresponding field strengths are ${\cal F}^i_{\hat{\mu} \hat{\nu}}$.

The full Lagrangian is completely determined by the choice
of the scalar manifold $\hat{M}$, which must
be a so-called very special real manifold, i.e., a 
cubic hypersurface \cite{GST}.
The hypersurface is characterised
by a cubic function, the prepotential ${\cal V}$:
\be
\cV :=  
c_{ijk}h^{i}h^{j}h^{k} \ = \
1~,
\label{5Dgeo}
\ee
where $c_{ijk}$ is a real symmetric constant tensor and 
$h^i$ are embedding coordinates 
of the scalar manifold. The physical scalars $\phi^x$ are
obtained by solving the hypersurface constraint
(\ref{5Dgeo}). It turns out to be convenient to work
with constrained fields $h^i$. 
When we refer to them as `five-dimensional scalars',
the constraint (\ref{5Dgeo}) is understood.

In order to identify the scalar geometry of the four-dimensional 
theories obtained by dimensional reduction, we only need 
to reduce the bosonic terms.
Therefore we start from the bosonic part of the 
five-dimensional Lagrangian for supergravity coupled to 
an arbitrary number of vector multiplets \cite{GST}:
\bea
\hat {\bf e}^{-1}\hat\cL &=&  \frac12\hat R-\frac34
a_{ij}\prt_{\hat\mu}h^{i}\prt^{\hat\mu}h^{j}-\frac14
a_{ij}\cF^{i}_{\hat\mu\hat\nu}\cF^{j\hat\mu\hat\nu}
+\frac{\hat {\bf e}^{-1}}{6\sqrt{6}}\,c_{ijk}
\epsilon^{\hat\mu\hat\nu\hat\rho\hat\sigma\hat\lambda}
\cF^{i}_{\hat\mu\hat\nu}\cF^{j}_{\hat\rho\hat\sigma}
\cA^{k}_{\hat\lambda}~.
\label{5Daction}
\eea
Here $\hat{\bf e}$ is the determinant of the f\"unfbein
and $\hat{R}$ the space-time Ricci scalar.
The terms quadratic in the matter
fields contain the field dependent coupling matrix 
$a_{ij}$, which is determined by the prepotential through
\be
a_{ij}\ = \ -\frac13\prt_{h^i}\prt_{h^j}\ln \cV _{|\cV=1}~.
\ee
The explicit expression is:
\be
a_{ij} = -2 \left( \frac{ (ch)_{ij} }{ chhh } - \frac32
\frac{ (chh)_i (chh)_j }{ (chhh)^2 } \right) \;,
\label{a-matrix}
\ee
where we introduced the following notation:
\be
chhh := c_{ijk} h^i h^j h^k \;, \;\;\;
(chh)_i :=  c_{ijk} h^j h^k \;, \;\;\;
(ch)_{ij} :=  c_{ijk} h^k \;.
\ee
The coefficients $c_{ijk}$ of the Chern-Simons terms 
are proportional to the third derivatives of the
prepotential. Note that the sigma model metric for the 
physical scalars $\phi^x$ is the pullback of the tensor 
field $a_{ij}dh^i dh^j$
to the hypersurface ${\cal V}=1$. However, for the purpose of dimensional
reduction it turns out to be convenient to work with 
the constrained scalars $h^i$. 

While the scalar manifold is determined by the constants
$c_{ijk}$, it is understood that the range of the scalars 
$h^i$ has been chosen such that both $a_{ij}$ and its
pull back onto $\hat{M}$ are positive definite. This is
needed in order to ensure that the scalars and gauge fields have
well defined (positive definite) kinetic terms. 

We close this section by pointing out that an interpretation 
of very special real geometry in the framework of affine differential
geometry has been given in \cite{ACDVP}. In that construction the
metric and very special real structure on $\hat{M}$
are induced through a centroaffine embedding into 
$\mathbb{R}^{n_V^{(5)}+1}$, equipped with its 
standard affine structure. The embedding is encoded in 
the real prepotential ${\cal V}$, which plays a similar role
as the holomorphic prepotential in the $\epsilon$-complex
case. We refer to \cite{ACDVP} for more details.

\subsection{Dimensional reduction of the bosonic terms}

We now perform the dimensional reduction of the bosonic
Lagrangian (\ref{5Daction}) with respect to a 
time-like ($\epsilon =1$) or space-like ($\epsilon =-1$) 
direction. A standard Ansatz for the 
f\"unfbein is:
\be
\hat e_{\hat\mu}{}^{\hat m} \ = \ \left(\begin{array}{cc} e^{\sigma} &0\\
e^\sigma \cA_\mu^0 & 
e^{-\sigma/2}e_\mu{}^m\end{array}\right)~,\quad 
\eta_{\hat m\hat
n}=\mbox{diag}\big(-\epsilon,\eta_{mn}=(+,+,+,\epsilon)\big)~.
\ee 
We introduced four-dimensional world indices
$\mu, \nu, \ldots = 1,\ldots, 4$ and four-dimensional tangent
space indices $m, n, \ldots= 1, \ldots, 4$. The compactified direction
is taken to be the 0-direction, for both $\epsilon=\pm1$. 
${\cal A}^0_{\mu}$ is the Kaluza-Klein gauge field,
$\sigma$ is the Kaluza-Klein scalar.

The four-dimensional epsilon tensor is:
\be
\epsilon_{mnpq} \ := \ \epsilon_{0\hat{m} \hat{n} \hat{p} \hat{q}} \;,
\;\;\;{\rm with}\;\;\;
\epsilon^{mnpq} \epsilon_{mnpq} \ = \ 4 ! \; \epsilon \;.
\ee

The 0-components of the five-dimensional 
gauge fields are four-dimensional  scalar fields, 
$m^i :=  \cA^i_0$.

We obtain the following bosonic Lagrangian:
\bea
{\bf e}^{-1}\cL^{\epsilon}&= & \frac12 R
-\frac34 (\prt_\mu\sigma)^2 -\frac34 
\,a_{ij}\prt_\mu h^{i}
\prt^\mu h^{j} +\epsilon \frac12 e^{-2\sigma}
a_{ij}\prt_\mu m^{i}\prt^\mu m^{j}\nnu\\
&&{} + \epsilon  \frac18 e^{3\sigma} (\cF^0_{\mu\nu})^2-\frac14
e^{\sigma}
a_{ij}\cF^{i}_{\mu\nu}\cF^{j\mu\nu} -e^{\sigma}\,\cA^{0\mu}\prt^\nu
a_{ij} m^i\,\cF^j_{\mu\nu}
-\frac12 e^{\sigma}a_{ij} \prt^\mu m^i\prt_\mu m^j
\cA^{0\mu}\cA^0_\mu\nnu\\
&&{}+\frac12 e^{\sigma}a_{ij}\prt^\mu m^i\prt^\nu
m^j\,\cA^0_\mu\cA^0_\nu 
- \epsilon  \frac{{\bf e}^{-1}}{2\sqrt6}c_{ijk}m^{k}
\epsilon^{\mu\nu\rho\sigma}
\cF^{i}_{\mu\nu}\cF^{j}_{\rho\sigma}
~.\label{lpm1}
\eea
Here ${\bf e}$ is the determinant of the vierbein $e_\mu^{\;\;m}$, 
and $R$ is the
four-dimensional Ricci scalar.
We remark that, as expected, the metric of the scalar manifold 
has split signature for $\epsilon =1$. 
This is due to the fact
that the scalars $m^i$ come from the time-like components of the
five-dimensional gauge fields.

The reduced Lagrangian contains terms in which bare
gauge fields appear. Therefore the gauge invariances of the
four-dimensional Lagrangian are not manifest. 
Of course, gauge invariance has not been broken
by the Kaluza-Klein reduction, but it is not manifest in terms of
the gauge fields ${\cal A}^i_{\mu}$. 
Also note that 
through dimensional reduction the reparametrisation symmetry of
the fifth direction has become an additional internal Abelian
gauge symmetry. The corresponding gauge field is the Kaluza-Klein
gauge field ${\cal A}^0_{\mu}$. Therefore 
the number of vector multiplets is increased by
one in dimensional reduction: $n_V^{(4)} = n_V^{(5)} +1$. Since the
four-dimensional ${\cal N}=2$ supergravity multiplet contains one gauge
field, the graviphoton, we expect to find  $n_V^{(4)} +1$
abelian gauge symmetries. 
To make the four-dimensional gauge symmetries manifest
we introduce redefined gauge fields:
\be
 A^{i}_\mu \ := \ \cA^{i}_\mu - m^{i}\!\cA^0_\mu~,\quad
 A^0_\mu \ :=  \ -\cA^0_\mu~.
\ee
We also introduce a new index $I=(0,i)$ and denote the
four-dimensional gauge fields by
$ A^I_{\mu}$. The corresponding field strength $ F^I_{\mu \nu}$ are
invariant under all the $n_V^{(4)}+1$
four-dimensional gauge transformations. The Hodge dual
field strengths are defined as
$\tilde{ F}_{\mu \nu} = \ft12 {\bf e} \epsilon_{\mu \nu \rho \sigma}  
F^{\rho \sigma}$, such that
$\tilde{\tilde{ F}}_{\mu \nu} = \epsilon  F_{\mu \nu}$.

Inserting the new field strengths into the Lagrangian we obtain 
the following terms for the gauge fields:
\bea
{\bf e}^{-1} {\cal L}^{\epsilon}_{\mscr{gauge}} &=& 
e^{3 \sigma} \big(
\ft18 ( \epsilon - 2\,e^{-2\sigma} (amm) )  F^0 \cdot F^0 
+ \ft12 e^{-2\sigma}(am)_i F^i \cdot F^0
- \ft14 e^{-2\sigma}\,a_{ij} F^i \cdot F^j \big)  \nonumber \\ 
 & &
- \epsilon 
\frac{1}{\sqrt{6}} \big( (cm)_{ij} F^i \cdot \tilde{ F}^j 
-  (cmm)_i F^i \cdot \tilde{ F}^0 
+ \frac13\,(cmmm) F^0 \cdot \tilde{ F}^0 \big) \label{top} \;,
\eea
where we suppressed contracted Lorentz indices on the field
strengths, i.e., $F^I \cdot G^J  := F^I_{\mu \nu} G^{J \; \mu \nu}$.
Note that only field strengths appear, so that the four-dimensional
gauge symmetries are manifest. There are 
two types of terms, generalised Maxwell terms in the first line
 and generalised
$\theta$-terms in the second line.

In order to make contact with the conventions of four-dimensional
special geometry, we now perform the following rescaling:
\bea
\label{rescalings}
&&
\qquad h^i \ =: \ 6^{-1/3}\,e^{-\sigma} y^i~,
\quad m^i\ =:  \ \frac{6^{1/6}}2\,x^i~,
\quad a_{ij} \ =: \ - \epsilon g_{ij} \cdot 8 \cdot 6^{-1/3}\,e^{2\sigma}\,,
\nnu\\ 
&& F^i_{\mu \nu} \ =: \ \frac{6^{1/6}}{\sqrt{2}}\,F^i_{\mscr{(new)}\mu \nu}~,
\quad F^0_{\mu \nu} \ =: 
\ \sqrt{2}\,F^0_{\mscr{(new)}\mu \nu}~, 
\nnu \\
&&
\quad \tilde{F}^i_{\mu \nu} \ =: \ \epsilon \frac{6^{1/6}}{\sqrt{2}}\,
\tilde{F}^i_{\mscr{(new)}\mu \nu}~,
\quad \tilde{F}^0_{\mu \nu} \ =: 
\ \epsilon  \sqrt{2}\, \tilde{F}^0_{\mscr{(new)}\mu \nu}~.
\eea
In four dimensions we adapt the range of our indices to 
the usual conventions, i.e., $\mu, \nu = 0,\ldots, 3$ for
$\epsilon = -1$ and $\mu, \nu = 1, \ldots, 4$ for $\epsilon =1$.
Since we prefer to normalise the four-dimensional $\epsilon$-tensor
such that $\epsilon^{0123} = 1$, we had to redefine
the dual field strength by an extra factor
$\epsilon$ to compensate for this redefinition. We note that
$chhh = 1 $ implies $Cyyy = 6 e^{3\sigma}$. Therefore the fields
$y^i$ are unconstrained, in contrast to the $h^i$, because the 
Kaluza Klein scalar has been scaled in. 
To avoid cluttered notation, we drop the subscript on the new field strength, 
$F^I_{\mscr{(new)}\mu \nu} =: F^I_{\mu \nu}$. 

The four-dimensional bosonic Lagrangian 
takes the following form in terms of the rescaled fields:
\bea
{\bf e}^{-1} \cL^{\epsilon} &=&  \frac{1}{2} R - g_{ij} 
\left( \prt_\mu x^i \prt^\mu x^j - \epsilon 
\prt_\mu y^i \prt^\mu y^j \right) \nnu \\
&& + \epsilon \, \left(
\frac{1}{4}  cyyy \, \left( \frac{1}{6} + \frac{2}{3} \, gxx \,
\right) F^0 \cdot F^0 
- \frac{1}{3}  \, cyyy \, (gx)_i F^0 \cdot F^i 
+ \frac{1}{6} \, cyyy \, g_{ij} \, F^i \cdot F^j \right)\nnu \\
&& - \frac{ 1 }{12}  \left( 
 cxxx \, F^0 \cdot \tilde{F}^0 
- 3 (cxx)_i \, F^i \cdot \tilde{F}^0 
+ 3  (cx)_{ij} \, F^i \cdot \tilde{F}^j \right)\;.
\label{4dLagr_from_reduction}
\eea
The explicit form of $g_{ij}$ is
\be
g_{ij} = \epsilon \, \frac{3}{2} \left( \frac{ (cy)_{ij}}{cyyy}
- \frac{3}{2} \frac{ (cyy)_i (cyy)_j }{ (cyyy)^2 } 
\right) \;,
\label{g_from_reduction}
\ee
and the metric of the scalar manifold of the four-dimensional
theory is $g_{ij} \oplus (-\epsilon) g_{ij}$. Introducing 
$\epsilon$-holomorphic coordinates $z^j = x^j + i_\epsilon y^j$ 
we observe that the metric is 
$\epsilon$-K\"ahler with $\epsilon$-K\"ahler potential 
$K=-\ln {\cal V}(y)$. 
The signature
is determined by the signature of the five-dimensional
scalar metric $a_{ij}$. To have standard kinetic terms in the
five-dimensional theory, $a_{ij}$ needs to be positive definite,
and then (\ref{a-matrix}) implies that
$g_{ij}$ is positive (negative) definite for
$\epsilon=-1$ ($\epsilon=1$). Thus for space-like reduction ($\epsilon=-1$)
the scalar metric $g_{ij} \oplus (-\epsilon) g_{ij}$ is positive definite,
while for time-like reduction ($\epsilon=1$) it has split signature.
In the latter case the scalars $x^i$, which descend from five-dimensional
gauge fields, have a non-standard negative definite kinetic term.
We will investigate and comment on this feature in due course.

\section{The four-dimensional Lagrangian and its special
$\epsilon$-K\"ahler geometry}

We will now show that the scalar geometry of the 
dimensionally reduced Lagrangian is
projective special $\epsilon$-K\"ahler. Moreover, we will show
that for space-like dimensional reduction it agrees with the
standard form \cite{dWvP84} of a four-dimensional vector
multiplet Lagrangian, and that the Euclidean vector multiplet 
Lagrangian is obtained from this by replacing the 
complex structure by a para-complex structure. 
We will work in local $\epsilon$-complex coordinates
and write all formulae such that they apply simultanously to 
both cases $\epsilon = \pm 1$. The $\epsilon$-complex unit  is
denoted $i_\epsilon$ and has the property that $i_\epsilon^2 = \epsilon$.
Thus $i_\epsilon=i$ with $i^2 =-1$ for $\epsilon=-1$, and
$i_\epsilon =e$ with $e^2 =1$ for $\epsilon =1$. 

Given the form of the scalar term in (\ref{4dLagr_from_reduction}), it 
is natural to introduce
$\epsilon$-complex scalar fields $Z^i = x^i + i_\epsilon y^i$.
Then the scalar term takes the form
\begin{equation}
\label{ScalarTerm}
{\bf e}^{-1} \cL^{\epsilon}_{\rm scalar} =
- \bar{g}_{ij} \partial_\mu Z^i \partial^\mu \overline{Z}^j \;,
\end{equation}
and we see that the scalar metric is $\epsilon$-Hermitean.
We will now elaborate on this
observation and make the geometry underlying 
(\ref{4dLagr_from_reduction}) manifest.

This section is organised as follows. In subsection 
\ref{GeneralPrePo} we generalise various standard
formulae used in the physics literature on special geometry
to the $\epsilon$-complex case. We work in local coordinates,
but mention the geometrical interpretation of various objects, 
where helpful. The details are postponed to subsection \ref{Reform}.
The main
result of subsection \ref{GeneralPrePo} is the
$\epsilon$-complex generalisation of the bosonic
part of the Lagrangian for four-dimensional ${\cal N}=2$ 
vector multiplets. The prepotential is required to be 
$\epsilon$-holomorphic
and homogenous of degree 2, but unconstrained otherwise. 
For $\epsilon=-1$ we show that we recover the bosonic part of
the ${\cal N}=2$ vector multiplet Lagrangian, as given in 
\cite{deWKapLouLue95}.\footnote{This reference uses the so-called
`new conventions', which differ from the conventions used in 
\cite{dWvP84}, \cite{deWLauVP85}. Most of the recent supergravity and
string theory literature uses the new conventions (or closely related
conventions). }

In subsection \ref{VSprepot}
we specialise to the case of so-called very special
prepotentials and show that the resulting Lagrangian
agrees with the one obtained by dimensional reduction 
over time (for $\epsilon=1$) and space (for $\epsilon=-1$),
respectively. 

In subsections \ref{Reform} and \ref{Sections} 
we return to the case of a  general prepotential 
and relate the formalism of subsection \ref{GeneralPrePo} to the
results of sections \ref{ASeKM} - \ref{UBdl}, thus providing the
geometrical interpretation. In section \ref{Reform} we show
that the scalar term of the four-dimensional Lagrangian has two 
gauge-equivalent formulations: one as a gauged sigma models
with scalars $X^I$ taking values in $M$, the other as a 
sigma model with scalars $Z^i$ taking values in $\bar{M}$. 
The second formulation is obtained by gauge-fixing the local
$\bC^*_\epsilon$ symmetry of the gauged sigma model. 
For $\epsilon=-1$ this is of course part of the well known
construction of ${\cal N}=2$ vector multiplet based on the
superconformal calculus. This constuction makes use of the 
gauge equivalence between $n+1$ superconformal vector multiplets
coupled to conformal supergravity (the Weyl 
multiplet)\footnote{For completeness we mention that one
further `compensating' multiplet needs to be added, which, however,
is not relevant for the purpose of this paper.} 
with $n$ vector multiplets coupled to Poincar\'e supergravity. 
While we do not fully develop the superconformal 
calculus for $\epsilon=1$, we cover its  most relevant 
aspect for vector multiplets, namely the underlying geometry.
As we will see in detail, the respective scalar manifolds
$M$ and $\bar{M}$ are precisely related by the geometrical
construction of section \ref{PSeKM}.

\subsection{The four-dimensional Lagrangian for general 
prepotentials \label{GeneralPrePo}}

We start from a prepotential $F(X)$, which
is $\epsilon$-holomorphic and homogenous of degree 2 in its
$\epsilon$-complex variables $X^I$, where $I =0,
\ldots, n_V^{(4)}$. 
The supergravity variables $X^I$ are scalar fields which 
take values in the conical special $\epsilon$-K\"ahler manifold $M$,
as we will see in more detail in section \ref{Reform}. They 
are the 
components of a map ${\cal X}$ from space-time $N$ into $M$, which
is parametrised in terms of the (conical holomorphic) special coordinates
introduced used in section \ref{CSeKM}: 
\be
X^I: N \stackrel{\cal X}{\rightarrow} M 
\stackrel{\phi^I}{\rightarrow} \bC_\epsilon \;.
\ee
Here $\phi^I$ denotes the $I$-th coordinate map with
respect to a system of (local conical holomorphic) special coordinates
on $M$. For convenience we will follow common usage in the physics
literature and refer to the fields 
$X^I$ simply as `special coordinates on $M$'.

Derivatives of the prepotential with respect
to the variables $X^I$ are denoted $F_{I}, F_{IJ}, \ldots$, and the
$\epsilon$-complex conjugated quantities are denoted by
$\bar{F}, \bar{F}_I,
\ldots$.  
We define
\be
Z^I = \frac{X^I}{X^0} \;,
\ee
so that $Z^0=1$, while $Z^i, i=1, \ldots, n_V^{(4)}$ 
`are' special coordinates on the projective special
$\epsilon$-K\"ahler manifold $\bar{M}$
defined by the prepotential.\footnote{
Here the same terminological simplification is applied as for
the $X^I$.}
The real and imaginary parts of $Z^i$ are denoted by
$x^i$ and $y^i$ respectively:
\be
Z^i = x^i + i_\epsilon y^i \;.
\label{Decomp}
\ee
Using that $F$ is homogenous of degree 2 we 
define a `rescaled, non-homogeneous prepotential' ${\cal F}(Z)$ by
\[
F(X^0,X^1, \ldots)=
(X^0)^2 F\left( 1, \frac{X^1}{X^0}, \ldots \right) = 
(X^0)^2 {\cal F}(Z^1, \ldots, Z^n) \;.
\]
Now we can rewrite $F$ and its derivatives in terms
of special coordinates $Z^i$:
\begin{equation}
\label{homogRewrite}
\begin{array}{lll}
F(X) = (X^0)^2 \cF(Z) \;,\;\;\;&
F_0(X) = X^0 ( 2 \cF - Z^i \cF_i) \;,\;\;\;&
F_i(X) = X^0 \cF_i \;,\;\;\;\\
F_{ij}(X) = \cF_{ij} \;,\;\;\;&
F_{0i}(X) = \cF_i - Z^j \cF_{ij} \;,\;\;\;&
F_{00}(X) = 2 \cF - 2 Z^i \cF_i + Z^i Z^j \cF_{ij} \;. \\
\end{array}
\end{equation}
We use a notation where 
${\cal F}_i= \frac{\partial {\cal F}}{\partial Z^i}$, etc.

The metric $\bar{g}$ on $\bar{M}$ is given by
\[
\bar{g}_{ij} = \frac{\partial^2 K}{\partial Z^i \partial \bar{Z}^j} \;,
\]
where 
\begin{equation}
K = -\log Y \;,\;\;\;
Y= i_\epsilon \left( 2 ({\cal F} - \bar{\cal F}) - 
(Z^i - \bar{Z^i})({\cal F}_i + \bar{\cal F}_i) \right) 
\end{equation}
is the $\epsilon$-K\"ahler potential.
For $\epsilon=-1$ this is the standard formula for the K\"ahler
potential of the metric  on $\bar{M}$ in terms of special coordinates. 
We will verify in subsection \ref{Reform} that this is 
an $\epsilon$-K\"ahler potential for the metric defined in section 
\ref{PSeKM}.

Following supergravity conventions, the metric $g$ of $M$ is
given by the matrix
\be
N_{IJ} = - i_\epsilon ( F_{IJ} - \bar{F}_{IJ} ) \;.
\ee
This quantity enters into the definition of the gauge field
coupling matrix
\be
\bar{\cal N}_{IJ} = F_{IJ}(X) + i_\epsilon \epsilon
\frac{(N \bar{Z})_I  (N \bar{Z})_J}{\bar{Z}N
\bar{Z}} \;.
\ee
For $\epsilon=-1$ this agrees with the 
standard definition of ${\cal N}_{IJ}$
in the `new conventions' of \cite{deWKapLouLue95}.\footnote{
The matrices ${\cal N}_{IJ}$ and $\bar{\cal N}_{IJ}$ are 
related by $\epsilon$-complex conjugation.}
Now consider the following 
four-dimensional bosonic Lagrangian:
\be
{\bf e}^{-1} \cL^{(4)}
=  \frac{1}{2} R - \bar{g}_{ij} \der_\mu Z^i \der^\mu 
\bar{Z}^j + \frac{1}{4} \mbox{Im} {\cal N}_{IJ} 
F^I \cdot F^J + \frac{1}{4} \mbox{Re} {\cal N}_{IJ} F^I \cdot 
\tilde{F}^J \;,
\label{4dLagr}
\ee
where $F^I_{\mu \nu}$ are field strengths, and we suppressed 
the Lorentz indices in the Lagrangian. For $\epsilon=-1$ this
is the bosonic part of the 
standard four-dimensional Lagrangian of ${\cal N}=2$
supergravity coupled to vector multiplets \cite{dWvP84}, written
in terms of  the `new conventions'
of \cite{deWKapLouLue95}. The bosonic Lagrangian
for $\epsilon=-1$ can be found, for example, in \cite{LopesCardoso:2000qm}
or \cite{TMhabil}.\footnote{Note that in these references 
the space-time Riemann tensor is defined
with a relative minus sign compared to the definition used in this paper.}
For $\epsilon =1$ we get the para-complex version of the standard
Lagrangian. 
While we have only derived a bosonic Lagrangian
here, it is known for $\epsilon=-1$, and expected for $\epsilon=1$,
that this is the bosonic part of an ${\cal N}=2$ supersymmetric
Lagrangian. The explicit study of fermionic terms for $\epsilon=1$
is left to future work. Since for rigid Euclidean ${\cal N}=2$
vector multiplets the full Lagrangian and supersymmetry rules
were constructed in \cite{CMMS}, it is clear that this is a
straightforward task. Moreover, for prepotentials which can be
obtained by dimensional reduction, the supersymmetry of the 
corresponding Lagrangian holds by construction.

\subsection{Very special prepotentials and comparison to 
the dimensionally reduced Lagrangian\label{VSprepot}}

We will now show that for a suitable choice of prepotential
the Lagrangian (\ref{4dLagr}) takes the form of the Lagrangian
(\ref{4dLagr_from_reduction}), which we obtained by dimensional 
reduction.  It is know from \cite{GST} that a space-like
dimensional reduction from five to four dimension
gives rise to 
a `very special prepotential':
\be
F(X) = \frac{1}{6} C_{ijk} \frac{X^i X^j X^k}{X^0} \;,
\label{VerySpecialF}
\ee
where $C_{ijk}$ are real. Such prepotentials are sometimes
referred to as `cubic', which 
alludes to the fact that they are in one-to-one correspondence
with the cubic prepotentials  of five-dimensional vector 
multiplets.
We anticipate that the case of time-like
reduction can be obtained by replacing the holomorphic coordinates
$X^I$ by para-holomorphic coordinates. \\

To compare the Lagrangians 
(\ref{4dLagr_from_reduction}) and (\ref{4dLagr}) to one another,
we need to 
compute the derivatives of a very special prepotential 
(\ref{VerySpecialF}) with respect to the $X^I$:
\begin{equation}
\begin{array}{lll}
F_0 = - \frac{1}{6} C_{ijk} \frac{X^i X^j X^k}{(X^0)^2} \;,\;\;\;&
F_i = \frac{1}{2} C_{ijk} \frac{ X^j X^k}{ X^0} \;,\;\;\;& 
F_{00} = \frac{1}{3} C_{ijk} \frac{X^i X^j X^k}{(X^0)^3} \;,\;\;\;\\
F_{0i} = - \frac{1}{2} C_{ijk} \frac{X^j X^k}{ (X^0)^2} \;,\;\;\;&
F_{ij} = C_{ijk} \frac{X^k}{X^0} \;.& \\
\end{array}
\end{equation}
Using (\ref{homogRewrite}) we can replace the $X^I$ by the special coordinates
$Z^i$ and obtain:
\begin{equation}
\begin{array}{lll}
\cF = \frac{1}{6} CZZZ \;,\;\;\;&
\cF_i = \frac{1}{2} (CZZ)_i \;,\;\;\;&
\cF_{ij} = (CZ)_{ij} \;,\;\;\;\\
\cF_{0i} = - \frac{1}{2} (CZZ)_i \;,\;\;\;&
\cF_{00} = \frac{1}{3} CZZZ \;,& \\
\end{array}
\end{equation}
where we suppressed indices which are summed over. 
To compute the scalar metric, we need $Y$, where 
$K=- \log Y$ is the $\epsilon$-K\"ahler potential.
The explicit 
expression for $Y$ is:
\be
Y = \frac{i_\epsilon}{3} ( CZZZ - C \bar{Z}\bar{Z}\bar{Z} ) - 
\frac{i_\epsilon}{2} (Z - \bar{Z})^i ( CZZ + C\bar{Z}\bar{Z})_i 
= - \frac{4}{3} Cyyy  \;,
\ee
where $y^i$ is the imaginary part of $Z^i$.
To compute the metric, the following form of $Y$ is convenient:
\be
Y = - \frac{i_\epsilon}{6} C(Z-\bar{Z})(Z-\bar{Z})
(Z-\bar{Z}) \;.
\ee
The resulting projective special $\epsilon$-K\"ahler metric is 
\bea
g_{i j} &=& \frac{ \der^2 K}{\der Z^i \der \bar{Z}^j}
= \frac{6}{Y} C(Z-\bar{Z})_{ij} - \frac{9}{Y^2}
C(Z-\bar{Z})(Z-\bar{Z})_i C (Z-\bar{Z})(Z-\bar{Z})_j
\nonumber \\
&=& \epsilon \left(
\frac{3}{2} \frac{ Cy_{ij} }{ Cyyy } - \frac{9}{4} 
\frac{ Cyy_i Cyy_j}{ (Cyyy)^2 } \right) \;.
\label{epsilon_metric}
\eea
Next, we evaluate the components of 
$N_{IJ} = - i_\epsilon ( F_{IJ} - \bar{F}_{IJ} )$:
\begin{equation}
N_{00} = - \frac{i_\epsilon}{3} ( CZZZ - C \bar{Z}\bar{Z}\bar{Z} ) 
\;,\;\;\;
N_{0i} = \frac{i_\epsilon}{2} ( CZZ_i - C \bar{Z}\bar{Z}_i )
\;,\;\;\;
N_{ij} = - i_\epsilon (CZ_{ij} - C \bar{Z}_i) 
\;.
\end{equation}
For later use we compute
\bea
N_{0I} Z^I &=& \frac{i_\epsilon}{6} CZZZ - \frac{i_\epsilon}{2} 
CZ\bar{Z}\bar{Z}
+ \frac{ i_\epsilon}{3} C\bar{Z}\bar{Z}\bar{Z}  \;,
\nonumber \\
N_{iI} Z^I &=& - \frac{i_\epsilon}{2} CZZ_i + i_\epsilon CZ\bar{Z}_i 
- \frac{i_\epsilon}{2} C \bar{Z}\bar{Z}_i \;,
\nonumber \\
ZNZ &=& - \frac{i_\epsilon}{3} \left( 
CZZZ - 3 CZZ\bar{Z} + 3 CZ\bar{Z}\bar{Z} - C \bar{Z}\bar{Z}\bar{Z}
\right) \;.
\eea
Note that for very special prepotentials we have 
\be
ZNZ = \bar{Z}N\bar{Z} = -2 ZN\bar{Z} = 2 Y \;.
\ee
Finally, we use our results to evaluate $\bar{\cal N}_{IJ}$:
\begin{equation}
\begin{array}{ll}
\bar{\cal N}_{00} = \frac{1}{3} Cxxx + i_\epsilon \epsilon \,
Cyyy \, \left( \frac{2}{3}  \bar{g}xx + \frac{1}{6} \right) \;,\;\;\;&
\bar{\cal N}_{0i} = - \frac{1}{2} (Cxx)_i - \frac{2}{3} i_\epsilon
\, \epsilon  \, Cyyy \, (\bar{g}x)_i \;,\\
\bar{\cal N}_{ij} = Cx_{ij} + \frac{2}{3} i_\epsilon \epsilon
\,Cyyy \,\bar{g}_{ij} \;, & \\
\end{array}
\end{equation}
where $\bar{g}_{ij}$ is the metric (\ref{epsilon_metric})
and $\bar{g}xx$ and $(\bar{g}x)_i$ denote the obvious contractions. 

We now have all the data required to compare 
(\ref{4dLagr_from_reduction}) and (\ref{4dLagr}) to one another.
Using  (\ref{g_from_reduction}) and (\ref{epsilon_metric})
together with (\ref{Decomp}) we see 
that the scalar terms agree for $C_{ijk} = \pm c_{ijk}$. 
To compare the gauge field terms we have to substitute the
components of  $\bar{\cal N}_{IJ}$ into 
(\ref{4dLagr}):
\bea
{\bf e}^{-1} \cL^{(4)}_{\rm gauge} &=& 
\frac{1}{4} \mbox{Im} {\cal N}_{00} F^0 F^0 +
\frac{1}{2} \mbox{Im} {\cal N}_{i0} F^i F^0 +
\frac{1}{4} \mbox{Im} {\cal N}_{ij} F^i F^j \nnu \\
&& +
\frac{1}{4} \mbox{Re} {\cal N}_{00} F^0 \tilde{F}^0 +
\frac{1}{2} \mbox{Re} {\cal N}_{i0} F^i \tilde{F}^0 +
\frac{1}{4} \mbox{Re} {\cal N}_{ij} F^i \tilde{F}^j  \nnu \\
&=& 
- \epsilon \left(  \frac{1}{4} Cyyy \left( \frac{1}{6} + \frac{2}{3}
gxx \right) F^0 F^0 -
\frac{1}{3} \,Cyyy \, (gx)_i F^0 F^i +
\frac{1}{6} \, Cyyy \, g_{ij} \, F^i F^j \right) \nnu \\
 && + \frac{1}{12} \left( 
Cxxx \,  F^0 \tilde{F}^0 - 3 (Cxx)_i \, F^i \tilde{F}^0 
+ 3 (Cx)_{ij} \, F^i \tilde{F}^j \right) \;.
\eea
Comparing this to the gauge field part of (\ref{4dLagr_from_reduction})
we see that the gauge field terms match if we set
$C_{ijk} = - c_{ijk}$.

\subsection{Reformulation of the scalar  sector 
as a gauged sigma-model \label{Reform}}

We now return to the case of a general ($\epsilon$-holomorphic
and homogenous) prepotential and relate the formalism
presented in subsection \ref{GeneralPrePo} to the geometrical
construction of sections \ref{ASeKM}-\ref{UBdl}. At the same
time we adapt those parts of the superconformal construction 
of vector multiplets which are relevant for the scalar term
to the $\epsilon$-complex framework. 
For an introduction to 
the superconformal calculus and its use in constructing supergravity
Lagrangians we refer the reader to \cite{deW:84}.
A detailed review of the construction of the vector multiplet
Lagrangian in this formalism is contained in \cite{TMhabil}, which 
also contains extensive references.
A short summary of the relevant material can be found in \cite{TMln}.

The following
diagram is useful in summarising the relevant spaces and maps:
\begin{equation}
\xymatrix{
 & M \ar@<.5ex>[d]^{\pi} \ar[r]^{\phi} & V' \ar[d]^{\pi_V} \ar[r]^{\subset} & 
{\cal U} \ar[dl]^{\pi_{U}} 
 \\
N \ar[r]^{\cal Z} \ar[ru]^{\cal X} & \bar{M} \ar[r]^{\bar{\phi}} \ar@<.5ex>[u]^{s}&
P(V')  & 
}
\end{equation}
Here $N$ is space-time, which is Riemannian or Lorentzian depending
on $\epsilon$. 
$M$ is a regular conical special
$\epsilon$-K\"ahler manifold (see Definition \ref{regDef})  
and $\phi :M \rightarrow V'\subset V=T^*\bC^{n+1} \simeq \bC^{2n+2}_\epsilon$ 
is the conical holomorphic 
immersion of Theorem \ref{conicThm},
which induces the holomorphic immersion 
$\bar{\phi}: \bar{M} \ra P(V')$. The map ${\cal X} : N \rightarrow M$
is locally described by the $n+1$ $\epsilon$-complex scalar fields 
$X^I\circ {\cal X}: N \rightarrow \bC_\epsilon$,  where $X^I$, $I=0,\ldots ,n$, 
are special coordinates  on $M$. Similarly, the induced map 
${\cal Z} : N \rightarrow \bar{M}$
in the above commutative diagram is locally described by the $n$ scalar fields
$Z^i\circ {\cal Z}$, where  $Z^i$, $i=1,\ldots,n$, are special coordinates
on $\bar{M}$. As usual in the physical literature, we shall use a simplified
notation where 
the scalar fields on $N$ are simply denoted by $X^I$ and $Z^i$, instead of 
$X^I\circ {\cal X}$ and $Z^i\circ {\cal Z}$.  For a generic
choice of the immersion, $\phi$ comes from a prepotential $F$,
which is $\epsilon$-holomorphic and homogenous of degree two.
The homogeneity condition is needed within the superconformal
framework in order to couple the corresponding $n+1$ vector multiplets
to conformal supergravity. Geometrically, it implies that 
locally $\phi$ maps $M$ into a Lagrangian
cone in the $\epsilon$-complex symplectic vector space $V$.  By dividing out
the local group action generated by the commuting vector fields $\xi$ and 
$J\xi$ on $M$ (which corresponds to the $\bC_\epsilon^*$-action in 
$V$ via the immersion $\phi$) one arrives at the projective
special $\epsilon$-K\"ahler manifold $\bar{M}$. In supergravity
$\bC^*_\epsilon$ is a local gauge symmetry which is part 
of the superconformal group.\footnote{This is known for 
$\epsilon=-1$, and we expect it to be true for $\epsilon=1$
as well.} As we will see, the projection $\pi: M \rightarrow \bar{M}$
corresponds to gauge-fixing this symmetry. Further, we have 
included in our diagram that the space of non-isotropic 
vectors $V'$ projects similary to the corresponding projective
space $P(V')$ of non-isotropic lines, into which $\bar{M}$ 
is immersed by $\bar{\phi}$.

In this subsection we start from $M$ and obtain $\bar{M}$ and
the corresponding sigma model by projection. For simplicity 
(and without restriction of generality), we shall
assume that the immersion $\phi : M \rightarrow V'$ is an embedding of
$M$ into a Lagrangian cone. In particular, this implies that the
local group action generated by $\xi$ and $J\xi$ is induced from 
a global action of the group $\bC_\epsilon^*$ on $M$.  
$M$ can be regarded as the total space of a $\bC_\epsilon^*$-bundle 
over $\bar{M}$, and we can go from $\bar{M}$ to $M$ 
by choosing a section $s:\bar{M} \rightarrow M$ of this bundle.
Moreover, there is a corresponding line bundle\footnote{Here and
in the following it is understood that `line bundle' means
`$\epsilon$-complex line bundle'.} $\pi_U: {\cal U} \rightarrow
P(V')$ over $P(V')$. This is the so-called canonical line bundle
introduced in section \ref{UBdl}, which coincides with 
$V' \rightarrow P(V')$ on the image of $\phi$. This allows us
to reinterpret various maps as sections of line bundles obtained
as pull-backs of the universal bundle. We will come back to this
fact in subsection \ref{Sections}, where we briefly relate
our construction to an alternative formulation of special
geometry, which makes extensive use of these sections.

We start by constructing a gauged sigma model with target 
space $M$, adapting the standard procedure used in the
superconformal formalism to the $\epsilon$-complex framework.
The $\epsilon$-complex scalars $X^I$ are subject to 
$\epsilon$-complex scale transformations, under which they
transform as follows:
\[
X^I \rightarrow \lambda X^I \;,\;\;\;\lambda \in \bC^*_\epsilon \;.
\]
The group $\bC^*_\epsilon=GL(1,\bC_\epsilon)$ contains real dilatations, where
$\lambda \in \bR^{>0}$, and  
$U(1)_\epsilon$ gauge transformations, where 
$U(1)_\epsilon := \{ z\in \bC_\epsilon | z \bar{z} = 1\}$. 
The latter are chiral $U(1)=SO(2)$-transformations for $\epsilon =-1$ and 
chiral 
$\bR^*=SO(1,1)$-transformations for $\epsilon=+1$. 
For $\epsilon=1$ the group $\bC^*_\epsilon = GL(1,\bC_\epsilon)
=\bR^{>0}\times O(1,1) \supset
GL^+(1,\bC_\epsilon)=\bR^{>0}\times SO(1,1)=\bR^{>0}\times U(1)_\epsilon$ 
is obtained
by removing all isotropic elements (i.e. the lightcone of the
origin) from $\bC_\epsilon$. It has four connected components. 
(The `$+$'-index stands for positive determinant of the 
representing real $2\times 2$-matrix.)  
Comparing to section \ref{CSeKM}, we see 
that the dilatations are the 
homotheties generated by the vector field $\xi$, whereas the 
Killing vector field $J\xi$ generates the  
maximal connected subgroup in the group $U(1)_\epsilon$. 
In special holomorphic coordinates the homothety $\xi$
takes the form
\be
\label{HomoHol}
\xi = X^I \frac{\partial}{\partial X^I} + 
\bar{X}^I \frac{\partial}{\partial \bar{X}^I} \;.
\ee
This expression follows from the one given
in Lemma \ref{1stLemma}, section \ref{CSeKM}
by going from special affine to special
holomorphic coordinates, while using that the prepotential is
homogenous of degree 2. By applying the $\epsilon$-complex structure
tensor $J$ to (\ref{HomoHol}) we obtain the following
expression for the Killing vector field $J\xi$ in special
holomorphic coordinates:
\[
J\xi = i_\epsilon X^I \frac{\partial}{\partial X^I} - 
i_\epsilon \bar{X}^I
\frac{\partial}{\partial \bar{X}^I} \;.
\]
The conical affine special $\epsilon$-K\"ahler metric $g$ 
on $M$ is obtained from the prepotential $F(X)$ by
\begin{equation}
N_{IJ} = 2 {\rm Im} F_{IJ} = -i_\epsilon (F_{IJ} - \bar{F}_{IJ}) \;.
\label{DefN}
\end{equation}
Here and in the following we follow supergravity conventions
and denote the matrix representing the metric $g$ in terms
of holomorphic special coordinates by $N_{IJ}$. (More precisely, 
$g$ is the real part of the sesquilinear form $N_{IJ}dX^I\otimes d\bar{X}^J$.)

To write down a Lagrangian which is invariant under local
$\bC^*_\epsilon$-transformations, we introduce gauge fields
$b_\mu$ for dilatations and ${\cal A}_\mu$ for $U(1)_\epsilon$  
gauge transformations. The covariant derivatives of scalars
are
\begin{eqnarray}\label{connEqu}
{\cal D}_\mu X^I = (\der_\mu -b_\mu +i_\epsilon {\cal A}_\mu) X^I \;,\;\;\; && 
{\cal D}_\mu \bar{X}^I = (\der_\mu -b_\mu -i_\epsilon
 {\cal A}_\mu) \bar{X}^I \;.
\end{eqnarray}
Notice that homogeneous coordinates on projective space are not functions but 
are sections of the line bundle ${\cal U}^*$ which is 
dual to the universal bundle $\cal U$, discussed in \ref{UBdl}. 
Correspondingly, the 
scalar fields $X^I$ are sections of the pull back of ${\cal U}^*$ to
space-time $N$. It follows from this remark that ${\cal A}_\mu=-A_\mu$,
where $A_\mu$ is the $U(1)_\epsilon$-connection one-form of the pull back of
the universal bundle $\cal U$ to $N$ with respect to the 
section $(X^I,F_I)$, see \ref{Sections} for a detailed discussion.    
Then the gauged non-linear sigma model is
\[
{\bf e}^{-1}{\cal L}_{\rm scalar} = - N_{IJ} 
{\cal D}_\mu X^I {\cal D}^{\mu} \bar{X}^J \;.
\]

It is instructive to consider the Einstein-Hilbert term
in (\ref{4dLagr}) alongside the scalar sigma model. 
The space-time metric is invariant under $U(1)_\epsilon$-transformations, but 
carries weight $-2$ under dilatations.\footnote{In the superconformal
formalism, all fields
transform under dilatations according to their weight. Here we use that   
the vielbein $e^a_\mu$ has weight $-1$ and that space-time coordinates
have weight $0$, see \cite{deW:84,TMhabil,TMln}.} 
The Einstein-Hilbert action can be made invariant under
dilatations by multiplying the Ricci scalar by a scalar field
which acts as a compensator.\footnote{This is a variant 
of the St\"uckelberg mechanism, which is an essential part of 
the superconformal formalism. See for example \cite{deW:84,TMln}.}
Adapting standard results from the superconformal calculus, 
we take the following locally $\bC^*_\epsilon$-invariant 
Lagrangian ${\cal L}_{\rm grav+scalar}$ as our starting point:
\begin{eqnarray}
\label{SCsc+grav}
{\bf e}^{-1} {\cal L}_{\rm grav+scalar} 
 &=& -\frac{i_\epsilon}{2} (X^I \bar{F}_I  - F_I \bar{X}^I) R 
- N_{IJ} {\cal D}_\mu X^I {\cal D}^\mu \bar{X}^J
\label{ScLag1} \;.
\end{eqnarray}
Here the composite scalar $i_\epsilon(X^I \bar{F_I} - F_I \bar{X}^I)$
plays the role of the compensating field for the dilatations.
We will show that we recover the
scalar and gravitational terms of (\ref{4dLagr}) by 
gauge fixing the $\bC^*_\epsilon$ symmetry, which in turn
amounts to implementing the quotient described in 
section \ref{PSeKM}.

At this point it is convenient to use a fact which is well known
from the superconformal calculus:
it is consistent to set  $b_\mu=0$ in (\ref{SCsc+grav}), because
the terms containing $b_\mu$ have to cancel anyway.\footnote{
In the superconformal framework, the condition $b_\mu=0$ is known as the 
K-gauge. We refer to \cite{TMhabil,TMln} for details. In particular, 
local dilatation invariance is discussed in section 2
of \cite{TMln}.} Next,
the $U(1)_\epsilon$-gauge field ${\cal A}_\mu$ is
non-dynamical and can be eliminated by its algebraic equation of motion,
\begin{equation}
{\cal A}_\mu = -A_\mu = \frac{1}{2} \frac{
\bar{F}_I \stackrel{\leftrightarrow}{\der}_\mu X^I 
- \bar{X}^I \stackrel{\leftrightarrow} {\der}_\mu F_I}{
i_\epsilon(F_I \bar{X}^I - \bar{F}_I X^I)}.  
\label{EoMU1GF}
\end{equation}
Notice that this coincides with the formula   
(\ref{UniConn}) for the Chern connection $A_\mu$ with respect to a 
unitary frame (for which automatically $b_\mu=0$). 
For us it is useful to
rewrite (\ref{EoMU1GF}) in the form
\[
i_\epsilon {\cal A}_\mu = - \frac{1}{2} 
\frac{N_{IJ}  X^I \stackrel{\leftrightarrow}{\partial}_\mu \bar{X}^J}{
-XN\bar{X} } \;.
\]
Substituting this back into the Lagrangian, the scalar part
becomes an `ordinary' (rather than gauged) 
non-linear sigma model. For our purposes the following
form of the result is convenient
\begin{eqnarray}
- N_{IJ} {\cal D}_\mu X^I {\cal D}^\mu \bar{X}^J &=& 
- \left( N_{IJ} + 
\frac{ (N \bar{X})_I (NX)_J }{-XN\bar{X}} 
\right) \partial_\mu X^I \partial^\mu \bar{X}^J \nonumber \\
 && + \frac{1}{4} \frac{
[ \partial_\mu (XN\bar{X}) - X (\partial_\mu N) \bar{X}]
[(\partial^\mu X) N \bar{X} + XN (\partial^\mu \bar{X})]}
{-XN\bar{X}}  \;.\label{IntOut}
\end{eqnarray}

The expression 
for the metric simplifies, after imposing a gauge condition
which fixes the local dilatation symmetry. The natural gauge
condition is\footnote{This is known as the D-gauge in the 
superconformal literature.}
\begin{equation}
i_\epsilon (X^I \bar{F}_I - F_I \bar{X}^I) = - 1 \;, 
\label{Dgauge}
\end{equation}
because this turns the first term of (\ref{ScLag1}) 
into the standard Einstein-Hilbert term:
\begin{equation}
{\bf e}^{-1} {\cal L}_{\rm grav}  = 
-\frac{i_\epsilon}{2} (X^I \bar{F}_I - F_I \bar{X}^I) R
=  \frac{1}{2} R \;.
\end{equation}
To analyze the scalar term, first note that (\ref{Dgauge})
is equivalent to 
\[
- N_{IJ} X^I \bar{X}^J = 1\;.
\]
Since the scalar fields are constrained to the hypersurface
(\ref{Dgauge}), it follows that 
\[
\partial_\mu( N_{IJ} X^I \bar{X}^J ) = 0 \;.
\]
Moreover, homogeneity of degree two of the prepotential implies
$F_{IJK} X^K=0$, and therefore
\[
(\partial_\mu N_{IJ}) X^I \bar{X}^J = 0\;.
\]
As a consequence the second line of (\ref{IntOut}) vanishes, and
the scalar sigma model takes the following form after imposing
the gauge condition (\ref{Dgauge}):
\begin{equation}
{\bf e}^{-1} {\cal L}_{\mscr{scal}} =
- (N_{IJ} + (N \bar{X})_I (N X)_J ) \der_\mu X^I \der^{\mu}
\bar{X}^J =: - M_{IJ} \der_\mu X^I \der^\mu \bar{X}^J  \;.
\label{ScLag2}
\end{equation}
This is a sigma model with 
`metric' $M_{IJ}$, which we need to relate to a sigma model
with values in $\bar{M}$ and with 
metric $\bar{g}_{ij}$, as it  occurs in (\ref{4dLagr}).

At this point it is useful to connect our discussion with 
the construction of $\bar{M}$ used in section \ref{PSeKM}.
First note that
\[
i_\epsilon(X^I \bar{F}_I  - F_I \bar{X}^I) =  N_{IJ}
X^I \bar{X^J} =  g(\xi, \xi)
\]
is the length-squared 
of the homothetic vector $\xi$. The gauge condition (\ref{Dgauge})
sets $g(\xi,\xi)=-1$, which, according to section \ref{PSeKM}, defines
a smooth hypersurface $S \subset M$. Moreover, since 
(\ref{Dgauge})
is $U(1)_\epsilon$ invariant,
it is manifest that the isometries generated by 
the Killing vector field $J\xi$ act on $S$, and we know from 
section \ref{PSeKM} that the 
projective special $\epsilon$-K\"ahler manifold $\bar{M}$
is obtained by taking the quotient of $S$ by this isometry.
We should therefore expect that $M_{IJ}$ is related to
the tensor field $h$ defined in (\ref{h-tensor}), which 
induces the $\epsilon$-K\"ahler metric on $\bar{M}$.

The following observation turns out to be helpful. The tensor field
$h$ is defined on $M$, and while $M_{IJ}$ is originally defined 
on $S$, we can  extend it to a tensor field 
on  $M$ in the following way. Take the function
\be
\label{KpotM}
K = - \log \left(- i_\epsilon ({X}^I \bar{F}_I - {F}_I \bar{X}^I)
\right) \;,
\ee
and define
\begin{eqnarray}
\label{M-tensor}
{M}_{IJ} = \frac{\partial^2 K}{\partial X^I \partial \bar{X}^J}
&=& \frac{-i_\epsilon (F_{IJ} - \bar{F}_{IJ})}{i_\epsilon
(F_K \bar{X}^K - X^K \bar{F}_K)}
+ \frac{ i_\epsilon (F_{IL} - \bar{F}_{IJ}) \bar{X}^L
i_\epsilon (F_{JK} - \bar{F}_{JK} ) X^K}{[ i_\epsilon
(F_K \bar{X}^K - X^K \bar{F}_K)]^2} \nonumber \\
&=& \frac{N_{IJ}}{-XN\bar{X}} + \frac{ (N\bar{X})_I (NX)_J}{
[ -XN\bar{X}]^2} \;.
\end{eqnarray}
This coincides with the original $M_{IJ}$ defined in 
(\ref{ScLag2}) when restricting to $S$, and can be shown to 
be proportional to the tensor field (\ref{h-tensor}). In order
to verify this we only have to use that the scalar
product $g(U,V)$ of two vectors $U,V$ on $M$ is given by
\[
g(U,V) = \frac{1}{2}\left( U^I N_{IJ} \bar{V}^J  +
V^I N_{IJ} \bar{U}^J\right) = \frac{1}{2}
\left( UN\bar{V} + V N \bar{U} \right) \;.
\]
Then it is straightforward to show that 
\[
h(U,V) = - \frac{1}{2} \left( UM\bar{V} + VM \bar{U} \right) \;,
\]
and therefore, up to an overall sign, $M_{IJ}$ is 
the representative of $h$ in special holomorphic coordinates.
While ${M}_{IJ}$ can be obtained by taking the second 
derivatives of the `$\epsilon$-K\"ahler potential'
(\ref{KpotM}), this tensor field  
is not a metric on $M$ because it is
degenerate along the directions generated by the vector fields
$\xi, J\xi$. This can be shown either by evaluating
(\ref{h-tensor}) on $\xi$ and $J\xi$ with the result
\[
h(\xi,\xi) = h(\xi, J\xi) = h(J\xi, J\xi) =0 \;,
\]
or by an equivalent calculation in local coordinates,
using that 
\begin{equation}
X^I {M}_{IJ} =0= {M}_{IJ} \bar{X}^J \;.
\end{equation}
However, according to section \ref{PSeKM} the tensor field
$h$  projects onto a non-degenerate metric on $\bar{M}$,
and therefore $M_{IJ}$ must be non-degenerate on the horizontal space
of the submersion $\pi : M \rightarrow \bar{M}$. These directions
are spanned by vectors which are orthogonal to the plane 
${\rm span}\{ \xi, J\xi \}$ with respect to the (non-degenerate)
metric $g$ on $M$. In local coordinates, vectors $W$ orthogonal
to ${\rm span}\{ \xi, J\xi \}$ satisfy:
\[
WN\bar{X} + X N \bar{W} =0 \;,
\]
which implies
\begin{equation}
\label{Signature}
W M  \bar{W} = \frac{WN\bar{W}}{-XN\bar{X}} \;.
\end{equation}
Since $N_{IJ}$ is non-degenerate and $X^I N_{IJ} \bar{X}^J$ is 
non-vanishing, it is clear that 
$M_{IJ}$ is non-degenerate on the horizontal space.
In fact from (\ref{M-tensor}) and (\ref{Signature}) we can easily
read off the signature of ${M}_{IJ}$ on the horizontal space.  
$M_{IJ}$ is
invariant under $N_{IJ} \rightarrow - N_{IJ}$, so that
the signature of $M_{IJ}$ is independent of the overall
sign of $N_{IJ}$. Now consider first $\epsilon=-1$, where
$N_{IJ}$ is either positive definite or negative definite along 
the complex direction spanned by $\xi, J\xi$. 
Then, by inspection of
(\ref{M-tensor}) and (\ref{Signature}),
if $N_{IJ}$ is either positive or
negative definite, then $M_{IJ}$ is negative
definite on the horizontal space. However, for a supergravity theory in Lorentzian
space-time we want $M_{IJ}$ to be positive definite
along these directions, which can be arranged by taking 
$N_{IJ}$ to have signature $(2,2n)$ or $(2n,2)$.\footnote{
Thus the K\"ahler metric $N_{IJ}$ on $M$ must have indefinite
signature. Such metrics are usually called pseudo-K\"ahler in 
the literature. In this paper we suppress the 
prefix `pseudo-' most of the time, 
but we stress that all the results obtained
for $\epsilon=-1$ apply irrespective of the metric being 
definite or indefinite.}
Next, consider the case $\epsilon=1$, where $N_{IJ}$ 
has always split signature $(n+1,n+1)$. Since the direction 
spanned by $\xi, J\xi$ is para-complex, it has signature
$(1,1)$, and therefore $M_{IJ}$ must have split signature 
$(n,n)$ on the horizontal space. Of course, this already follows
from $\bar{M}$ being para-K\"ahler.

Imposing the gauge (\ref{Dgauge}) has brought us from $M$ 
to the real hypersurface $S$ (a level set of the moment map $g(\xi ,\xi )$), 
on which $U(1)_\epsilon$ acts
isometrically. $\bar{M}$ is then obtained
by taking the quotient of $S$ with respect to $U(1)_\epsilon$. 
This is precisely the $\epsilon$-K\"ahler quotient of $M$ 
with respect to the isometric and holomorphic $U(1)_\epsilon$-action.
The submersion $M \rightarrow \bar{M}$ is $\epsilon$-holomorphic and 
a homothety on horizontal spaces, whereas $S\rightarrow \bar{M}$ is 
even a Riemannian submersion. The crucial point is that
the vector field $J\xi$ on $M$ is not only Hamiltonian, which is sufficient to
induce a symplectic structure on $\bar M$ (that is to perform the 
symplectic quotient), but that it is also a Killing vector field 
with respect to the $\epsilon$-K\"ahler metric. Therefore, $\bar M$
inherits not only a symplectic structure but also a pseudo-Riemannian metric.
Combining the two yields the $\epsilon$-complex structure.  

To descend from $S$ to $\bar{M}$ we could impose a 
condition which gauge-fixes the $U(1)_\epsilon$ transformations.
However, it is more convenient to express everything in terms of 
 $U(1)_\epsilon$-invariant objects. Therefore we introduce
para-complex scalar fields, 
\begin{equation}
Z^I = \frac{X^I}{X^0}\;,
\end{equation}
which are invariant under $\bC^*_\epsilon$ and therefore in particular 
under $U(1)_\epsilon$. The $Z^I$ are defined on the open set where $X^0\neq 0$. 
Note that $Z^0=1$, so that there are $n$ independent fields $Z^i$, which 
we will show to be the scalar fields 
in the Lagrangian (\ref{4dLagr}). We remark that 
$X^0, Z^i$ provide local coordinates on $M$.

Using the homogeneity properties of the prepotential 
and the formulae (\ref{homogRewrite}) from section 
\ref{GeneralPrePo}, we can rewrite (\ref{KpotM})
as a function of $X^0$ and $Z^i$:
\begin{equation}
K = - \log \left( 
2 i_\epsilon ({\cal F} - \bar{\cal F}) - i_\epsilon
(Z^i - \bar{Z}^i) ({\cal F}_i + \bar{\cal F}_i) \right)
- \log \left(X^0 \bar{X}^0 \right) \;.
\end{equation}
We now observe that the second term can be removed by a K\"ahler
transformation. Therefore ${M}_{IJ}$ only depends
on the $\bC^*_\epsilon$-invariant variables $Z^i$. 
To obtain the metric $\bar{g}_{ij}$ 
we need to project $M_{IJ}$ onto $\bar{M}$. 
We take the $Z^i$ as coordinates on $\bar{M}$, and
interprete the $X^I$ as functions of the $Z^i$, by 
picking a holomorphic non-vanishing function $h(Z)$ and setting $X^0=h(Z)$.
We can now pull back $M_{IJ}$ to $\bar{M}$, and the
result does not depend on our choice of $h(Z)$, because changing
this function amounts to a K\"ahler transformation.
The resulting scalar Lagrangian is
\begin{equation}
{\bf e}^{-1} {\cal L}_{\mscr{scal}} = 
- \bar{g}_{i {j}} \der_\mu Z^i \der^\mu \bar{Z}^j
\;,\;\;\;
\mbox{where} 
\;\;\;
\bar{g}_{ij} = \frac{\der^2 K}{\der Z^I \der \bar{Z}^j} \;,
\end{equation}
with $K$ given by 
\begin{eqnarray}
K = -\log \left( 
2i_\epsilon ({\cal F} - \bar{\cal F}) - i_\epsilon
(Z^i - \bar{Z^i})({\cal F}_i + \bar{\cal F}_i)
\right) \;.
\;\;\; 
\label{KP}
\end{eqnarray}
This agrees with the scalar term in (\ref{4dLagr}), and completes
the proof that the scalar and gravitational part of the 
Lagrangian (\ref{4dLagr})
is gauge-equivalent to the Lagrangian (\ref{ScLag1}).
Moreover, it is clear
that the signature of $\bar{g}_{ij}$ is the same as the
signature of ${M}_{IJ}$ on the horizontal space. For $\epsilon=-1$,
we have a theory with Lorentzian space-time, and therefore
impose that $\bar{g}_{ij}$ is positive definite. Thus we
need to choose the metric $g$ of $M$ such that it has
signature $(2n,2)$ or $(2,2n)$. For $\epsilon=1$ 
the metrics of both $M$ and $\bar{M}$ necessarily have
split signature. The relevance of this feature will 
become clear when we discuss instanton solutions.


\subsection{Reformulation in terms of line bundles \label{Sections}}

In the previous subsection we presented the field-theoretic
implementation of the projection $\pi:M \rightarrow \bar{M}$
by adapting methods taken from the superconformal calculus. 
The special geometry of vector multiplets can be reformulated
in various ways. One such reformulation, which is frequently used
in the literature, focusses on $\bar{M}$ rather than $M$,
and reinterprets various quantities which we already encountered as
sections of a line bundle over $\bar{M}$ \cite{Strom:90,CasDauFer:90,CRTV:97}. 
We refer the reader
to \cite{AndriEtAl:96} for a detailed review of ${\cal N}=2$ supergravity in 
this formalism. In the following we will
briefly indicate how our results can be expressed
from this alternative point of view. Moreover, we will also provide
a geometrical interpretation for the $U(1)_\epsilon$ gauge field
${\cal A}_{\mu}$ and of the associated covariant derivative
${\cal D}_\mu X^I = \partial_\mu X^I + i_\epsilon {\cal A}_\mu X^I$ introduced 
in subsection \ref{Reform}.

In order to proceed, it is useful to summarise the results
of section \ref{UBdl} in the following diagram:
\[\xymatrixcolsep{5pc} \xymatrixrowsep{5pc}
\xymatrix{
{\cal U}^{N} \ar@<.5ex>[d]  \ar[r]& 
{\cal U}^{\bar{M}} \ar[r] \ar@<.5ex>[d] & 
{\cal U}^M \ar@<.5ex>[d] \ar[r]& 
{\cal U}  \ar[d]^{\pi_U}  
\\
N \ar[r]^{\cal Z} \ar@/_1pc/[rr]_{\cal X}  \ar@<.5ex>[u]^{{\cal X}^* \phi}& 
\bar{M} \ar[r]^s  \ar@/_1pc/[rr]_{\bar{\phi}} \ar@<.5ex>[u]^{s^*\phi} 
& M \ar[r]^{\bar{\phi} \circ \pi =\pi_V \circ
\phi} \ar@<.5ex>[u]^{\phi} & P(V')  
}
\]
Here $\pi_U: {\cal U}\rightarrow P(V')$ is the universal line bundle
introduced in section \ref{UBdl}. Since $M$ and $\bar{M}$
are mapped into $P(V')$ by $\bar{\phi} \circ{\pi}$ and
by $\bar{\phi}$ respectively, one obtaines line bundles
${\cal U}^M$ over $M$ and ${\cal U}^{\bar{M}}$ over $\bar{M}$
by pulling back the universal line bundle. Space-time
$N$ is mapped into $M$ and $\bar{M}$ by ${\cal X}$ and
by ${\cal Z}$, respectively, so that one also obtains
a line bundle ${\cal U}^N$ over $N$. The immersion $\phi: M \rightarrow V'$
can be interpreted as a section of ${\cal U}^M$, and sections
of the line bundles ${\cal U}^{\bar M}$ and ${\cal U}^N$ 
are obtained by pull back. Finally, the universal line bundle
comes equipped with the Chern connection ${\cal D}$ described in 
Lemma \ref{ChernL}. 
Connections on the other lines bundles are
obtained by pull back and are likewise denoted by ${\cal D}$. 
Notice that the canonical 
maps ${\cal U}^N\ra {\cal U}^{\bar{M}}\ra {\cal U}^M \ra {\cal U}$
restrict to isomorphisms on the fibers.  

To make contact with the supergravity formalism, we note that
$\phi$, which can be interpreted as an $\epsilon$-holomorphic
section of ${\cal U}^M$,
takes the following form in terms of special coordinates:
\[
\phi: (X^I) \rightarrow (X^I, F_I(X))  \;.
\]
If we take a non-vanishing 
$\epsilon$-holomorphic section $s:\bar{M}\rightarrow M$
of the $\bC_\e^*$-bundle $\pi: M \rightarrow \bar{M}$, we can pull back
$\phi$ to an $\epsilon$-holomorphic section $s^* \phi$ of 
${\cal U}^{\bar M}$. If $\zeta^a$, $a=1,\ldots, n$
are $\epsilon$-holomorphic coordinates on ${\bar M}$, then
\begin{eqnarray}
s: (\zeta^a) &\rightarrow&  (X^I (\zeta)) \nonumber  \;,\\
s^*\phi: (\zeta^a) &\rightarrow&  (X^I (\zeta), F_I(\zeta)) \;. 
\end{eqnarray}
Finally, this pulls back to a section of ${\cal U}^N$, which
takes the form
\[
(s \circ {\cal Z})^* \phi:
(x^\mu) \rightarrow (X^I (\zeta(x)), F_I(\zeta(x)))   \;,
\]
where $x^\mu$ are coordinates on space-time $N$. 

One particular choice of $\epsilon$-holomorphic coordinates
on $\bar{M}$ are the special coordinates $Z^i=\frac{X^i}{X^0}$.
In the previous subsection we found the expression 
(\ref{KP}) for the $\epsilon$-K\"ahler potential of the metric
$\bar{g}$ of $\bar{M}$ in terms of special coordinates. We also
noted that the $X^I$ could be interpreted as functions on $\bar{M}$
by picking (locally) a smooth non-vanishing function $h$ on 
$\bar{M}$ and
setting   $X^0=h(Z)$. If we take this function
to be $\epsilon$-holomorphic, then $s: (Z^i) \rightarrow X^I(Z)$ is an 
$\epsilon$-holomorphic
section of $\pi: M \rightarrow \bar{M}$, expressed in terms 
of special coordinates. By an $\epsilon$-holomorphic change of
coordinates we can go from the special coordinates $Z^i$ to 
general $\epsilon$-holomorphic coordinates $\zeta^a$. In terms of these
the K\"ahler potential (\ref{KP})  takes the form
\begin{equation}
K = - \log \left( -i_\epsilon ( X^I(\zeta) \bar{F}_I(\bar{\zeta})
- F_I (\zeta) \bar{X}^I(\bar{\zeta}) ) \right) \;,
\label{KPhol}
\end{equation}
where $s^*(\phi): (\zeta^a) \rightarrow (X^I(\zeta), F_I(\zeta))$ is
the $\epsilon$-holomorphic section of $U^{\bar{M}}$, which is 
obtained by pulling back $\phi$ using $s$. This can
be rewritten in a coordinate free way as
\[
K = -\log |\gamma(\phi(s), \phi(s))| \;,
\]
where $\gamma$ is the $\epsilon$-Hermitean form on $V=T^*\bC^{n+1}$. 
Note that the resulting metric $\bar{g}$ on $\bar{M}$, which is given
by
\[
g_{ab} = \frac{\partial^2 K}{\partial \zeta^a \partial \bar{\zeta}^b} \;,
\]
does not depend on the choice of the section $s$. Locally, 
any other non-vanishing $\epsilon$-holomorphic section is
of the form $e^f s$, where $f$ is an $\epsilon$-holomorphic
function. Replacing $s$ by $e^f s$ changes the $\epsilon$-K\"ahler 
potential by a K\"ahler transformation, and therefore the
metric is invariant.


Another useful quantity is 
\begin{equation}
\partial_a K = - \frac{\partial_a X^I(\zeta) \bar{F}_I(\bar{\zeta}) 
- \partial_a F_I(\zeta)  \bar{X}^I (\bar{\zeta})}
{X^I(\zeta) \bar{F}_I(\bar{\zeta})  - F_I(\zeta)  \bar{X}^I(\bar{\zeta})}\;.
\label{HolConnPhys}
\end{equation}
By comparing to (\ref{HolConn}) we see that $\partial_a K = -i_\epsilon A_a^h$, 
where $i_\epsilon A_a^h$ is the connection one-form $i_\epsilon A_a^h$ 
of the Chern connection on ${\cal U}^{\bar M}$, 
evaluated on a holomorphic section. Therefore $\partial_a K$ is the 
connection one-form of the dual connection in the dual bundle with respect
to the dual section $s^*$.  
In terms of coordinates
the equivalent statement is that
\begin{eqnarray}
{\cal D}_a X^I(\zeta) &=& (\partial_a  + (\partial_a K)) X^I(\zeta) \;,
\nonumber \\
{\cal D}_{\overline{a}} X^I(\zeta) &=& \partial_{\overline{a}} X^I(\zeta)  
(=0)  \;, \label{HolCovDer}
\end{eqnarray}
is a covariant derivative with respect to $\epsilon$-holomorphic
transformations $X^I \rightarrow e^f X^I$, where $f$ is an 
$\epsilon$-holomorphic function on $\bar{M}$. Here covariant 
derivative means that ${\cal D}_a X^I$ transforms homogenously, i.e., 
\[
{\cal D}_a X^I \rightarrow e^f {\cal D}_a X^I \;.
\]
The formulae (\ref{KPhol}), (\ref{HolConnPhys}), (\ref{HolCovDer})
are the key formulae for expressing special geometry in terms of 
holomorphic sections of ${\cal U}^{\bar M}$. In particular, note that 
our  expression (\ref{HolCovDer}) for the covariant derivative
on the holomorphic line bundle 
agrees with the formula (4.17) of \cite{AndriEtAl:96}.

Another, closely related reformulation of special geometry 
is obtained by rewriting these formulae in terms of 
unitary sections. Given the holomorphic 
section $\phi: M \rightarrow {\cal U}^M$, we can obtain 
a unitary section $\phi_1$ by normalising it:
\[
\phi_1 = \frac{\phi}{||\phi||}  \;,
\]
where $||\phi|| = \sqrt{|\gamma(\phi,\phi)|}$. Since 
the D-gauge can be expressed as $\gamma(\phi,\phi)=-1$,
the formalism based on unitary sections is closely related 
to the gauged sigma model discussed in subsection \ref{GeneralPrePo}.

In terms of coordinates, a unitary section of 
${\cal U}^{\bar M}$ is obtained from the holomorphic section 
$(\zeta^a) \rightarrow (X^I(\zeta), F_I(\zeta))$ by
\[
(\zeta^a) \rightarrow (X^I, F_I) \;,
\]
where 
\[
X^I = e^{\frac{1}{2} K} X^I(\zeta) \;,\;\; F_I = e^{\frac{1}{2}K} F_I(\zeta)\;,
\]
and where $K$ is the $\epsilon$-K\"ahler potential.  
Under holomorphic transformations $X^I(\zeta)  \rightarrow e^f X^I(\zeta)$ 
the components of the unitary section transform by a $U(1)_\epsilon$
transformation:
\[
X^I \rightarrow e^{i_\epsilon {\rm Im} f} X^I \;.
\]
Therefore $(X^I,F_I)$ can also be interpreted as a section of
the principal $U(1)_\epsilon$ bundle associated to the line bundle
${\cal U}^{\bar M}$. The covariant derivative
(\ref{HolCovDer}) induces the $U(1)_\epsilon$-covariant derivative
given by 
\begin{eqnarray}
{\cal D}_a X^I = ( \partial_a + (\frac{1}{2} \partial_a K)) X^I \;,
\nonumber \\
{\cal D}_{\overline a} X^I = ( \partial_{\overline{a}} -
(\frac{1}{2} \partial_{\overline{a}} K)) X^I \;.
\label{CovHolSect}
\end{eqnarray}
By comparing to (\ref{UniConn}) we see that,  up to sign, 
$\frac{1}{2} (\partial_a K, -\partial_{\overline{a}}K)$ 
is equal to the connection one-form
$i_\epsilon A_a, i_\epsilon A_{\overline{a}}$ of the 
Chern connection evaluated on a unitary section 
of ${\cal U}^{\bar M}$. Therefore we find that 
$\frac{1}{2} (\partial_a K, -\partial_{\overline{a}}K)$ is again the
connection one-form of the dual connection. 
This shows that the formulation of special geometry in 
terms of unitary sections can be obtained by replacing
holomorphic sections of the pulled back universal bundle
by the corresponding unitary sections. In particular, note that
our formula  (\ref{CovHolSect}) for $U(1)_\epsilon$-covariant
derivatives agrees with the formula (4.15) of 
\cite{AndriEtAl:96}.

Finally, we would like to interprete the $U(1)_{\epsilon}$
gauge field $A_\mu$ of the gauged sigma model discussed
in subsection \ref{GeneralPrePo} within this framework. Since
$A_{\mu}$ is defined on space-time $N$, we need to 
consider the pullback ${\cal U}^N$ of
the universal bundle to space-time $N$. 
Equation
(\ref{EoMU1GF}) expresses $A_{\mu}$ in terms of the pull back
of the section $\phi$ of ${\cal U}^M$ to space-time $N$. 
Imposing the D-gauge amounts
to taking a unitary section, which is equivalent to working
with the associated $U(1)_\epsilon$-principal bundle.
The pull back of the Chern connection to ${\cal U}^N$ 
evaluated on a unitary section 
is given by (\ref{U1connST}). 
Comparing this to (\ref{EoMU1GF}), evaluated in the D-gauge,
we see that the pull back of the Chern connection to ${\cal U}^N$ is dual to 
the $U(1)_\epsilon$-connection 
used in the gauged sigma model.

\section{Scalar solutions of the Euclidean field equations 
\label{Section:InstGeneral}}
In this section we will discuss solutions of ${\cal N}=2$
supergravity coupled to vector multiplets in four dimensions.
The action is completely determined by  the projective special $\e$-K\"ahler
target, which for simplicity from now on is denoted by $(M,g)$ instead of
$(\bar{M},\ol{g})$. 
We will restrict ourselves to solutions where all field strengths and
all fermions are set to zero. The remaining fields are the metric
and the scalar fields. If the action can be obtained from a five-dimensional
action by dimensional reduction over time, then solutions of the
Euclidean action lift to stationary solutions of the five-dimensional
theory which involve the metric, the five-dimensional scalars, and the
electric components of the five-dimensional gauge fields. The use of
dimensional reduction over time as a solution generating technique
dates back to \cite{NeuKra:69}, where it was applied to four-dimensional
Einstein-Maxwell theory. Later, the method was adapted to construct
four-dimensional black hole solutions in Kaluza-Klein theories
\cite{GibBreMai:88}. Then this was extended to $p$-brane 
solutions \cite{CleGal},
and it was realised, as reviewed in \cite{Stelle}, that dimensional 
reduction and lifting provided a viable approach to generating
and classifying solitonic solutions in string theory. More recently 
dimensional reduction over time has been used to explore extremal
black holes (both supersymmetric and non-supersymmetric) 
\cite{GaiLiPadi:08,BCPTvR08,PSVV:08}.

In this section we give a self-contained account of the structure
of the Euclidean field equations of scalars coupled to gravity,
its relation to harmonic maps, and provide an overview of the
classes of solutions which can be constructed through harmonic
maps onto totally geodesic submanifolds of the scalar manifold.
We give a coordinate-free definition of the relevant maps, which 
applies to the case where the totally geodesic submanifold is
totally isotropic, and we analyse one family of symmetric 
spaces in detail. A concrete example chosen form this family
is worked out in the two following sections.
 
After truncating out the gauge fields and the fermions of the four-dimensional
Euclidean theory,
the remaining field equations 
are the Euler-Lagrange equations of the following truncated 
action: 
\begin{eqnarray} \label{truncEqu} S &=& \int d^4x{\cal L}=
\int {\rm dvol}(h)\left( \frac{1}{2}R -\langle df,df\rangle 
\right)\nonumber\\
 &=&
\int d^4x\sqrt{|\det h |}\left(\frac{1}{2}R(h) -
\sum g_{ab}\partial_\mu \zeta^a\partial^\mu
\ol{\zeta}^b\right),
\end{eqnarray}
where $R=R(h)$ stands for the scalar curvature of the  space-time metric $h$
and the projective special $\e$-K\"ahler metric $g=(g_{ab})$ is 
evaluated along the map $f: N \ra M$.
$\zeta^a$ are holomorphic coordinates on $M$. 
\bp \label{fieldProp} 
The   Euler-Lagrange equations of (\ref{truncEqu}) are given by
the harmonic map equation for $f$ 
$$ {\rm tr} Ddf = 0$$
and the Einstein equation 
$$Ric-\frac{1}{2}Rh = T,\quad  
T= 2f^*g -\langle df,df\rangle h,$$
where $D$ is the covariant derivative
induced by the Levi-Civita connections of the source and target
manifolds of $f: N\ra M$. 
\ep 
In components, the harmonic map equations reads 
$$ \Delta_h\zeta^a + \sum \Gamma^a_{bc}\partial_\mu \zeta^b\partial^\mu 
\zeta^c =0$$
and the energy momentum tensor  
$$T_{\mu \nu} = \frac{-2}{\sqrt{|\det h |}}
\frac{\delta {\cal L}}{\delta h^{\mu \nu }}=2\sum 
g_{ab}\partial_\mu \zeta^a\partial_\nu\ol{\zeta}^b -
h_{\mu \nu}\sum g_{ab}\partial_\mu \zeta^a\partial^\mu
\ol{\zeta}^b.$$

\subsection{Analysis of the field equations}
The harmonic map equation can be simplified if the target manifold
possesses totally geodesic submanifolds. Let $\iota: M'
\rightarrow (M,D)$ be an embedding of $M'$ into $M$, where $M$
is equipped with a connection $D$. \bd The embedding $\iota: M'
\rightarrow (M,D)$ is called {\cmssl 
totally geodesic} if for any two vector fields $X,Y$ which are tangent
to $M'$ the covariant derivative $D_X Y$ is again tangent 
to $M'$.\ed In this case the embedded submanifold $M'$ is called 
totally geodesic. 
Let $X_1, \ldots, X_n$ be a local frame for $M$ defined 
on a neighbourhood of a point $p\in M'$, such that 
the restriction of the vector fields $X_1, \ldots, X_m$ to $M'$
is a local frame for $M'$. Here $m$ and $n$ are the 
dimensions of $M'$ and $M$, respectively. 

Then $M'$ is totally geodesic
if the equation 
\[
D_{X_i} X_j = \sum_{k=1}^m \Gamma_{ij}^k X_k 
\]
holds along $M'$ for all $i,j \in \{1, \ldots, m\}$. 
If $M'$ is a totally geodesic submanifold,
then the connection $D$ on $M$ induces a connection $D$ on $M'$
such that $Dd\iota =0$. This can be verified by noting that
in terms of the local frame $X_i$ the differential of $\iota$ takes
the form
\[
d \iota = \sum_{i=1}^m X^*_i \otimes X_i
\]
where $X_i^*$ is the dual frame. Using the relation between
the connection coefficients of the connection $D$ on $TM'$ 
and the 
dual connection on $T^*M'$, we find
\[
D_{X_j} d \iota = \sum_{i,k=1}^m \left[ (- \Gamma_{ji}^k X_k^*) \otimes X_i 
+ X_i^* \otimes \Gamma_{ji}^k X_k \right] = 0 \;.
\]
If $(M,g)$ is pseudo-Riemannian with Levi-Civita connection $D$
and if $M'$ is a non-degenerate submanifold,
then the induced connection $D$ on $M'$ coincides with the Levi-Civita
connection of the induced metric $g_{|M'}$. 
Note that we have formulated the
notion of totally geodesic embedding in sufficient generality 
in order to include isotropic submanifolds.

\bd
A smooth map
$f:N \rightarrow M$ from a pseudo-Riemannian manifold $(N,h)$
to a manifold $M$ endowed with a connection $D$ is called
{\cmssl harmonic}, if it satisfies the harmonic map equation
\[
\mbox{tr} Ddf= \sum_i \varepsilon_i (D_{e_i} df) (e_i)  = 0 \;,
\] 
where $D$ stands for the connection on $T^*N \otimes
f^* TM$ induced by the Levi-Civita connection on $N$ and the
connection $D$ on $M$, and the summation is over an orthonormal
basis, such that $h(e_i, e_i) = \varepsilon_i$.     
\ed

\bp
Let $\iota : M'\ra M$ be a totally geodesic embedding. Then a map 
$\varphi : N \ra M'$ is harmonic if and only if $f=\iota \circ \varphi :
N \ra M$ is harmonic. \ep
\pf To see this we first note that the chain rule
implies that
\[
d f = d(\iota \circ \varphi) = d \iota \circ d \varphi \;.
\]
Given that $\iota$ is totally geodesic, the connection $D$ of $M$
and the Levi-Civita connection of $N$ 
induce connections on $T^*N \otimes f^* TM$, $T^*N \otimes \varphi^* TM'$
and $\varphi^* T^* M' \otimes f^* TM$, which we also denote by $D$, and 
which are compatible with the composition of maps between the
underlying manifolds: 
\[
D df = D(d\iota) \circ d \varphi + d \iota \circ D d \varphi =
d \iota \circ D d \varphi \;,
\]
since $\iota$ is totally geodesic.

This implies 
\[
\mbox{tr} D df = d \iota \left(  \mbox{tr} D d \varphi \right) \;.
\]
which, by the injectivity of $d\iota$, shows that 
$f: N\rightarrow M$ is harmonic if and only if 
$\varphi: N \rightarrow M'$ is harmonic.
\qed

This means that we can reduce the problem
of solving the harmonic map equation for $f:N \ra M$ to the
following two problems:
\begin{enumerate}
\item Find all totally geodesic embeddings $\iota : M' \subset M$. 
\item Solve the harmonic map equation for $\varphi : N \ra M'$.
\end{enumerate}
For instance, any totally geodesic embedding $\iota : N\ra M$
defines a particular solution with $M'=N$ and $\varphi = id$.  
Another special case is to consider flat totally geodesic submanifolds 
$M' \subset M$. In that case the harmonic map equation for
$\varphi : N \ra M'$ reduces to a system of linear equations
for the components of $\varphi$ with respect to affine 
coordinates $\sigma^{a}$, $a=1,\ldots, m$, on $M'$:    
$$\Delta_h \sigma^{a}=0.$$ 

In the simplest case, the projective special $\e$-K\"ahler manifold $M$  
is a pseudo-Riemannian symmetric space $M=G/K$. For instance, we
can take $G=G_1\times G_2={\rm SL}_2(\bR )\times {\rm SO}_0(p+1,q+1)$, $K=
K_1\times K_2= {\rm SO}_0(1,1) 
\times {\rm SO}_0(1,1) \times {\rm SO}_0(p,q)$, $M=M_1\times M_2 = G_1/K_1
\times G_2/K_2$. 
For any  symmetric space we have 
a so-called symmetric decomposition 
$$\mathfrak{g}=
\mathfrak{k}+\mathfrak{m},\quad [\mathfrak{k},\mathfrak{k}]\subset 
\mathfrak{k},\quad [\mathfrak{k},\mathfrak{m}]\subset \mathfrak{m},\quad
 [\mathfrak{m},\mathfrak{m}]\subset \mathfrak{k},$$
where $\mathfrak{g}=Lie\, G$, $\mathfrak{k}=Lie\, K$ and the 
subspace 
$\mathfrak{m}\subset \mathfrak{g}$ is complementary to  $\mathfrak{k}$. 
The pseudo-Riemannian metric of $M=G/K$ is completely determined by 
an $Ad_K$-invariant scalar product on $\mathfrak{m}\cong T_oM$, where 
$o=eK$ is the canonical base point. The corresponding curvature tensor 
is given by 
\begin{equation}
\label{CurvatureEquation}
R(X,Y)=-ad_{[X,Y]}: \mathfrak{m} \rightarrow \mathfrak{m} \;.
\end{equation}

There is a one-to-one correspondence between (complete) totally geodesic 
submanifolds $M'\subset M$ and {\it Lie triple systems}, that is 
subspaces $\mathfrak{m}' \subset \mathfrak{m}$
such that 
$$[[\mathfrak{m}',\mathfrak{m}'],\mathfrak{m}']\subset \mathfrak{m}'.$$
Putting $\mathfrak{k}':=[\mathfrak{m}',\mathfrak{m}']$ one can easily check
that 
\be \label{g'Equ}\mathfrak{g}':=\mathfrak{k}'+\mathfrak{m}'
\subset \mathfrak{g}\ee
is a 
Lie subalgebra and that (\ref{g'Equ}) is again a symmetric decomposition.
The corresponding symmetric submanifold $M'=G'/K' \subset M=G/K$ is
totally geodesic.  The induced connection of $M'$ coincides 
with the Levi-Civita connection of the induced metric, provided that 
the restriction of the metric of $M$ to $M'$ is nondegenerate.  
$M'$ is flat with respect to the induced connection 
if and only if 
\be \label{flatEqu} [[\mathfrak{m}',\mathfrak{m}'],\mathfrak{m}']=0,\ee
as follows from (\ref{CurvatureEquation}). 
The latter statement holds even for isotropic submanifolds.
For Riemannian symmetric spaces (that is those with a positive definite
metric) the condition (\ref{flatEqu}) is equivalent to 
$$[\mathfrak{m}',\mathfrak{m}']=0.$$ 
In that case $G'=M'$ is an Abelian Liegroup. 

We have the following examples of totally geodesic submanifolds
of $M_2:=\frac{{\rm SO}_0(p+1,q+1)}{{\rm SO}_0(1,1)\times {\rm SO}_0(p,q)}$:
$$\frac{{\rm SO}_0(p'+1,q'+1)}{{\rm SO}_0(1,1)\times {\rm SO}_0(p',q')},\quad
\frac{{\rm SO}_0(p',q'+1)}{{\rm SO}_0(p',q')}\times 
\frac{{\rm SO}_0(p''+1,q'')}{{\rm SO}_0(p'',q'')},$$
where $p'+p''\le p$ and $q'+q''\le q$.  
In particular, 
$$\frac{{\rm SO}_0(p,q+1)}{{\rm SO}_0(p,q)}\quad \mbox{and}\quad 
\frac{{\rm SO}_0(p+1,q)}{{\rm SO}_0(p,q)}$$
are maximal totally geodesic submanifolds of non-zero constant curvature of
$M_2$ 
and we have a totally geodesic Riemannian sphere $S^r\subset M_2$ 
and hyperbolic space  $H^r\subset M_2$ of maximal dimension
$r=\max (p,q)$. 

A flat Lorentzian totally geodesic surface $M'\subset M_2 \subset M$ 
is given by 
$$\mathfrak{m'}= {\rm span}\{ e_1'\otimes e_1'',e_2'\otimes e_2''\}
\subset \mathfrak{m}_2 = E' \otimes E'' \;,$$ 
where $(e_1',e_2')$ is an orthonormal basis of $E'=\bR^{1,1}$
and  $(e_1'',e_2'')$ is an orthonormal basis of a two-dimensional
nondegenerate subspace of $E''=\bR^{p,q}$.

A flat totally isotropic and totally geodesic submanifold of $M_2$ of
maximal dimension is associated to the Lie triple system
$$\mathfrak{m'}= e'\otimes E'',$$
where $e'\in E'$ is a non-zero null vector.

Similarly, a flat totally isotropic and totally geodesic curve 
$M' \subset M_1 = {\rm SO}_0(1,2)/{\rm SO}_0(1,1)\subset M$ is given by
\[
\mathfrak{m}' =
e' \otimes E'' \subset 
\mathfrak{m}_1 = E' \otimes E'' = \bR^{1,1} \otimes \bR^{0,1} 
\]
where $e'$ is a non-zero null vector in $E'=\bR^{1,1}$ and
$E'' = \bR^{0,1}$.
The example discussed in the next section is of this type.

Next we analyse the Einstein equation
\begin{equation}
\label{Einstein}
2f^*g =Ric-\frac{1}{2}Rh+\langle df,df\rangle h.
\end{equation}
In two dimensions $Ric - \frac{1}{2} Rh=0$, and the Einstein
equation reduces to the statement that $f$ is conformal with
conformal factor $\frac{1}{2} \langle df \;, df \rangle$.
If the dimension of $N$ is $n\not=2$, and 
under the assumption that $h$ is an Einstein metric, i.e.
$Ric=\frac{R}{n}h$, (\ref{Einstein}) simplifies to
\be \label{EinsteinEqu}f^*g = \frac{1}{2}\left( \langle df,df\rangle 
-  \frac{n-2}{2n}  R \right) h.\ee 
\bp Let $(N,h)$ be an  
Einstein manifold of dimension $n>2$ and $f$ a solution of 
(\ref{EinsteinEqu}). 
Then either
\begin{enumerate}
\item $\langle df,df\rangle = 
\frac{n-2}{2n} R$, in which case  
$Ric=0$ and $f$ is totally isotropic or
\item  $\langle df,df\rangle \neq \frac{n-2}{2n} R$ and $f$ is a conformal
immersion with conformal factor \linebreak
$\frac{1}{2}(\langle df,df\rangle -\frac{n-2}{2n} R)$. 
\end{enumerate}
\ep 

\pf 1. This follows from  $f^*g=0 \Longrightarrow 
\langle df,df\rangle = {\rm tr}_h f^*g=0\Longrightarrow R=0\Longrightarrow 
Ric=0$.\\
2. Equation (\ref{EinsteinEqu}) shows that $f^*g$ is nondegenerate,
hence that $f$ is an immersion.   \qed


\section{Instanton solutions of the Euclidean 
STU model \label{Sect:InstExamples}}

In this section we consider explicit instanton solutions
for a particular choice of the prepotential  
in detail. This does not only
illustrate the general results of the previous section, but
also allows us to discuss various physical 
properties of Euclidean actions and their instanton solutions.

\subsection{The Euclidean STU model}

The model which we consider is the Euclidean version of the
so-called STU model.\footnote{Part of our results on the Euclidean STU model
were reported already in the proceedings contribution \cite{MohProc08}.
The Euclidean STU model has also been studied in \cite{BCPTvR08}.} 
This is a model with three vector multiplets
which arises from dimensional reduction of the heterotic string
on $K3 \times T^2$.\footnote{This model also has hypermultiplets, which
are not relevant for the following discussion.} We only consider
the classical limit of this model, which contains the leading
(tree-level) part in both the expansion in the string coupling
$g_S$ and in the string scale $\sqrt{\alpha'}$. The corresponding
prepotential is of the very special form (\ref{VerySpecialF})
and 
can be obtained by starting with the effective Lagrangian of 
the compactification on $K3 \times S^1$ and reducing further on a 
circle. We arrive at the Euclidean STU-model by taking this
circle to be time-like.

The prepotential of the STU model is obtained by setting 
$c_{123}=-C_{123} =-1$ in (\ref{VerySpecialF}), while all
other independent $C_{ijk}$ vanish. Following conventions used
in the supergravity literature, we parametrise the scalar fields as follows:
\begin{equation}
S =  \epsilon i_\epsilon z^1 \;,\;\;\;
T =  \epsilon i_\epsilon z^2 \;,\;\;\;
U =  \epsilon i_\epsilon z^3 \;.
\end{equation}
The resulting $\epsilon$-K\"ahler potential takes the form
\begin{equation}
K =- \log \left( (S + \bar{S}) (T + \bar{T}) (U + \bar{U})
\right) \;.
\end{equation}
For space-like compactifications this is a K\"ahler potential
for the projective special K\"ahler manifold
\begin{equation}
{M}_{(\epsilon = -1)} =
\left( \frac{ SU(1,1) }{U(1)}  \right)^3 = 
 \left( \frac{ SL(2,\mathbb{R}) }{SO(2)}  \right)^3 \;.
\end{equation}
For time-like compactifications this becomes the projective 
special para-K\"ahler manifold
\begin{equation}
{M}_{(\epsilon = 1)} =
 \left( \frac{ SL(2,\mathbb{R}) }{SO_0(1,1)}  \right)^3 \;.
\end{equation}
In the notation of section \ref{Section:InstGeneral}, this is of the form
$M_1 \times M_2$, with $M_1 = SL(2,\bR)/SO_0(1,1)$ and
$M_2=SO_0(2,2)/(SO_0(1,1) \times SO_0(1,1)) \simeq
SL(2,\bR)/SO_0(1,1) \times SL(2,\bR)/SO_0(1,1)$.  

Since the scalar manifold factorises, we can focus on a single
factor $SL(2,\bR)/SO(2)$ or 
$SL(2,\bR)/SO_0(1,1)$. This is parametrised by one $\epsilon$-complex 
scalar field, which we take to be the field $S$
for definiteness. The corresponding 
sigma-model takes the form
\begin{equation}
\label{SigmaSflat}
{\bf e}^{-1} {\cal L}_S = 
- g_{S \bar{S}} \der_\mu S \der^\mu \bar{S} = 
- \frac{ \der_\mu S \der^\mu \bar{S}}{(S + \bar{S})^2} \;.
\end{equation}
For space-like compactifications we immediately recognize that 
the sigma model metric is proportional to the Poincar\'e metric 
on the upper half plane by setting $\tau = iS$.

It will turn out to be useful to 
decompose $S$ into its real and imaginary part. The real part of $S$ must 
be non-vanishing, and choosing it to be positive we set:
\begin{equation}
S = e^{-2 \phi} + i_\epsilon a \;,
\end{equation}
where $\phi$ and $a$ are real scalar fields. In heterotic string theory
the field $S$ is the four-dimensional complex dilaton.
Its real part is related to the four-dimensional heterotic string
coupling $g_S$ by
\begin{equation}
e^{\langle \phi \rangle} = g_S \;,
\end{equation}
where $\langle \phi \rangle$ is the vacuum expectation value
of the real dilaton $\phi$.
The Lagrangian (\ref{SigmaSflat}) is invariant under shifts 
in the imaginary part $a$, which 
is called the universal string axion. This shift symmetry 
is preserved in perturbation theory, but broken by non-perturbative
corrections. We will see explicitly that instanton solutions 
break the continuous shift symmetry to a discrete one.
The permutation symmetry between  
the three $\epsilon$-complex scalar fields $S$, $T$ and $U$ 
is already broken by perturbative corrections.
This implies that in the full
theory the relation of the  field $S$ to the string coupling 
is unambigous.

For later use we 
rewrite the sigma model Lagrangian for $S$ in terms of the real fields:
\begin{equation}\label{realfieldsLagr}
{\bf e}^{-1} {\cal L}_S = 
-\der_\mu \phi \der^\mu \phi - (-\epsilon)
\frac{1}{4} e^{4 \phi} \der_\mu a \der^\mu a \;.
\end{equation}

\subsection{Instantons in the scalar picture \label{Sec:InstScalarPict}}

We would like to find instanton solutions of the same type
as the ten-dimensional IIB D-instanton \cite{GibGrePer:95}
and the hypermultiplet
instantons in type-II Calabi-Yau compactifications 
\cite{BehGaiLueMahMoh:97,Beckers:99,GutSpa:00,VanEtAl1}.
As solutions of the bosonic field equations, such instantons
are characterised by the property that the scalar fields
have a non-trivial profile, while the gauge fields vanish
and the metric is flat  (in the Einstein frame\footnote{This is the frame where
the Einstein Hilbert term takes its `usual' form, as in the
previous sections. Other frames, such as the so-called string frame
are obtained by conformal rescalings of the metric, with the
conformal factor being a function of the scalar fields (usually
the dilaton). We will discuss such other frames later on.}). 
Moreover, they have four Killing spinors and preserve $\frac{1}{2}$
of the Euclidean supersymmetry. 

In this paper we have focussed on the bosonic part of the theory,
and we did not derive the Euclidean supersymmetry
transformations. However, the supersymmetry transformations
for rigid Euclidean vector multiplets have been derived 
in \cite{CMMS}, and one can check that for purely scalar
backgrounds with a flat Einstein frame metric the conditions
for the existence of Killing spinors are the same for rigidly
and for locally supersymmetric vector multiplets. 
In the following we use the formalim of \cite{CMMS}, take
the supersymmetry parameters to be symplectic Majorana spinors,
and  work with 
para-complex linear combinations of spinors. In this notation,
the condition for a purely scalar field configuration to be
invariant under Euclidean supersymmetry is 
\begin{equation}
\gamma^m \partial_m Z^i (\epsilon^a + i e \gamma^0 \epsilon^a) =0 \;,
\end{equation}
where $Z^i$ are the para-complex scalar fields corresponding 
to special coordinates, $\epsilon^a$ are the supersymmetry 
transformation parameters, and $a=1,2$ is the $SU(2)_R$ 
index.\footnote{One can verify that this condition is
related by dimensional lifting with respect to time
to the Killing spinor equations of
\cite{ChamSab},
which characterise
supersymmetric static black holes in five dimensions.}
When taking the $\epsilon^a$ to be eigenvectors of $i\gamma^0$,
$i\gamma^0 \epsilon^a = \pm \epsilon^a$, then field configurations
of the form
\begin{equation}
\partial_m \mbox{Re} Z^i = \pm \partial_m \mbox{Im} Z^i
\end{equation}
are $\frac{1}{2}$-BPS, i.e. they admit four independent 
Killing spinors. These field configurations are `isotropic'
in the sense that the scalar fields vary along an isotropic
submanifold $M'$, and we will see below this condition
implies that the energy-momentum tensor vanishes, which 
makes the assumption of a flat space-time metric consistent. 
Furthermore, the `bulk' action (\ref{4dLagr}) vanishes when evaluated
on such solutions, thus raising the question of how to obtain 
a non-vanishing instanton action. We will come back to this question
later. In the following we will restrict ourselves to solutions involving
one para-complex scalar field. A discussion of more general 
solutions will be given in \cite{MohWai}.

The Lagrangian 
(\ref{4dLagr}) can be truncated consistently 
by setting all gauge field strengths to zero and
two of the scalar fields, say $T$ and $U$, to constant values.
To get a consistent solution with a flat space-time metric
we must impose that the energy-momentum tensor vanishes. Since
only the field $S$ is non-trivial, the relevant part of the
energy momentum tensor is: 
\begin{equation}
T_{\mu \nu}^{(S)} = -\frac{2}{{\bf e}} \frac{\delta {\cal L}}{\delta
h^{\mu \nu}} =
2 \der_\mu \phi \der_\nu \phi - \epsilon \frac{1}{2} e^{4 \phi} \der_\mu a
\der_\nu a  -  h_{\mu \nu} 
\left( \partial_\alpha \phi \partial^\alpha \phi
- \epsilon \frac{1}{4} e^{4 \phi} \partial_\alpha a \partial^\alpha a \right)\;.
\label{CondFlat}
\end{equation}
Now we take (\ref{realfieldsLagr}) with $\epsilon=1$,
set $h_{\mu \nu} =\delta_{\mu \nu}$, and obtain the following
flat-space Euclidean scalar action for the dilaton:
\begin{equation}
S_{(0,4)}^{(\rm indef)}[\phi,a] = \int d^4 x \left( \der_\mu \phi \der^\mu \phi 
- \frac{1}{4} e^{4 \phi} \der_\mu a \der^\mu a \right) \;.
\label{action1}
\end{equation}
For later convenience we have taken the Euclidean action to 
be {\em minus} the integral of the Euclidean Lagrangian.
In the following we use a notation for actions which 
specifies the space-time signature ($(0,4)$ for Euclidean space,
$(1,3)$ for Minkowski space) and whether the action is 
positive definite or indefinite.\footnote{In Euclidean signature
the label definite/indefinite refers to the action itself, 
in Minkowski signature it refers to the kinetic terms (the terms
quadratic in the time derivatives).} The relation between the various
actions which we consider in the following is summarised in 
Figure \ref{Dia1}.

The equations of motion obtained by variation of (\ref{action1}) are:
\begin{eqnarray}
\Delta \phi &=& - \frac{1}{2} e^{4 \phi} \der_\mu a \der^\mu a \;,
\label{EoMphi} \\
\Delta a & =& - 4  \der_\mu \phi \der^\mu a  \;.
\label{EoMa} 
\end{eqnarray}
Here $\Delta$ is the four-dimensional Laplace operator.
Solutions of these equations are only solutions of the full theory
defined by (\ref{4dLagr}) if we impose 
the vanishing of (\ref{CondFlat}) as a constraint:
\begin{equation}
\label{CondFlat2}
\left. 
T_{\mu \nu}^{(S)}
\right|_{h_{\mu \nu} = \delta_{\mu \nu}} =
2\der_\mu \phi \der_\nu \phi - \frac{1}{2} e^{4 \phi} \der_\mu a
\der_\nu a  - \delta_{\mu \nu} 
\left( \partial_\alpha \phi \partial^\alpha \phi
-  \frac{1}{4} e^{4 \phi} \partial_\alpha a \partial^\alpha a 
\right) = 0\;.
\end{equation}
Similar constraints appear in the literature on extremal 
black hole solutions, where they are usually referred to as 
Hamiltonian constraints. 
Equation (\ref{CondFlat2}) is equivalent to
\be
\der_\mu \phi = \pm \frac{1}{2} e^{2 \phi} \der_\mu a \,,
\label{instanton_ansatz}
\ee
where we take the same sign for all $\mu$.
To see that (\ref{CondFlat2}) implies (\ref{instanton_ansatz}),
one takes the trace of $T_{\mu \nu}^{(S)}$ to show that 
$\partial_\alpha \phi \partial^\alpha \phi - \frac{1}{4}
e^{4 \phi} \partial_\alpha a \partial^\alpha a=0$, which implies 
that  $\der_\mu \phi \der_\nu \phi - \frac{1}{4} e^{4 \phi} \der_\mu a
\der_\nu a =0$ for all $\mu, \nu$. This shows that the four 
two-component vectors
$(\partial_\mu \phi, \partial_\mu a)$ are isotropic an colinear.

We refer to (\ref{instanton_ansatz}) as the instanton ansatz. 
Since $S=e^{-2\phi} + e a$, the instanton ansatz implies the Euclidean
 $\frac{1}{2}$-BPS condition 
\[
\partial_m \mbox{Re} S = \pm \partial_m \mbox{Im} S \;,
\]
and the resulting field configurations are supersymmetric.

Note that the instanton 
ansatz does not work in Minkowski signature, $\epsilon=-1$.
In this case one would have to set 
$\der_\mu \phi = \pm \frac{i}{2} e^{2 \phi} \der_\mu a$, both 
to obtain a vanishing energy-momentum tensor, and to
have a supersymmetric field configuration.
For real fields $\phi$ and $a$ this forces one to set all scalars to 
constant values, resulting in a vacuum solution.
Instanton solutions of the
type considered here require target spaces of indefinite signature,
which allow non-constant scalar supersymmetric field configurations with 
vanishing energy-momentum tensor. The indefinite signature of
the target space is an automatic consequence of Euclidean supersymmetry.
More precisely, the existence of an action which is invariant under
Euclidean supersymmetry transformations requires for vector multiplets
that the target space is special para-K\"ahler and hence has
indefinite signature \cite{CMMS}.
The indefiniteness of the Euclidean action is an unusual feature,
which we will further investigate below. We now continue with 
solving the field equations.

Given that we impose the instanton ansatz, the system 
(\ref{EoMphi}), (\ref{EoMa}) is reduced to
\be
\Delta \phi + 2 \der_\mu \phi \der^\mu \phi = 0 \;,
\ee
which is equivalent to
\be
\Delta e^{2 \phi} = 0 \;.
\label{Solphi}
\ee
Thus by imposing the instanton ansatz and performing the field 
redefinition $\phi \rightarrow e^{2\phi}$, 
we have reduced the non-linear harmonic map equation
to an ordinary harmonic equation on $\mathbbm{R}^4$. 
This corresponds to the fact that $e^{2\phi}$ is the affine
coordinate on the null geodesic $M'\subset M$. 
This solution
illustrates one of the cases discussed
in section \ref{Section:InstGeneral}, namely harmonic maps into flat
totally isotropic and totally geodesic submanifolds of $M_1\subset M$.

We have seen that the field
$e^{2 \phi}$ must be harmonic, while $a$ is
fixed in terms of $\phi$ up to an integration constant.
Single-instanton solution are obtained by further imposing
spherical symmetry, which implies
\be
e^{2\phi} = e^{2\phi_\infty} + \frac{C}{r^2} \;.
\label{single_instanton}
\ee
Here we use four-dimensional spherical coordinates,
with $r$ as the radial variable. The string coupling
at infinity $g_S = e^{\phi_\infty}$ can take any value
$0\leq g_S < \infty$. 
To obtain solutions where the real part of the field $S=e^{-2\phi} + e a$ 
is positive for positive $r$, we need
to impose that the constant $C$ is non-negative. A vanishing $C$ corresponds
to the trivial special case where the field $S$ is constant.
We will  see later that $C$ is proportional to the absolute value of 
the instanton charge.  Multi-instanton solutions are 
obtained by choosing multi-centered harmonic functions. 

In the single-centred case, the axion
takes the following form:
\be
a = \mp e^{-2 \phi} + D = \mp \left( e^{2 \phi_{\infty}}
+ \frac{C}{r^2} \right)^{-1} + D \;.
\label{Sola}
\ee
We will argue later that the integration constant
$D$ should be chosen to be zero.

The solution (\ref{single_instanton})
is singular at $r=0$ which we interpret as
the position of a source for the field $S$. In string theory pointlike
objects localised in space and (Euclidean) time are called
$(-1)$-branes. The most prominent example is 
the interpretation of the D-instanton of IIB supergravity as
a D-$(-1)$-brane in type-IIB string theory \cite{GibGrePer:95}.
While the geometry is flat in the Einstein frame, it 
takes the form of a wormhole in the string frame:
\begin{equation}
\label{semi-infinite1}
ds^2_{\rm String} = e^{2 \phi} ds^2_{\rm Einstein} =
\left( e^{2\phi_\infty} + \frac{C}{r^2} \right) \delta_{\mu \nu}
dx^\mu dx^\nu =
\left( e^{2\phi_\infty} + \frac{C}{r^2} \right) 
\left( dr^2 + r^2 d\Omega_{(3)}^2 \right)
 \;.
\end{equation}
This is a semi-infinite wormhole with a throat approaching
a finite size for $r \rightarrow 0$. The asymptotic three-sphere
at $r \rightarrow 0$ has radius $R= \sqrt{C}$ and volume
$2 \pi^2 C^{3/2}$. 
In contrast, the ten-dimensional
D-instanton is a finite-neck wormhole, which approaches flat space
for $r\rightarrow 0$ and has a minimal size for an intermediate 
`critical' value of $r$, which corresponds to the fixed point set of the
discrete isometry which exchanges the two asymptotic regimes. 
This difference between the four-dimensional and the ten-dimensioal case 
has nothing to do with the dimensionality but
is caused by the different coupling of the 
axion to the dilaton. In four dimensions we could obtain
a finite neck wormhole by replacing $e^{4\phi}$ by
$e^{2\phi}$ in the Lagrangian (\ref{realfieldsLagr})
\cite{BehGaiLueMahMoh:97}. 
Instanton solutions supported by hypermultiplet scalars
involve axions with both types of couplings to the dilaton, and
the corresponding wormholes 
can be finite(-neck), semi-infinite or have a more complicated structure
\cite{BehGaiLueMahMoh:97,VanEtAl1}.

Let us now point out some remarkable features of the instanton solution 
(\ref{instanton_ansatz}, \ref{Solphi}, \ref{Sola}) and of 
the underlying Euclidean action (\ref{action1}).
\begin{itemize}
\item
For an instanton we expect that the action is non-zero and 
proportional to $\frac{1}{g_S^2}$
(or proportional to $\frac{1}{g_S}$ for D-instantons). However, if we evaluate
the action (\ref{action1}) on the instanton solution, we get zero. 
\item
The Euclidean action (\ref{action1}) is 
indefinite: while the kinetic term for $\phi$ is 
positive definite, the kinetic term for $a$ has a relative minus
sign and is negative definite. This is necessary for the existence
of scalar instanton solutions, since it allows the energy momentum
tensor to vanish on a non-trivial scalar field configuration.
But it also implies that the action is not bounded from 
below, so that the functional integral measure defined by
$\exp(-S[\phi,a]_E)$ is not damped.\footnote{The Euclidean Einstein-Hilbert
action exhibits the same feature. This is known as the 
`conformal factor problem', and we refer to \cite{Kiefer}
for a discussion of the problem and proposals of its solution.
Leaving the Einstein-Hilbert term aside, one expects 
that the matter action is positive definite, as this seems to 
be required for a well-defined functional integral in the limit
where gravity is decoupled.}
\item
The Euclidean action (\ref{action1}), and, more generally, the
scalar part of (\ref{4dLagr}), is different from the Euclidean
action obtained by a Wick rotation of the corresponding Lorentzian
action. Both differ by an analytic continuation in field space.
Restricting our attention to the case of a single scalar field $S$,
the Wick rotation of the Lorentzian version of (\ref{SigmaSflat})
yields:
\begin{equation}
S^{\rm (def)}_{(0,4)}[\phi,a] 
= \int d^4 x \left(\der_\mu \phi \der^\mu \phi 
+\frac{1}{4} e^{4 \phi} \der_\mu a \der^\mu a \right) \;.
\label{action1a}
\end{equation}
This action is positive definite.\footnote{We take the Wick rotation 
to be $t\rightarrow - it$. The Minkowskian action $S$ and the
rotated action $S_{\rm Wick}$ are related by 
$i \left. S \right|_{t \rightarrow -it} =- S_{\rm Wick}$. With this
convention Minkowski signature matter actions continue into 
positive definite Euclidean actions.} 

To obtain the action (\ref{action1}) 
one needs to combine the Wick rotation with the
analytic continuation $a \rightarrow ia$ of the axion. 
For more complicated
target space geometries one has to perform an analytic 
continuation of all the axionic scalars. 
\end{itemize}
These observations give rise to the question whether
the `correct' Euclidean action is the indefinite action
(\ref{action1}) or the positive definite action (\ref{action1a})
with its standard, positive definite scalar kinetic term.
The answer depends on which properties of the Euclidean action
we decide to insist on. Note that
the Euclidean action obtained by Wick rotation also has some 
undesirable features:
\begin{itemize}
\item
The instanton solution (\ref{instanton_ansatz}, \ref{Solphi}, \ref{Sola})
is not a solution of the field equations of the Wick rotated
action (\ref{action1a}). This is clear, because the 
energy-momentum tensor obtained from the definite 
Euclidean action (\ref{action1a})
has the same form as in Minkowski signature, namely 
\[
T_{\mu \nu}^{(S)} = \partial_\mu \phi \partial_\nu \phi + \frac{1}{4}
e^{4\phi} \partial_\mu a\partial_\nu a - \frac{1}{2} \delta_{\mu \nu}
\left( \partial_\alpha \phi \partial^\alpha \phi + \frac{1}{4}
e^{4\phi} \partial_\alpha a \partial^\alpha a \right) \;.
\]
Then $T_{\mu \nu}=0$ cannot be achieved when   
$\phi$ and $a$ are (non-constant) real fields. 
In other words, the instanton can only
be realised as  a complex rather than a real saddle point of the Wick rotated
action. 
\item
The action (\ref{action1a}) cannot be extended to an action 
invariant under Euclidean supersymmetry. The dimensional
reduction from five Lorentzian to four Euclidean dimensions
preserves supersymmetry
and leads to a scalar sigma model with split signature. 
In the rigid case it was shown that the split signature
and para-complex (rather than complex) structure of the scalar
manifold is determined by the subgroup $SO_0(1,1)$ of the R-symmetry 
group of the Euclidean supersymmetry algebra \cite{CMMS}. 
The same reasoning applies to the supergravity case.
\end{itemize}

The difference between (\ref{action1}) and (\ref{action1a})
illustrates the general fact that 
dimensional reduction over space followed by Wick rotation is
different from dimensional reduction over time.
Similarly, 
Wick rotation and (Hodge-)dualisation do not commute.
This brings into play a 
third type of Euclidean action, which can be obtained by 
dualising the axion field $a$ into a two-form gauge field
$B_{\mu \nu}$. We will see that this leads to a Euclidean 
action for $\phi$ and $B_{\mu \nu}$ which is positive definite and has 
instanton solutions.

\subsection{Instantons in the scalar-tensor picture}

We start with the following Euclidean action:
\begin{equation}
S_{(0,4)}^{\rm (def)}[\phi, B] = \int d^4 x \left( \der_\mu \phi \der^\mu \phi 
+ \frac{1}{2 \cdot 3!} e^{-4 \phi} H_{\mu \alpha \beta} 
H^{\mu \alpha \beta} \right)
\;.
\label{action2}
\end{equation}
This action can be obtained in two ways. One way is to start
from (\ref{action1}) and to dualise the axion field $a$ into
an antisymmetric tensor field $B_{\mu \nu}$. We will 
investigate  the relation between 
(\ref{action2}) and (\ref{action1}) in detail below. 
The second way to obtain (\ref{action2}) is to start from an 
$N=2$ vector-tensor multiplet, to truncate it to the two
fields $\phi$ and $B_{\mu \nu}$, and then to perform a 
Wick rotation. 

Since any supersymmetric
string theory contains the ten-dimensional metric $G_{MN}$,
dilaton $\Phi$ and tensor field $B_{MN}$, the dimensionally
reduced theory always contains the four-dimensional metric
$g_{\mu \nu}$, dilaton $\phi$ and tensor field $B_{\mu \nu}$.
In four dimensions $B_{\mu \nu}$ can be dualised into the
universal axion $a$. However, there are subtleties when 
one wants 
to perform this dualisation while preserving off-shell
$N=2$ supersymmetry. One expects that the vector multiplet
containing the dilaton $\phi$ and axion $a$ can be dualised into
an $N=2$ vector-tensor supermultiplet containing 
$\phi$ and $B_{\mu \nu}$ \cite{deWKapLouLue95}. 
But though an off-shell description for vector-tensor multiplets
is known, vector-tensor multiplets are only dual to vector multiplets when 
certain conditions are met \cite{Sieb,ClausEtAl}. The off-shell
dualisation of the dilaton vector multiplets is not possible if the 
prepotential depends linearly on the dilaton.
Under dualisation,
the off-shell dilaton vector multiplet mixes with the gravitational
multiplet, which prevents one from identifying a dual off-shell 
vector-tensor multiplet.
However, one can at least identify an
on-shell heterotic dilaton vector-tensor multiplet when 
going to the Einstein frame. This is the vector-tensor multiplet
we take as our starting point. More precisely we take
the string frame Lagrangian (5.40) of \cite{Sieb}, 
transform it to the Einstein frame, truncate it to the two 
fields $\phi$ and $B_{\mu \nu}$, and perform a Wick rotation.
Modulo constant rescalings, the result is 
(\ref{action2}). Later we will dualise this part of the action
back into an action involving two scalars.

The action (\ref{action2}) is positive definite and therefore 
$\exp(-S[\phi,B])$ 
could be used
to define a functional measure which is damped. We will now
find instanton solutions of (\ref{action2}), and then,
by dualising (\ref{action2}) into (\ref{action1}) we will show
that these instantons are identical to the ones found in section
\ref{Sec:InstScalarPict}.

Since we want the solution to be consistent with a flat Euclidean space-time
metric,
we need to impose that the energy-momentum tensor vanishes when evaluated
on the 
solution. Therefore we compute the energy-momentum tensor\footnote{
This is done by re-installing the space-time metric and varying it.}:
\begin{equation}
T_{\mu \nu} = 
2\der_\mu \phi \der_\nu \phi + \frac{1}{2} e^{-4\phi}
H_{\mu \alpha \beta} H_{\nu}^{\;\;\alpha \beta} 
- \delta_{\mu \nu} \left(
\der_\alpha \phi \der^\alpha \phi + \frac{1}{2 \cdot 3!}
e^{-4 \phi} H_{\alpha \beta \gamma} H^{\alpha \beta \gamma}
\right) \;.
\end{equation}
To obtain a field configuration with $T_{\mu \nu} =0$, 
we make the instanton ansatz
\begin{equation}
H_{\mu \nu \rho} = A e^{2 \phi} \epsilon^\alpha_{\;\;\mu \nu \rho}
\der_\alpha \phi \;,
\label{instanton_ansatz2}
\end{equation}
where $A$ is a real constant.
By a straightforward calculation we find 
\[
T_{\mu \nu} = \left(1 - \frac{1}{2}A^2\right) 
\left( \partial_\mu \phi \partial_\nu \phi - \frac{1}{2}
\delta_{\mu \nu} \partial_\alpha \phi \partial^\alpha \phi \right) \;.
\]
This implies that 
$T_{\mu \nu}=0$ if we choose $A^2 =2$, i.e. $A=\pm \sqrt{2}$. 

The equations of motion resulting from the action (\ref{action2}) are
\begin{eqnarray}
\Delta \phi + \frac{1}{3!} e^{- 4 \phi} H_{\mu \nu \rho}
H^{\mu \nu \rho} &=& 0 \;, \label{EoMphi2} \\
\der^\mu \left( e^{-4 \phi} H_{\mu \nu \rho} \right) &=& 0 \;.
\label{EoMH}
\end{eqnarray}
The equation (\ref{EoMH}) for $H_{\mu \nu \rho}$ is 
satisfied identically if we impose the ansatz (\ref{instanton_ansatz2}).
The equation (\ref{EoMphi2}) leads to the condition
\begin{equation}
\label{HarmonicE2phi}
\Delta e^{2 \phi} = 0 \;,
\end{equation}
which is identical to the equation (\ref{Solphi}) that we found in 
the scalar picture. Note that  (\ref{HarmonicE2phi}) follows
already from the instanton ansatz (\ref{instanton_ansatz2}),
because the tensor field $H_{\mu \nu \rho}$ must satisfy the
Bianchi identity $\epsilon^{\sigma \mu \nu \rho} 
\partial_{\sigma} H_{\mu \nu \rho} =0$. Substitution of the
instanton ansatz (\ref{instanton_ansatz2}) into the Bianchi
identity implies (\ref{HarmonicE2phi}) due to the identity
\[
\epsilon^{\sigma \mu \nu \rho} \epsilon_{\alpha \mu \nu \rho}
= 3! \delta^\sigma_\alpha \;.
\]
Conversely, by the same identity, (\ref{HarmonicE2phi}) and
the instanton ansatz (\ref{instanton_ansatz2}) imply the
Bianchi identity for $H_{\mu \nu \rho}$. The fact that
upon imposing the instanton ansatz an equation of motion
becomes equivalent to an Bianchi identity is analogous
to Yang-Mills instantons. Also note that the instanton 
ansatz (\ref{instanton_ansatz2}) can be viewed as a 
variant of the (anti-)self-duality constraint of Yang-Mills
instantons. Apparently, these analogies between scalar
instantons and Yang-Mills instantons become manifest
in the scalar-tensor picture, because $B_{\mu \nu}$ 
is a gauge field.

To find explicit solutions for $H_{\mu \nu \rho}$ one can choose
any harmonic function for $e^{2\phi}$ and inserts the result
into (\ref{instanton_ansatz2}). This fixes 
$B_{\mu \nu}$ up to a closed two-form. Later, we will compare
this to the solution (\ref{Sola}) for the axion $a$.

We now compute the instanton action by inserting the scalar-tensor
instanton 
solution back into (\ref{action2}). For any field
configurations satisfying the instanton ansatz (\ref{instanton_ansatz2})
we have
\begin{equation}
S = 2 \int d^4 x \der_\mu \phi \der^\mu \phi \;,
\end{equation}
i.e. the contributions of the two terms in the action (\ref{action2})
are equal. We can express $\der_\mu\phi$ in terms of $e^{2\phi}$:
\begin{equation}
\der_\mu \phi = \frac{1}{2} e^{-2\phi} \der_\mu e^{2 \phi} \;.
\end{equation}
Since we evaluate the action on instanton configurations we can
use that $\Delta e^{2 \phi} = 0$:\footnote{More precisely, we 
only require this for $r>0$ and admit a source term at $r=0$.
As we will see below the boundary at $r=0$ does not
contribute to the integral.}
\begin{equation}
S[\phi,B]_{\rm inst.} 
= 2 \int d^4 x \frac{1}{4} e^{-4 \phi} \der_\mu e^{2 \phi}
\der^\mu e^{2 \phi} 
= \frac{1}{2} \int d^4 x e^{-4 \phi} \der_\mu \left( e^{2 \phi}
\der^\mu e^{2 \phi}  \right) \;. 
\end{equation}
This is a total derivative, up to terms which vanish for 
$\Delta e^{2\phi}=0$:
\begin{equation}
\label{InstAction2}
S[\phi,B]_{\rm inst.} = - \frac{1}{2} \int d^4 x \der_\mu \left(
e^{-2 \phi} \der^\mu e^{2 \phi} \right) \;.
\end{equation} 
We then use Stoke's theorem to write this as an integral over the
boundary of the integration region
\begin{equation}
\label{InstAction3}
S = - \frac{1}{2} \oint d^3 \Sigma_\mu e^{-2 \phi}
\der^\mu e^{2 \phi}  = - \oint d^3 \Sigma_\mu \partial^\mu \phi \;.
\end{equation}
We evaluate this expression 
on a single instanton solution (\ref{single_instanton}).
Since the solution is singular at $r=0$, the 
integration region is $\mathbb{R}^4 
- \{0 \}$, and the boundaries can be taken to be  asymptotic three-spheres
$S^3_r$ with $r\rightarrow \infty$ and $r \rightarrow 0$. For $r\rightarrow
\infty$ the solution approaches a ground states, because $\phi$ goes to the 
constant value $\phi_\infty$. Since $e^\phi$ is the (field-dependent)
heterotic string coupling, $e^{\phi_\infty}$ is the (constant)
value $g_S$ of the heterotic string coupling in this ground state.
With the specified boundaries, the instanton action is
\begin{equation}
S[\phi,B]_{\rm inst.} = - \frac{1}{2} \lim_{R \rightarrow \infty} \oint_{S^3_R}
d^3 \Omega \; r^3 e^{-2 \phi} \der_r e^{2 \phi} 
+ \frac{1}{2} \lim_{R' \rightarrow 0} \oint_{S^3_{R'}}
d^3 \Omega \; r^3 e^{-2 \phi} \der_r e^{2 \phi} \;.
\end{equation}
We compute
\begin{equation}
r^3 e^{-2 \phi} \der_r e^{2 \phi}  = 
\frac{ - 2 C }{e^{2 \phi_\infty} + \frac{C}{r^2}} =
\frac{ -2 C r^2}{e^{2 \phi_\infty} r^2 + C}\;.
\end{equation}
This approaches a constant value for $r\rightarrow \infty$, but
vanishes for $r\rightarrow 0$. The resulting instanton action is
\begin{equation}
S[\phi,B]_{\rm inst.} 
= - \frac{1}{2} \Omega_3 \lim_{r \rightarrow \infty}
\frac{-2C}{e^{2 \phi_\infty} + \frac{C}{r^2}} =
\Omega_3 C e^{-2 \phi_\infty} \;,
\end{equation}
where $\Omega_3 = 2\pi^2$ is the volume of the unit three-sphere.
Using the relation between $\phi$ and the heterotic string 
coupling, we see the typical dependence of an instanton
action on the coupling:
\begin{equation}
S[\phi,B]_{\rm inst.} \sim \frac{1}{g_S^2} \;.
\end{equation}
In fact the factor of proportionality is proportional
to the absolute value of the instanton charge. To 
define the instanton charge, remember that the Bianchi
identity $\epsilon^{\mu \nu \rho \sigma} \partial_{\mu} H_{\nu \rho
\sigma} =0$ for the field strength $H_{\mu \nu \rho}$ is violated
in the presence of magnetic sources. 
The magnetic current is
\[
j = \epsilon^{\mu \nu \rho \sigma} \partial_{\mu} H_{\nu \rho \sigma}
\]
and the associated conserved charge is obtained by integrating $j$ over 
the full Euclidean space. As usual for gauge theories, this 
charge can be rewritten as a surface charge, because the current
$j$ is a total derivative:
\begin{equation}
j = \partial_\mu Q^\mu \;,\;\;\;
Q^\mu = \epsilon^{\mu \nu \rho \sigma} H_{\nu \rho \sigma} \;.
\label{TopCurrent}
\end{equation}
We define the instanton charge to be proportional to the
 magnetic charge obtained
by integrating the magnetic current, and include a conventional
factor for later convenience:
\begin{equation}
\label{TopCharge}
Q_{\rm inst} = \frac{1}{\sqrt{2} \,3!} \int d^4 x j =
\frac{1}{\sqrt{2} \,3!} \oint d^3 \Sigma_\mu \epsilon^{\mu \nu \rho
\sigma} H_{\nu \rho \sigma} \;.
\end{equation}
We used Stoke's theorem to rewrite the volume integral as a surface
integral, where the surface encloses all the magnetic charges. 
For multicentered harmonic functions, $j$ is a linear combination of delta 
functions concentrated at the centers. Here we restrict ourselves
to single-centered instanton solutions. Using 
(\ref{instanton_ansatz2}) we obtain:
\begin{equation}
Q_{\rm inst} = \pm \frac{1}{2}
\oint
d^3 \Sigma_\mu \der^\mu e^{2\phi} \;.
\end{equation}
The sign depends on the choice of the constant $A=\pm \sqrt{2}$
in (\ref{instanton_ansatz2}). Let us take a single-instanton solution
(\ref{single_instanton}) and choose the surface to be the three-sphere 
of radius $R>0$, centered at the singularity of the harmonic function:
\begin{equation}
Q_{\rm inst} = \pm \frac{1}{2} \oint_{S^3_R} d^3 \Sigma_{\mu}
\partial^\mu e^{2\phi} = \pm \frac{1}{2} \Omega_3 R^3 \left( \partial_r 
e^{2\phi} \right)_{r=R} = \mp \Omega_3 C \;.
\end{equation}
Remember that the constant $C$ must be positive, because we require
that instanton solutions are regular outside $r=0$. The constant
$C$ is thus proportional to the instanton charge. Instantons
with $Q_{\rm inst}>0$ correspond to taking $A=-\sqrt{2}$, while 
anti-instantons, i.e. solutions with $Q_{\rm inst}<0$ correspond
to taking $A=\sqrt{2}$.

As in the case of type-IIB D-instanton, there are also dual 
solutions which carry electric charge with respect to $H_{\mu \nu \rho}$.
This electric charge is related to the Noether current associated
with the abelian two-form gauge symmetry 
$B_{\mu \nu} \rightarrow B_{\mu \nu} + \der_{[\mu} A_{\nu]}$. By 
Dirac quantisation generalised to $p$-form gauge fields
the allowed spectrum of charges is therefore
discrete. The sources of magnetic $B$-charge have a 
zero-dimensional Euclidean world volume and
are therefore ($-1$)-branes. Their electric duals have a 
two-dimensional Euclidean world volume. To keep terminology
consistent with using the term ($-1$)-branes for zero-dimensional
Euclidean world volume, they should be called 1-branes.
The analogous objects in ten-dimensional type-IIB  
string theory are D7-branes.

Using the instanton charge, we can now express the instanton 
action as:
\begin{equation}
S[\phi, B]_{\rm inst.} = \frac{|Q_{\rm inst.}|}{g_S^2}
\;.
\end{equation}
Next, we show that instanton solutions have minimal action
for given charge. This is done by deriving a Bogomol'nyi 
bound. 
The action (\ref{action2}) is bounded from below by zero, and it 
can be re-written as the sum of a perfect square 
and a remainder:
\begin{equation}
S_{(0,4)}^{\rm (def)}[\phi,B] = \int d^4 x \left(
\der_\mu \phi \pm \frac{1}{\sqrt{2}\cdot 3!} e^{-2 \phi}
\epsilon_{\mu \nu \rho \sigma} H^{\nu \rho \sigma} \right)^2 
\mp 2 \int d^4 x \frac{1}{\sqrt{2}3!} 
\der_\mu \phi e^{-2\phi}
\epsilon^{\mu \nu \rho \sigma} H_{\nu \rho \sigma} \;.
\end{equation}
The perfect square vanishes, if and only if we impose the 
instanton ansatz (\ref{instanton_ansatz2}). 
\begin{equation}
S_{(0,4)}^{\rm (def)}[\phi,B] \geq 
\mp 2 \int d^4 x \frac{1}{\sqrt{2}3!} 
\der_\mu \phi e^{-2\phi}
\epsilon^{\mu \nu \rho \sigma} H_{\nu \rho \sigma} 
= 2 \int d^4x \der_\mu \phi \der^\mu \phi \;.
\end{equation}
This observation provides an
alternative  way of deriving the instanton ansatz, instead of
requiring $T_{\mu \nu}=0$ or Euclidean supersymmetry. As noticed
above, the instanton ansatz implies, when combined with the
Bianchi identity, already the equations of motion. 
Hence 
\begin{equation}
S[\phi, B]_{\rm inst} = 
\frac{|Q_{\rm inst}|}{g_S^2} 
\geq 0  \;, 
\end{equation}
which shows explicitly that instantons solutions have minimal 
action for given charge.

Let us summarise the properties of the scalar-tensor instanton
(\ref{instanton_ansatz2}), (\ref{Solphi})
and of the underlying Euclidean action (\ref{action2}):
\begin{itemize}
\item
The action is positive definite. 
\item
The instanton is a solution of the field equations, with 
finite, minimal action $\frac{|Q_{\rm inst}|}{g_S^2}$. 
\end{itemize}

We close this section by relating our results to the literature. 
The ten-dimensional D-instanton can also be obtained from 
a scalar-tensor Lagrangian \cite{GibGrePer:95}. 
The main difference is that the 
instanton action is proportional to $g_S^{-1}$ rather than
to $g_S^{-2}$. This is due to a different coupling of 
the axion to the dilaton and is related to the different
wormhole geometries obtained in the string frame: finite neck
instantons have an action proportional to $g_S^{-1}$, while
semi-infinite wormholes have action proportional to $g_S^{-2}$.
These remarks also apply to instantons in the
hypermultiplet sector of four-dimensional $N=2$ compactifications 
\cite{BehGaiLueMahMoh:97,VanEtAl1}.

We would also like to mention that the 
bosonic action (\ref{action2})
coincides with the bosonic part of the action of
an $N=1$ tensor multiplet. In other words our 
scalar-tensor instanton solution can be interpreted as
a solution of $N=1$ supergravity, which coincides with
the solution found in \cite{Rey:1991}.

\subsection{Back to the scalar picture}

Let us now dualise the scalar-tensor action (\ref{action2}) and
show that this leads to the scalar action (\ref{action1}), plus
a boundary term accounting for the correct instanton action.
As a by-product we will see that the instanton solutions obtained
from both actions are indeed identical.

The dualisation proceeds in the standard way. First we 
promote the Bianchi identity of $H_{\mu \nu \rho}$
to a field equation by introducing a Lagrange multiplier field
$a$:
\begin{equation}
\label{MasterAction}
\hat{S}[\phi,H,a] = \int d^4 x \left(
\der_\mu \phi \der^\mu \phi +
\frac{1}{2 \cdot 3!} e^{-4 \phi} H_{\mu \nu \rho}
H^{\mu \nu \rho} + \lambda a \epsilon^{\mu \nu \rho \sigma}
\der_\mu H_{\nu \rho \sigma} \right) \;.
\end{equation}
Here $\lambda$ is a {\em real} constant, which we will fix later by
imposing that the axion is normalised in the same way 
as in (\ref{action1}).
The dualisation proceeds by eliminating the field
$H_{\mu \nu \rho}$, which can now be treated as an independent
tensor field, by its equation of motion. This entails that we have
to integrate the third term in the above action by parts. 
Following the analogous analysis of the type-IIB D-instanton
\cite{GibGrePer:95}, we keep the resulting boundary term, 
despite that it does not
contribute to the equations of motion. 

We can now eliminate $H_{\mu \nu \rho}$ by its equation of
motion 
\be
H_{\nu \rho \sigma} = 3! \lambda e^{4 \phi} 
\epsilon_{\mu \nu \rho \sigma} \der^\mu a \;.
\label{Heom}
\ee
Substituting this back into (\ref{MasterAction}), and performing
an integration by parts on the last term we obtain
\begin{equation}
\hat{S}[\phi, a] = \hat{S}_{\rm bulk} [\phi, a] + \hat{S}_{\rm bound}[\phi, a]\;.
\end{equation}
The bulk term 
has the form
\be
\hat{S}[\phi,a]_{\rm bulk} = \int d^4 x \left( 
\der_\mu \phi \der^\mu \phi - \frac{1}{2} (3! \lambda)^2  
e^{4 \phi} \der_\mu a \der^\mu a
\right) \;.
\ee
If we choose $\lambda^2 = \frac{1}{2} \cdot \frac{1}{(3 !)^2}$,
this agrees with (\ref{action1}):
\begin{equation}
\hat{S}[\phi,a]_{\rm bulk}  =  S[\phi,a]_{(0,4)}^{\rm (def)} \;.
\end{equation}
By combining (\ref{Heom}) with (\ref{instanton_ansatz2}) we obtain 
(\ref{instanton_ansatz}). Since we already saw that the condition 
(\ref{Solphi}) on 
$\phi$ is the same for both solutions, it follows that the 
two instanton solutions are the same.

The boundary term of $\hat{S}[\phi,a]$ is
\be
\hat{S}_{\rm bound}[\phi,a] = (3!\lambda)^2
\oint d^3 \Sigma_\mu   a \der^\mu a 
e^{4 \phi} \;.
\label{boundaryaction}
\ee
If we set $(3!\lambda)^2 = \frac{1}{2}$, and evaluate
the boundary term on the instanton solution 
(\ref{Solphi}), (\ref{Sola}) we obtain
\begin{eqnarray}
\hat{S}_{\rm bound}[\phi,a] &=& 
\frac{1}{2} \oint d^3 \Sigma_\mu e^{2\phi} \partial^\mu e^{-2\phi}
\mp \frac{D}{2} \oint d^3 \Sigma_\mu e^{4\phi} \partial^\mu e^{-2\phi} 
\nonumber \\
&=& \hat{S}_{\rm inst.} \pm \frac{\Omega_3 D}{2} \;,
\label{boundaryaction2}
\end{eqnarray}
where $\hat{S}_{\rm inst.}$ is the instanton action,
$\Omega_3$ is the volume of the unit three-sphere, and $D$ is
the integration constant in the solution (\ref{Sola}) for the axion. 
When comparing to (\ref{InstAction2}), ({\ref{InstAction3}),
it is useful to note that
\[
\frac{1}{2} \oint d^3 \Sigma^\mu e^{2\phi} \partial_\mu e^{-2\phi}
= - \oint d^3 \Sigma^\mu \partial_\mu \phi 
= -\frac{1}{2} \oint d^3 \Sigma^\mu e^{-2\phi} \partial_\mu e^{2\phi} \;.
\]
Thus the boundary action gives precisely the instanton action,
provided we set the integration constant $D=0$. We have no other
way of fixing this integration constant, since the axion only
enters into the bulk action and into the equations of motion 
through its first derivatives. Thus there is no obvious 
contradiction in setting $D=0$. When we add the boundary
term to the bulk action (or, in other words, if we keep it
after dualisation), then the improved action 
\[
\hat{S}_{(0,4)}[\phi,a] = 
S[\phi,a]_{(0,4)}^{\rm (indef)} + \hat{S}_{\rm bound}[\phi,a]
\]
agrees with the scalar-tensor action $S[\phi,B]_{(0,4)}^{\rm (def)}$
when evaluated on instanton solutions. However, the improved
action also has one feature which is different from the
scalar-tensor action. Since the boundary term contains 
the axion field $a$ explicitly, the axionic shift symmetry 
is broken, in contrast to the manifest gauge invariance
of the $B$-field in the scalar-tensor action. At the classical
level the breaking of axionic shift invariance by the boundary
term is not an issue, because this term does not contribute
to the equations of motion. The implications on the quantum
theory need to be investigate in a different set-up, e.g., by
the investigation of instanton corrections to quantum transition 
amplitudes. This will be left to future work. Also note
that there are other boundary terms which evaluate to
the correct instanton action but do not break axionic
shift symmetry. Explicit examples will be given when we
consider the dimensional lifting of instanton solutions 
to black holes.

We should also provide an interpretation for the instanton charge
in the scalar picture. Since the tensor field $B_{\mu \nu}$
and the axion $a$ are related by Hodge duality, magnetic
(electric) $B$-charge corresponds to electric (magnetic) 
charge for the $a$-field.
A non-vanishing `electric' charge density with respect
to the axionic shift symmetry $a \rightarrow a + \mbox{const.}$ 
corresponds to adding a source term to the equation of motion 
for $a$:
\begin{equation}
j = \der^\mu \left( e^{4\phi} \der_\mu a \right)  \;.
\end{equation}
For instanton solutions a delta-function type charge density 
is located at the centers of the harmonic functions. 
This density is indeed proportional to the
`magnetic' density associated with the tensor field $B_{\mu \nu}$,
as expected.
The associated charge is obtained by integration over
four-dimensional space. Since the density is a total
derivative, it can be rewritten as a surface charge,
which we can normalise such that it is equal to the
instanton charge  (\ref{TopCharge}):
\begin{eqnarray}
Q_{\rm inst.} & =& \frac{1}{2}\lim_{r \rightarrow \infty} 
\oint_{S^3_r} d^3 \Sigma_\mu e^{4 \phi} \der^\mu a \;.
\end{eqnarray}

\subsection{Discussion of instantons, Euclidean actions and boundary 
terms}

One particular feature of the Euclidean action (\ref{action1})
is that it is indefinite. While this is necessary for the existence 
of instanton solutions, it prevents us from using the expontial
of the action $\exp \left( - S[\phi,a]_{(0,4)}^{\rm (indef)} \right)$
to define a functional measure. Here the natural candidate is
the definite action (\ref{action1a}), which leads to a damped
measure factor $\exp \left( - S[\phi,a]_{(0,4)}^{\rm (def)} \right)$,
but does not have instanton solutions. Thus regarding instanton
corrections at the quantum level, we seem to be stuck with 
two actions which both are deficient. This problem is not unique
to our class of models, but occurs generally if one wants to construct
non-trivial Euclidean finite action solutions involving axionic
scalars. Examples which have been discussed previously in the literature
include scalar field wormholes \cite{BrownEtAl:89}, the D-instanton
solution of type-IIB supergravity \cite{GibGrePer:95}, and 
instanton solutions involving hypermultiplets 
\cite{BehGaiLueMahMoh:97,VanEtAl1,ChiGut:09}.

Since the scalar-tensor action (\ref{action2}) is both positive 
definite and has instanton solutions, one option is to base
the quantum theory exclusively on it. There are several potential
problems with this. One is that the complete theory involves
vector multiplets or vector-tensor multiplets, and, as already 
mentioned there are problems and subtleties with the Hodge dualisation
of the full supermultiplets. Another, more general point is the
question whether and how precisely the duality between axions and
antisymmetric tensors works at the quantum level. This cannot be
answered by just looking at actions, but requires the investigation
of instanton contributions to quantum amplitudes. Studies performed
on similar models in the literature show that boundary conditions
play an important role \cite{BrownEtAl:89,BergshoeffEtAl:05,ChiGut:09}.
A central question is the fate of the
axionic shift symmetry, which corresponds to the gauge symmetry
of the $B$-field under duality. Here we encounter an asymmetry
between the scalar picture and the scalar-tensor picture.
The boundary term generated in the dualisation, which is needed to
obtain the correct instanton action, contains the axion explicitly
and breaks the axionic shift symmetry. The corresponding 
measure factor $\exp\left( - S[\phi,a]_{(0,4)}^{\rm (indef)} - 
\hat{S}_{\rm bound}[\phi,a]\right)$ is still invariant under 
discrete imaginary shifts. In contrast, the corresponding 
gauge symmetry in the scalar-tensor picture cannot be broken.
The general expectation is that instanton effects break the
continuous axionic shift symmetry to a discrete subset, 
and it is not obvious how this can be expressed in the 
scalar-tensor picture. 
Therefore, a better understanding
of the scalar picture and of its relation to the scalar-tensor
picture is required. Note that it is not completely clear to us whether
the boundary term found by dualisation is responsible for
the expected breaking of axionic shift symmetries in the quantum 
theory. As we will see later, one can motivate other boundary terms,
which provide the correct instanton action, but do not break
the axionic shift symmetry. Within the classical realm, we are not
aware of a criterion which could allow us to single out one of
these candidate boundary terms as the correct one.

While the investigation of quantum amplitudes is left to
future work, we can already make a few observations. The two
scalar actions are related by analytic continuation of 
the axion, but so far we have only related the indefinite
scalar action directly to the scalar-tensor action. Let us
now display the dualised action, including the boundary term,
without fixing the parameter $\lambda$:
\begin{eqnarray}
\hat{S}[\phi, a] &=& \int d^4 x \left(
\partial_\mu \phi \partial^\mu \phi - \frac{1}{2}
e^{4\phi} (3!\lambda)^2 \partial_\mu a \partial^\mu a \right)
\nonumber \\
&+& (3!\lambda)^2 \oint d^3 \Sigma_\mu \left( a \partial^\mu a e^{4\phi}
\right) \;.
\end{eqnarray}
Here it is manifest that 
for real $\lambda$ with 
$(3! \lambda)^2 = \frac{1}{2}$ the bulk terms equals the indefinite
scalar action
(\ref{action1}), while for imaginary $\lambda$ with 
$(3! \lambda)^2 = -\frac{1}{2}$ we obtain the definite scalar action
(\ref{action1a}). Thus we can either preserve the saddle points
of the scalar-tensor action by choosing $\lambda$ real,
or preserve its definiteness by choosing $\lambda$ imaginary,
but not both. The choice of an imaginary Lagrange multiplier
is unconventional from the classical point of view, because
it does not preserve the equations of motion, but natural within
the context of Euclidean functional integrals, because it corresponds
to implementing the Bianchi identity for the $B$-field through
a functional delta function. It is a particular feature of Euclidean
signature that definiteness and saddle points cannot be preserved
simultanously. In Minkowski signature a real Lagrange multiplier
preserves both properties, and corresponds to implementing the
Bianchi identity through a functional delta function.

Thus the definite scalar action seems to be correct choice
for defining the quantum theory dual to the scalar-tensor theory.
While the instanton is not a saddle point in the strict sense,
it can be regarded as a complex saddle point, and there 
are several examples of path integrals and functional integrals
in quantum mechanics and quantum field theory which are dominated
by complex saddle points \cite{BrownEtAl:89,BergshoeffEtAl:05}. 
In this interpretation both scalar
actions play a role: the definite action defines the measure, 
the indefinite action identifies the saddle point. In fact,
it is convenient to regard $a$ as a complex field, and 
to view the two real scalar actions as arising from a single
complex scalar action. Note that not only the bulk actions 
but also the boundary actions obtained for real and imaginary
$\lambda$ respectively, are related by the analytic continuation
$a \rightarrow ia$. The boundary term is needed to obtain
the correct instanton action, irrespective of whether we
work with the definite or the indefinite real action.

Since $a$ and
$\phi$ are in the same supermultiplet, we could 
also promote $\phi$ (and all the other fields which 
have been truncated out) to complex fields, and view
different real Euclidean actions as different real forms
of one underlying complex `master action'. 
In the scalar sector this corresponds to the complexification
of the (pseudo-)Riemannian target space, resulting in 
a complex-Riemannian space. From such a `complex point of view'
it is natural to work with complex saddle points.
A necessary and sufficient condition for a pseudo-Riemannian manifold
to admit a complexification is that the manifold and the metric
are \emph{real analytic}. Contrary to complex manifolds, which
are complex analytic and a fortiori real analytic, para-complex
manifolds are not always analytic. Therefore it is not possible 
to obtain every para-K\"ahler manifold as a real section of a 
complex-Riemannian manifold. This implies that a para-K\"ahler
manifold cannot be in general Wick rotated into a (pseudo-)K\"ahler
manifold.
Viewing  target space geometries, which are related by dimensional reduction
over either space or time, or by analytic continuation 
of axions, as different real sections of one underlying 
complex space should lead to a
more unified picture of instantons, solitons and 
other solutions of supergravity theories, since these are often 
related by analytical continuation in either time or
target space. We also expect that the relation between
Minkowskian and Euclidean supersymmetry, and their relation
to the concepts of pseudo- or fake-supersymmetry 
can 
be understood systematically in such a framework.
A similar point of 
view has been taken recently with regards to ten-dimensional
supergravity in \cite{BerEtAl:07}.

\subsection{Summary of the relation between actions}

In this subsection we summarise the properties and mutual
relations between the various actions occuring in this
paper. For concreteness we refer to the truncated version 
of the actions, which contains one scalar together
with one axion, or tensor field or (five-dimensional)
gauge field. However, the same properties and relations
hold between the complete supersymmetric actions (modulo
subtleties with regards to the off-shell dualisation of
vector multiplets into vector-tensor multiplets). 

Figure \ref{Dia1} only involves the actions which we
actually encountered in previous sections. For each action
the fields are specified. All actions contain a scalar
$\phi$ while the second field is either a five-dimensional
gauge field $A=A_{\hat{\mu}}$, or an axion $a$ or an antisymmetric
tensor field $B=B_{mn}$. For each action the space-time
dimension and signature is specified as a lower label. 
For Euclidean actions an additional upper label provides
the information whether the action positive definite or
indefinite. 

The basic operations relating the actions are: 
dimensional reduction/lifting with respect to space or
time, denoted $D_S$, $D_T$, respectively, Wick rotation
between Minkowksi space and Euclidean space, denoted $W$,
and Hodge dualisation between an axion and an antisymmetric
tensor, denoted $H$. As apparent from the diagram, all
actions can be obtained by composing these basic operation.
There are two Euclidean actions involving $\phi$  and $a$,
for two related reasons: (i) in Euclidean signature, 
Wick rotation and Hodge dualisation do not commute, and
(ii) dimensional reduction over space followed by Wick 
rotation gives a result different from reduction over
space. We have also displayed further maps between the
actions, which are equivalent to compositions of the
basic operations $D_S,D_T,H,W$. These are the analytic continuation
of scalars $a\rightarrow ia$, denoted $A$, the modified
Wick rotation $W'$ (which combines analytic continuation 
of time with analytic contiunation of axions), and
the modified Hodge dualisation $H'$, which uses an
imaginary Lagrange multiplier and thus combines 
Hodge dualisation with analytic continuation of axions.
\begin{figure}
\[
\xymatrix{
 & S_{(1,4)}[\phi, A] \ar_{D_S}@{<->}[ld] \ar^{D_T}@{<->}[rd]& \\
S_{(1,3)}[\phi, a] \ar^{W}@{<->}[rd] \ar^{H}@{<->}[dd]  
\ar^{W'}@{<->}[rr]& & 
S_{(0,4)}^{(\rm indef)}[\phi, a] \ar^{A}@{<->}[ld] \ar^{H}@{<->}[dd]\\
 & S_{(0,4)}^{(\rm def)}[\phi, a] \ar^{H'}@{<-->}[rd] & \\
S_{(1,3)}[\phi,B] \ar^{W}@{<->}[rr]& & S_{(0,4)}^{(\rm def)}[\phi, B] \\
}
\]
\caption{This diagram summarises the relations between the
actions occuring in section 9. Further explanations are
given in the text. }
\label{Dia1}
\end{figure}
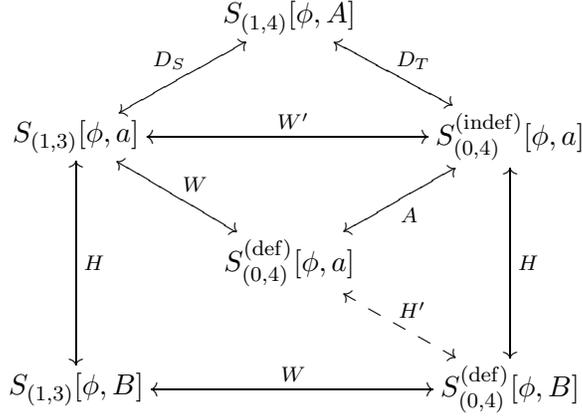
The first diagram is not complete, in the sense that
further actions can be obtained by composing the basic
transformations in different order. For completeness we 
present a second diagram \ref{Dia2} which contains all the eight
four-dimensional actions which can be obtained this way.
In this extended diagram the Minkowski space actions
also carry a label def/indef, which specifies whether the
kinetic terms (the terms quadratic in the first time derivatives)
are definite or indefinite. For actions involving an axion
this label specifies whether the target space metric is
definite or indefinite.
Lorentzian signature actions with
indefinite target space geometries occur in string theory
when performing T-duality transformations along a time-like
direction \cite{IIstar}. A particular example is provided by the
II${}^*$ string theories. 
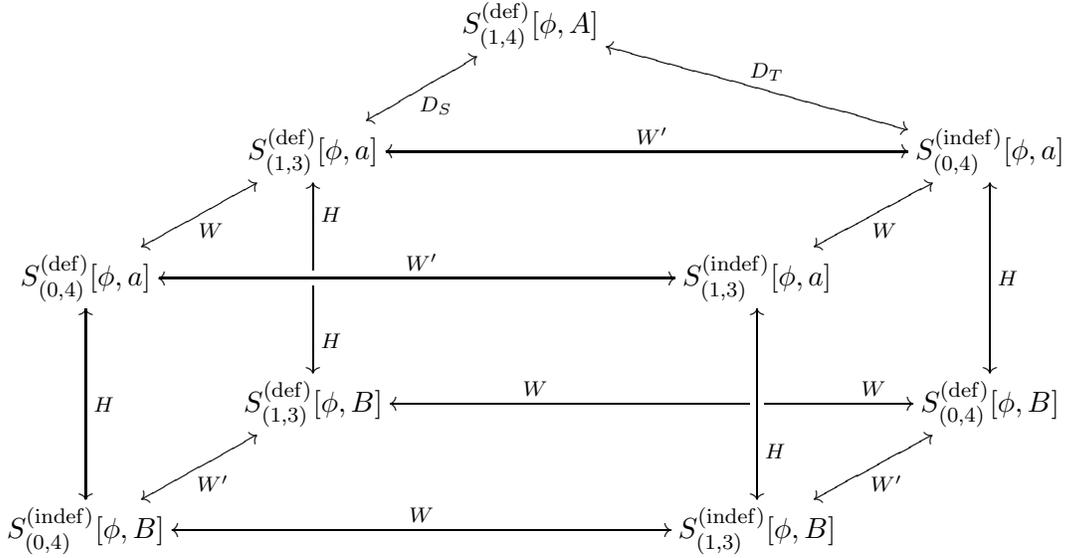
\begin{figure}
\[
\xymatrix{
 && S_{(1,4)}^{(\rm def)}[\phi, A]  \ar^{D_S}@{<->}[ld] \ar^{D_T}@{<->}[rrd]& & \\
&S_{(1,3)}^{(\rm def)}[\phi,a] \ar^{W}@{<->}[ld] \ar^{W'}@{<->}[rrr] 
\ar^{H}@{<->}'[d][dd]& & & 
S_{(0,4)}^{(\rm indef)}[\phi,a] \ar^{W}@{<->}[ld] \ar^{H}@{<->}[dd] \\
S_{(0,4)}^{(\rm def)}[\phi,a] \ar^{H}@{<->}[dd] \ar^{W'}@{<->}[rrr]& & & 
S_{(1,3)}^{(\rm indef)}[\phi,a] \ar^>>>>>>>{H}@{<->}[dd]& \\
& S_{(1,3)}^{(\rm def)}[\phi,B] \ar^{W'}@{<->}[ld] \ar^{W}@{<->}'[rr][rrr] & & & 
S_{(0,4)}^{\rm (def)}[\phi,B] \ar^{W'}@{<->}[ld] \\
S_{(0,4)}^{\rm (indef)}[\phi,B]\ar^{W}@{<->}[rrr] & & & 
S_{(1,3)}^{\rm (indef)}[\phi,B] & \\
}
\] 
\caption{This extended diagram contains all four-dimensional actions which 
can be generated from a given action containing one normal scalar and one 
axion by applying Wick rotations and Hodge dualisations. We have also
included the relation to a five-dimensional scalar-gauge field action 
via dimensional reduction/lifting. Further 
explanations are given in the text.}
\label{Dia2}
\end{figure}
The existence of precisely eight different four-dimensional actions
reflects a three-fold binary alternative: the action can either
contain an axion or an antisymmetric tensor, space-time signature
can be Euclidean or Minkowskian, the action (for Minkowski signature,
its kinetic terms) can be definite or indefinite. From the diagram 
it is clear that all eight theories can be related by using 
Wick roation $W$ and Hodge dualisation $H$. We have also included
the modified Wick rotations $W'$, but not the analytic continuations
$A$ and modified Hodge dualisations $H'$ in order to keep the
diagram transparent. The relation to the five-dimensional
Minkowksi space action has been included. While the diagram is
complete with respect to four-dimensional actions, 
further five- and three-dimensional actions could be 
obtained by applying dimensional reduction to three dimension
and Wick rotations and Hodge dualisations in five and three
dimensions.

\section{Dimensional lifting of  four-dimensional 
instantons \label{Section:5dBHs}}

\subsection{Five-dimensional black holes}

Instantons can be used as generating solutions for
a variety of higher-dimensional solitons. In this section
the one-charge instanton solution obtained 
previously will be lifted to 
five dimensions. We will show that we obtain an extremal
black hole, and that the ADM mass of the black hole
equals the instanton action. Both the ADM mass and the
instanton action are boundary terms, which agree on 
black hole/instanton solutions, and we observe that 
such a boundary term can be generated by transforming
the four-dimensional Einstein-Hilbert term from the
Einstein frame into another conformal frame, which
we call the Kaluza-Klein frame. In this frame, the
metric of the instanton solution agrees with the metric
of the black hole, restricted to a space-like hypersurface.

Since we
know the explicit relation between the five-dimensional 
action (\ref{5Daction}) and
the four-dimensional action (\ref{4dLagr}), it is straightforward 
to lift four-dimensional
instantons to five-dimensional 
space-times. Let us apply this to the
one-charge instanton solution (\ref{Solphi}), (\ref{Sola})
of the Euclidean STU-model. This model
lifts to five-dimensional supergravity coupled to two
vector multiplets, which is a subsector of the effective
field theory of the heterotic string theory compactified on
$K3 \times S^1$. 

The only field excited in the four-dimensional one-charge
instanton is the four-dimensional heterotic 
dilaton
\[
S = \epsilon i_\epsilon z^1 =  \epsilon i_\epsilon (x^1 + i_\epsilon y^1) \;.
\]
According to (\ref{rescalings}),
the relation between the $y^i$ and the five-dimensional scalars $h^i$
is $y^i = 6^{1/3} e^\sigma h^i$, 
while the $x^i$ lift to the temporal components of the 
five-dimensional gauge potentials. 
We can compute the Kaluza-Klein scalar using the constraint 
$c_{ijk} h^i h^j h^k =1$:
\[
y^1 y^2 y^3 =  \frac{1}{6} c_{ijk} y^i y^j y^k =
e^{3 \sigma} \;.
\]
In the one-charge solution $y_2,y_3$ are constant, $y_2 y_3 = B$,
and therefore
\[
e^{3 \sigma} = e^{-2\phi} B \;.
\]
The Kaluza-Klein vector is trivial, and therefore the
four-dimensional Einstein frame metric \\
$ds^2_{\rm Einstein}  = \delta_{\mu \nu} dx^\mu dx^\nu$ 
lifts to the five-dimensional static metric
\begin{equation}
\label{EinsteinFrame}
ds^2_{(5)} = - e^{2 \sigma} dt^2 
+ e^{-\sigma} \delta_{\mu \nu} dx^\mu dx^\nu\;.
\end{equation}
Since this metric is asymtotically flat, 
we impose that it approaches the canonically normalised Minkowski
metric $\eta_{\hat{\mu}\hat{\nu}}$ at infinity. This implies that
the constant $B$ is related to the value of the four-dimensional
dilaton at infinity by
\[
B = e^{2 \phi_\infty} \;.
\]
We can now express the five-dimensional metric in terms of the
four-dimensional dilaton:
\begin{eqnarray}
ds^2_{(5)} &=& 
- e^{-4/3 (\phi-\phi_\infty)} dt^2 +  e^{2/3 (\phi-\phi_\infty)} 
\delta_{\mu \nu} dx^\mu dx^\nu 
\end{eqnarray}
By comparing to \cite{ChamSab}, and using that 
$e^{2\phi}$ is harmonic, we immediately recognize this solution as
a supersymmetric extremal black hole, 
which is charged under a single
$U(1)$. In the single center case we have
\[
e^{-3\sigma} = e^{2 (\phi - \phi_\infty)} 
=
1 + \frac{e^{-2 \phi_\infty} C}{r^2} \;,
\]
and
\begin{equation}
\label{5dBH}
ds^2_{(5)} = - \left( 1 + \frac{e^{-2 \phi_\infty} C}{r^2} \right)^{-2/3}
dt^2 +  \left( 1 + \frac{e^{-2 \phi_\infty} C}{r^2} \right)^{1/3}
\delta_{\mu \nu} dx^\mu dx^\nu \;.
\end{equation}
If we fix a space-like hypersurface by setting  
$t=\mbox{const.}$, we obtain the four-dimensional positive definite
metric
\begin{equation}
\label{semi-infinite2}
ds^2_{t=const.} = \left(1+ \frac{e^{-2\phi_\infty}C}{r^2}
\right)^{1/3} \delta_{\mu \nu} dx^\mu dx^\nu =
\left(1+ \frac{e^{-2\phi_\infty}C}{r^2} \right)^{1/3}
\left( dr^2 + r^2 d \Omega^2_{(3)} \right)\;.
\end{equation}
This is a semi-infinite wormhole akin to (\ref{semi-infinite1}). However,
due to the different power of the harmonic function in front,
the volume of the three-sphere transverse to the throat 
goes to zero in the limit $r\rightarrow 0$. This is as expected,
because a supersymmetric five-dimensional black hole needs
to carry at least three charges in order to have a 
non-vanishing horizon area. Since the semi-infinite wormhole 
(\ref{semi-infinite2})
describes the spatial geometry of a degenerate black hole, we call it a
degenerate semi-infinite wormhole. 

From the four-dimensional point of view the conformal frame 
where we obtain the spatial geometry of the five-dimensional
black hole is neither the Einstein frame 
where four-dimensional
geometry is flat, nor the string frame (\ref{semi-infinite1}).
We call the conformal frame defined by 
(\ref{semi-infinite2}) the Kaluza-Klein frame. Its relation 
to the other two frames is given by
\begin{equation}
\label{KKframe}
ds^2_{\rm KK} = e^{-\sigma} ds^2_{\rm Einstein} = 
e^{-2\phi - \sigma} ds^2_{\rm String} \;.
\end{equation}

So far we have seen that the horizon area of the black hole is
given by the size of the asymptotic three-sphere of the instanton
in the Kaluza-Klein frame.
To extend our instanton--black hole dictionary, we will compare
the ADM mass of the black hole to the instanton action. The ADM
mass measures the flow generated by asymptotic time translations
through an asymptotic sphere at spatial infinity \cite{Poisson}. 
The 
relevant formulae for higher-dimensional black holes can be
found in \cite{MyersPerry}. Let 
\[
ds^2 = - h_{tt} dt^2 + 2 h_{t\mu} dt dx^\mu + h_{\mu \nu} dx^\mu dx^\nu
\]
be the line element of an $(n+1)$-dimensional space-time. We have
chosen a parametrisation where $t=\mbox{const}$ defines a foliation by 
spacelike hypersurfaces, and where the spatial part $h_{\mu \nu}$ 
of the metric approaches the 
flat Euclidean $n$-dimensional metric
$ds^2_{\rm flat} = {\delta}_{\mu \nu} dx^\mu dx^\nu = dr^2 + r^2 d\Omega^2_{n-1}$ 
at infinity, where $d \Omega^2_{n-1}$ is the line
element of the unit $(n-1)$-sphere. We choose one of the spatial
hypersurfaces and denote its asymptotic boundary by $S^{n-1}_\infty$. 
Then the ADM mass is given by 
\begin{equation}
\label{ADMmass}
16 \pi G_N M_{\rm ADM} = \oint_{S^{n-1}_\infty} d\Sigma^\mu \left(
\partial^\nu
h_{\mu \nu} - \partial_\mu 
(\delta^{\rho \sigma}
h_{\rho \sigma})  \right)
:= \lim_{r \rightarrow \infty} \oint_{S^{n-1}_r} d \Sigma^\mu
\left(
\partial^\nu
h_{\mu \nu} - \partial_\mu 
(\delta^{\rho \sigma}
h_{\rho \sigma})  \right) \;,
\end{equation}
where $G_N$ is Newton's constant, 
$d\Sigma^\mu$ is the vectorial volume element of the 
sphere $S^{n-1}_r$.\footnote{This means that 
$d\Sigma^\mu=n^\mu d\mbox{vol}$, where
$n^\mu$ is the Euclidean unit normal 
of $S^{n-1}_r$, and
where $d\mbox{vol}$ is the canonical volume element of 
$S^{n-1}_r$.} It is known that (\ref{ADMmass}) is independent
of the choice of the asymptotically flat coordinate system 
if the scalar curvature of the metric $h_{\mu \nu}$ is 
norm-integrable \cite{Bartnik}.

For the solutions obtained above, the 
spatial metric is conformally flat and takes the form 
\[
h_{\mu \nu} dx^\mu dx^\nu = \left( 1 + \frac{m}{r^{n-2}} + \cdots 
\right) (dr^2 + r^2 d \Omega^2_{n-1})\;,
\]
where $m$ is a constant.  
The  evaluation of the integral
(\ref{ADMmass}) gives
\[
16 \pi G_N M_{\rm ADM} = (n-1)(n-2) \Omega_{n-1} m \;,
\]
where $\Omega_{n-1}$ is the area of the unit $(n-1)$ sphere, which 
reproduces the result of  \cite{MyersPerry}. 

Using that $n=4$
and that  $h_{\mu \nu} dx^\mu dx^\nu = e^{-\sigma} \delta_{\mu \nu} dx^\mu
dx^\nu$, we find
\begin{equation}
16 \pi G_N M_{\rm ADM} =
- 3 \oint_{S^3_\infty} d \Sigma^\mu \partial_\mu e^{- \sigma} 
= \lim_{r \rightarrow \infty}
\oint_{S^3_r} \partial_r \left( 1 + \frac{e^{- 2\phi_\infty}C}{r^2}
\right)^{\frac{1}{3}} r^3 d \Omega^3_{S^3_1} =
\frac{2 A\Omega_3 C}{e^{2\phi_\infty}} \;.
\end{equation}
In the previous sections of this paper we have used units where 
$8 \pi G_N =1$. Since the instanton 
charge satisfies $|Q_{\rm inst}| = \Omega_3 C$, we see that the ADM mass
of the five-dimensional black hole equals the action of the four-dimensional
instanton:
\[
M_{\rm ADM}  = \frac{|Q_{\rm inst}|}{e^{2\phi_\infty}} = S_{\rm inst} \;.
\]

As we discussed previously, the bulk action (\ref{action1})
vanishes when evaluated on the instanton solution. To find
the instanton action, we either need to work in the scalar-tensor
picture, or to add a boundary term. One way to obtain a boundary
term which gives the same instanton action as the scalar-tensor
formulation of the theory is to apply Hodge dualisation. However,
the ADM mass of the lifted solution is an alternative
candidate for the boundary term. Besides the above observation, 
there is a general reason
to expect a relation between the ADM 
mass of a soliton and the action of the instanton obtained
by dimensional reduction. As is well known, $p$-brane
solitons can be obtained from $(p+1)$-branes by double dimensional
reduction, and in this case the respective brane tensions are
related by the volume of the internal dimension. One should expect
that this extends to $0$-branes (solitons) 
and $(-1)$ branes (instantons), where the brane tension is the mass
and the action, respectively.\footnote{When reducing the five-dimensional
action in Section 6, we did not include an explicit parameter for 
the volume of the internal circle. This volume factor, which 
controlls the ratio between the higher- and lower-dimensional
Newton constant, and, hence, sets the ratio between soliton
mass and instanton action, could of course be easily reinstated.}

In order to see that the relation between ADM mass and instanton
action is general rather than accidental, we take the formula which
expresses the ADM mass as a boundary term and re-write it in terms
of the four-dimensional dilaton instead of the Kaluza-Klein scalar:
\begin{equation}
M_{ADM} = - \frac{3}{2} \oint d^3 \Sigma^\mu \partial_\mu e^{-\sigma}
= - \frac{3}{2} \oint d^3 \Sigma^\mu \partial_\mu e^{\frac{2}{3}
(\phi - \phi_\infty)} =
- \oint d^3 \Sigma^\mu   e^{\frac{2}{3}(\phi-\phi_\infty)}  \partial_\mu \phi\;.
\end{equation}
This can now be compared to the boundary term obtained by
Hodge-dualisation of the scalar-tensor action:
\[
S_{\rm bd} = 
- \oint d^3 \Sigma^\mu  \partial_\mu \phi\;.
\]
Both boundary terms are different, but give the same result
whenever the additional factor $e^{\frac{2}{3}(\phi-\phi_\infty)}$
in the ADM boundary term approaches its constant limit value
fast enough. This is in particular the case when $e^{2\phi}$ is harmonic.
If one considers more complicated instanton solutions, which 
involve several scalar fields, the role of the four-dimensional
dilation is played by a particular combination of all scalar
fields, but the fall-off properties of the boundary terms
remain the same, and the relation between ADM mass and 
instanton action is seen to hold generally \cite{MohWai}.

When relating actions by dimensional reduction one usually
neglects boundary terms. This raises the question whether the
boundary term which accounts for the instanton action 
can be obtained by keeping the boundary terms 
occuring in the dimensional reduction of the action. 
In section 6 we have performed the reduction such that we went from the
five-dimensional Einstein frame to the four-dimensional
Einstein frame. More generally, we can use the following
family of parametrisations:
\[
ds^2 = - e^{2\beta \sigma} (dt + A_\mu dx^\mu)^2 + e^{2\alpha \sigma}
g_{\mu \nu} dx^\mu dx^\nu \;.
\]
While the choice $\alpha=\frac{1}{2}$, $\beta=1$ brings us to the
four-dimensional Einstein frame, the alternative choice
$\alpha=0$, $\beta=1$ brings us to the four-dimensional
Kaluza-Klein frame introduced above. The Ricci scalars
corresponding to the two frames are related by (see \cite{Wald},
Appendix D):
\begin{equation}
\label{Rkk}
R_{KK} = e^\sigma \left( R_E + 3 \nabla^\mu \partial_\mu \sigma
- \frac{3}{2} \partial_\mu \sigma \partial^\mu \sigma \right)\;.
\end{equation}
For $\alpha=0$, $\beta=1$, the temporal reduction of the five-dimensional
action 
\[
S_{(1,4)} = \frac{1}{2} \int d^5 x \sqrt{g_{(5)}} ( R_{(5)} + \cdots) 
\]
gives\footnote{Since we define Euclidean actions
with an explicit minus sign, the temporal reduction gives minus the
Euclidean action.}
\begin{eqnarray}
S_{(0,4)}^{KK}&& = - \frac{1}{2} \int d^4 x 
\sqrt{g_{KK}} (e^{\sigma} R_{KK} + \cdots) \nonumber \\
 & & = - \frac{1}{2} \int d^4 x \sqrt{g_E} ( R_E + 3 \nabla^\mu \partial_\mu
\sigma - \frac{3}{2} \partial_\mu \sigma \partial^\mu \sigma  + \cdots ) 
\nonumber\\
&& = S_{(0,4)}^{E} - \frac{3}{2} \oint d^3 \Sigma^\mu \partial_\mu \sigma 
 = S_{(0,4)}^{E} + \oint d^3 \Sigma^\mu \partial_\mu \phi \;.
\end{eqnarray}
Thus the boundary term obtained by transforming from the Kaluza-Klein 
frame to the Einstein frame is precisely the instanton action:
\begin{equation}
\label{ADMfromAction}
M_{ADM} = S_{\rm inst} = S_{(0,4)}^E - S_{(0,4)}^{KK} \;.
\end{equation}
As already noted, the two metrics entering into the ADM formula
can be identified with the four-dimensional Kaluza-Klein 
frame and Einstein frame metrics. This observation 
is interesting, as it relates the ADM mass formula to 
an action. 
Notice that the equation (\ref{Rkk}) shows that 
the boundary term satisfies 
\[
0< \frac{3}{2}\int  d^4x \sqrt{g_E} \nabla^\mu\partial_\mu \sigma =
- \oint d^3 \Sigma^\mu \partial_\mu \phi= M_{ADM} 
\]
if the scalar curvature satisfies $R_{KK}>0$, in accordance with the
relation between positivity of scalar curvature
and positivity of the mass, familiar from the positive mass
theorem.

\subsection{Ten-dimensional Five-branes}

In the context of string compactifications, five-dimensional
supersymmetric black holes can be interpreted in terms
of ten-dimensional components, which are wrapped $p$-branes or
other stringy solitons.\footnote{Here `wrapping' refers to embeddings
of the ($p+1$)-dimensional brane world volume $\Sigma$ 
into a total space-time of the form $N\times K$, where 
$N$ is not compact and interpreted as 
the `dimensionally reduced space-time', where 
$K$ 
is compact and interpreted as internal space, 
and where the image of the world volume is of the form 
$\Sigma_1 \times \Sigma_2 \subset N \times K$. A totally wrapped
brane corresponds to an embedding of $\Sigma$ into $K$.}  
The particular black hole we have 
obtained by lifting the four-dimensional one-charge instanton 
can be further lifted to a ten-dimensional five-brane. 

To see this, remember that 
the string-frame metric of a solitonic five-brane in 
ten dimensions is:
\begin{eqnarray}
ds^2_{\rm String} &=& -dt^2 + (dy^1)^2 + \cdots
+ (dy^5)^2 + H(x) \left( (dx^1)^2 + \cdots (dx^4)^2 \right) \;,
\nonumber \\
e^{2 (\Phi - \Phi_\infty)} &=& H(x) \;,\;\;\;
dB =   \star_4 d H(x) \;, \nonumber \\ 
\Delta_4 H &=& 0 \;.
\end{eqnarray}
Here $\Phi$ is the ten-dimensional dilaton, $B$ the 
universal two-form gauge field, $\star_4$ is the 
Hodge operator with respect to four transverse directions.
All fields are 
given in terms of a function $H(x)$, which is harmonic
in the four transverse coordinates $x^1, \ldots, x^4$. 
This solution only excites fields in the universal 
sector common to  any theory of closed oriented strings and 
exists for both heterotic and type-II
string theories. Dimensional reduction along five
spatial world volume directions results in the  
following five-dimensional string frame metric:
\[
ds_{{\rm String}(5)}^2 
= - dt^2 + H(x) \left( (dx^1)^2 + \cdots (dx^4)^2 \right) \;.
\]
The five-dimensional dilaton equals the ten-dimensional one, 
while the two-form reduces to a gauge potential, under which 
the solution is charged. The relation between the 
the five-dimensional string and Einstein frames is\footnote{
In general, the relation between string frame and Einstein frame
metric is $ds^2_{\rm String} = e^{m\Phi} ds^2_{\rm Einstein}$,
where $m$ is chosen such that $\sqrt{|h_{\rm String|}} e^{-2\Phi}
R_{h_{\rm String}} = \sqrt{|h_{\rm Einstein}|} R_{h_{\rm Einstein}}
+ \cdots$, where the omitted terms do not involve the space-time
curvature. Using the transformation properties of the metric 
under Weyl transformations (see for example \cite{Wald}, Appendix D),
one finds that $m=\frac{4}{d-2}$, where $d>2$ is the dimension of
space-time. } 
$ds^2_{\rm String} = e^{\frac{4}{3} \Phi} ds^2_{\rm Einstein}$.
Using that $e^{2\Phi}=H(x)$ is harmonic, we obtain
\[
ds^2_{{\rm Einstein}(5)} = 
-H^{-\frac{2}{3}} dt^2   + H^{\frac{1}{3}} 
\left( (dx^1)^2 + \cdots (dx^4)^2 \right) \;.
\]
For the single-center case this is precisely the five-dimensional
black hole (\ref{5dBH}), which can therefore be 
lifted to a wrapped five-brane. Further reduction 
along a time-like circle gives the four-dimensional
instanton which can thus be interpreted as 
a five-brane where all six world volume directions 
have been wrapped.

Further details depend on the
string theory into which one embedds the solution. 
Since we constructed instanton solutions in the vector multiplet
sector of an $N=2$ compactification, we need to pick a string
compactification which preserves $N=2$ supersymmetry, and where
the dilaton sits in a vector multiplet. 
This happens for the heterotic string, compactified on 
$K3 \times S^1$ to five dimensions. 
Therefore the microscopic description of the 
four-dimensional instanton (\ref{Solphi}),
(\ref{Sola}) is a completely wrapped heterotic 
five-brane. This observation strongly suggests that the
instanton solutions considered in this paper are the
supergravity approximations of string instantons. 
One difference compared to the string instanton calculus
is that we reduce over time instead of considering 
Euclidean wrappings (which implicitly assumes that the
world volume time has been Wick rotated). 

Other instanton solutions of Euclidean vector multiplets 
will have different microscopic interpretations. 
Consider for example the Euclidean STU-model. Since
this has, at the classical level, a permutation symmetry
between the fields $S$, $T$ and $U$, we can immediately
replace $S$ by any of the other two fields. From the
supergravity point of view this appears to be rather
trivial, but the microscopic interpretation of these
new solutions is completely different. Whereas $S$
is the dilaton, $T$ and $U$ are geometric moduli, and
the solutions do not involve the string coupling. 
Therefore they cannot be space-time instantons, but 
must be world sheet instantons (or more precisely
the effective supergravity description thereof).
The detailed investigation of the microscopic, stringy
aspects of vector multiplet instantons is left to future work.

\section{Outlook}

In this paper we have defined projective special para-K\"ahler
manifolds and shown that they arise as target manifolds for 
the scalars of Euclidean $N=2$ vector multiplets coupled to
gravity. A subset of these theories can be obtained by 
dimensional reduction of five-dimensional vector multiplets
over time, which defines a temporal version of the $r$-map.
To understand the geometry of the scalar sector it was
sufficient to focus on the bosonic sector of the theory.
For rigid vector multiplets the fermionic terms and supersymmetry
transformations rules were found in \cite{CMMS}, and it is
desirable to extend this to the local case in the future. 
To complete the programme of characterising the special
geometries of Euclidean $N=2$ supersymmetry, and relating the various
special geometries by geometric constructions,
which was started
in \cite{CMMS,CMMS2} and continued in this paper, we finally need
to explore para-quaternion-K\"ahler geometry of Euclidean 
hypermultiplets and its relation to projective special 
$\epsilon$-K\"ahler geometry through the $c$-map. 

Potential applications of our work include the systematic 
construction of instanton solutions and the generation of
solitonic solutions through dimensional lifting, which we
have illustrated with a detailed example. General solutions
involving an arbitrary number of scalar fields will be discussed
in \cite{MohWai}. We have also seen
that Euclidean actions and instanton solutions involving 
axionic scalars involve ambiguities and subtleties which deserve
further study. The geometric framework provided by 
\cite{CMMS,CMMS2} and this paper should be useful in this 
respect. Another question, which we only touched upon 
briefly, is the microscopic, `stringy' interpretation of 
Euclidean supergravity solutions.

\subsection*{Acknowledgements}

We would like to thank Gabriel Lopes Cardoso and Ulrich Theis
for various useful discussions. An early version of part of this
work was obtained some years ago in collaboration with Carl Herrmann.
V.C. thanks the Research Center in Mathematics and Modelling and
the Department of Mathematical Sciences of the University of Liverpool
for support and hospitality. 
T.M. thanks the Center for Mathematical Physics of the University of 
Hamburg for support and hospitality during 
several visits in Hamburg. He also thanks the LMU Munich for 
hospitality and the Royal Society for support of this work 
through the Joint Projects Grant `Black Holes, Instantons 
and String Vacua'.


\end{document}